\documentclass[10pt,journal]{IEEEtran}
\usepackage{graphicx}
\usepackage{subfigure}
\usepackage{graphics}
\usepackage{algorithm}
\usepackage{algorithmic}
\usepackage[algo2e,linesnumbered,ruled]{algorithm2e}
\usepackage{amssymb}
\usepackage{amsmath}
\usepackage{amsthm}
\usepackage{multirow}
\usepackage{textcomp}
\usepackage[numbers,sort&compress]{natbib}

\newtheorem{definition}{Definition}

\ifCLASSOPTIONcompsoc
  \usepackage[nocompress]{cite}
\else
  \usepackage{cite}
\fi
\ifCLASSINFOpdf
\else
\fi
\begin{document}
\title{Distributed Continuous Range-Skyline Query Monitoring over the Internet of Mobile Things}

\author{Chuan-Chi~Lai,~\IEEEmembership{Member,~IEEE},
	Zulhaydar~Fairozal~Akbar,
	Chuan-Ming~Liu,~\IEEEmembership{Member,~IEEE},
	Van-Dai~Ta,
	and~Li-Chun~Wang,~\IEEEmembership{Fellow,~IEEE}% <-this % stops a space
	\IEEEcompsocitemizethanks{\IEEEcompsocthanksitem C.-C. Lai and L.-C. Wang are with the Department of Electrical and Computer Engineering, National Chiao Tung University, Hsinchu, Taiwan.  (E-mail: cclai1109@nctu.edu.tw; lichun@g2.nctu.edu.tw)	 
		\IEEEcompsocthanksitem Z. F. Akbar is with the Department of Informatics Engineering, Electronic Engineering Polytechnic Institute of Surabaya (PENS), Surabaya, Indonesia. (E-mail: fyrozal.akbar@gmail.com)		\IEEEcompsocthanksitem C.-C. Liu and V.-D. Ta are with the Department of Computer Science and Information Engineering, National Taipei University of Technology, Taipei, Taiwan. (E-mail: cmliu@csie.ntut.edu.tw; daitv88@gmail.com) %\protect\\
		}% <-this % stops an unwanted space
}


\markboth{IEEE INTERNET OF THINGS JOURNAL,~Vol.~XX, No.~X, October~201X}%
{Lai \MakeLowercase{\textit{et al.}}: Bare Demo of IEEEtran.cls for Computer Society Journals}



\IEEEtitleabstractindextext{%
\begin{abstract}
A Range-Skyline Query (RSQ) is the combination of range query and skyline query. It is one of the practical query types in multi-criteria decision services, which may include the spatial and non-spatial information as well as make the resulting information more useful than skyline search when the location is concerned. Furthermore, Continuous Range-Skyline Query (CRSQ) is an extension of Range-Skyline Query (RSQ) that the system continuously reports the skyline results to a query within a given search range. This work focuses on the RSQ and CRSQ within a specific range on Internet of Mobile Things (IoMT) applications. Many server-client approaches for CRSQ have been proposed but are sensitive to the number of moving objects. We propose an effective and non-centralized approach, Distributed Continuous Range-Skyline Query process (DCRSQ process), for supporting RSQ and CRSQ in mobile environments. By considering the mobility, the proposed approach can predict the time when an object falls in the query range and ignore more irrelevant information when deriving the results, thus saving the computation overhead. The proposed approach, DCRSQ process, is analyzed on cost and validated with extensive simulated experiments. The results show that DCRSQ process outperforms the existing approaches in different scenarios and aspects.
\end{abstract}

}

\maketitle

\IEEEdisplaynontitleabstractindextext

% \fi
\IEEEpeerreviewmaketitle

%\IEEEraisesectionheading{\section{Introduction}\label{sec:introduction}}

\section{Introduction}\label{sec:introduction}

\IEEEPARstart{I}{n} recent years, skyline queries receive much attention in various applications such as multi-preference analysis and decision making. In such applications, a skyline set contains the most interesting objects or best objects and retrieves the objects that are not dominated by any other objects. In database systems, queries specialized to search for the non-dominated data objects are called \emph{skyline queries} and their corresponding result sets are known as skyline sets. The data objects in a skyline set are known as skyline objects. In tradition, the skyline query is discussed in a static environment, where all the data objects and query are static. Now it is progressively extended for dynamic or distributed environments. If the user is moving or the query is issued from a dynamic environment, such a case addresses the skyline problem in dynamic environments. The skyline query is also used in spatial networks and all the considered data objects are highly distributed. Thus, how to process skyline queries in distributed environments has become an important issue.

Conventional Location-Based Services (LBSs) focus on processing proximity-based queries, including range query~\citep{SeokJin:2006:IndexRangeQueryBroadcast,Jianting:2004:MultiDimensionRangeQueryBroadcast} and nearest neighbor (NN) query~\citep{Li2015,Liu:2008:kNNIndexTreeMultiDimension}. However, these queries are not sufficient for providing high-quality services to mobile users without considering both spatial and non-spatial information simultaneously. A typical scenario is finding a nearby hotel with a cheap price, in which the distance is a spatial attribute and the price is non-spatial. Clearly, in this case, a multi-criteria query is more appealing than a conventional spatial query that considers the distance only. Among various multi-criteria queries, the skyline query is considered as one of the most classical ones and receives a great deal of attention in LBS research. However, the resulting skyline may contain many useless data objects since the resulting data objects (hotels) may be too far away from the query (user). Some other works~\citep{6081864,Rahul:2012:ARQ:2424321.2424406} tend to solve the range-skyline query to improve the QoS of LBSs by considering the dynamic data objects and supporting the \emph{continuous range-skyline} query. The continuous range-skyline is a collection of range-skyline answers during a specific time interval that the query concerns. Such a query is applied in many LBSs whose environments are dynamic. For example, searching taxis is an application of the continuous range-skyline query. Users can use such a service to obtain some candidate taxis which are nearby, cheap, and high-ranked. Hence, this work focuses on processing the range-skyline query and the continuous case.

%\subsection{Motivation}
Most of the existing approaches~\citep{6081864,Rahul:2012:ARQ:2424321.2424406,Dimitris:2005:SkylineComputation,Tian:2007:CMS:1254850.1254861} process the skyline or range-skyline in a centralized way under the assumption that data objects are stored in a centralized fashion. They provide some algorithms that focus on how to index the spatial or non-spatial data and how to efficiently process queries with a large number of data objects. Although some of them have discussed spatial queries and moving data objects in the distributed and mobile environments like the Internet of Mobile Things (IoMT) or Mobile Wireless Sensor Networks (MWSNs), the computing model is still centralized. The collected data objects are stored distributively and the process of data sensing (collection) phase is not discussed. They discussed the changes (or updates) of data and treated such cases as moving data objects. In these approaches, each mobile node needs to continuously obtain the location information of itself by GPS and sends the information back to the server(s) for updating the database(s). However, the overhead of the data collection process was not mentioned and addressed. If we use the conventional methods in a specific scenario, the position of each data object changes frequently, the server(s) will receive a huge amount of information for updating the location of each data object in a short time. In this case, the system will be overloaded. Furthermore, a large number of messages for updates will occupy quite a lot of communication bandwidth.

In general, the IoMT applications based on MWSNs are self-configuring and infrastructure-less. IoMT consists of many mobile sensor nodes connected by wireless communication. In such environments, each node can move freely and independently in any direction, so the communication links between nodes will change frequently. Each node can forward the information unrelated to its owns and act as a router. In comparison with the client-server environment, IoMT applications may have no centralized server to handle the spatial queries. Accordingly, the information system for an infrastructure-less IoMT application must process queries in a distributed (or decentralized) way. Each mobile sensor node can cooperate and exchange data with each other and then derive the answers for spatial queries. Since most of the existing works consider multiple data sources but only a few works consider the fully distributed and dynamic computing environments, one of the main objectives of this work is to provide a fully distributed approach for processing \emph{Range-Skyline Queries} (RSQ) and \emph{Continuous Range-Skyline Queries} (CRSQ) in an infrastructure-less mobile environments, IoMT.

%\subsection{Objectives}
In this work, we propose a \emph{Distributed Range-Skyline Query process} (DRSQ process) in IoMT whose computing model is decentralized. We further extend DRSQ to \emph{Distributed Continuous Range-Skyline Query process} (DCRSQ process) for supporting continuous range-skyline query processing. Each mobile node can filter out the irrelevant data objects, derive a primary candidate answer set of the received query, and then report the candidate set to the query node.
For validating the DRSQ process, we perform the simulation experiments with the following measurements: the \emph{response time} and the \emph{number of messages} (\emph{I/O operations}). Furthermore, to validate the DCRSQ process, we consider four measurements: the number of accessed objects (the overhead on the query node), the number of messages, precision and recall. We also consider the effects on the number of sensor nodes, the number of queries, the transmission range of a node, and the query range in the DRSQ process. One additional effect, node speed, is considered in the DCRSQ process. Besides, we give the analysis on network cost for the proposed approach and compare the proposed approach with the centralized approach. As the results show, the proposed approach has a better performance in the simulation.

%\subsection{Contributions}
We address the distributed continuous range-skyline query (DCRSQ) processing over the IoMT and make the following contributions:
\begin{itemize}
	%\item We identify the range-skyline query in LBS applications built on mobile wireless sensor networks and categorize some existing works about this issue.
	\item We propose a distributed approach, DCRSQ, to make the process of range-skyline query appropriate to the Internet of Mobile Things environments.
	\item To the best of our knowledge, this is the first study for this problem simultaneously considering the computing process and information filtering in the data collection phase so that the performance of system is significantly improved in comparison with the conventional approach.
	\item With the mobility, DCRSQ can make each node predict the time when its neighboring mobile data objects fall in the query range and avoid periodically flooding messages for updating the information of neighbors.
	\item We give a formal analysis of the network cost on the proposed approach and conduct extensive simulations to evaluate the performance. The simulation results show that DCRSQ can save more than 15$\%$ network cost and achieve almost 90$\%$ accuracy in most of the scenarios.
\end{itemize}

The balance of this paper is organized as follows. In Section~\ref{related_work}, we introduce the background and review related research. Section~\ref{preliminary} presents the overview of problem and the notations used in this work. Section~\ref{DCRSQ} introduces the proposed solution and a breakdown of the data structures and algorithms. Some analysis on network cost will be discussed in Section~\ref{analysis}. Simulation experiments are presented in Section~\ref{simulation}. Finally, we make concluding remarks in Section~\ref{conclusion}.

\section{Related Work}
\label{related_work}
The IoMT has triggered a lot of emerging applications and services~\citep{7903653,7945539} in wireless communications, fog/edge computing, and (mobile) big sensor data processing. Ang et al.~\citep{7903653} identified some important research topics, like data analysis and processing, in smart city ecosystems which base on IoMT. They also investigate some use cases in IoMT (smart cities) such as real-time urban monitoring~\citep{5594641} and spatial decision support system for flood risk management~\citep{HORITA201584}. These use cases are spatio-temporal applications classified in~\citep{7945539}. In spatio-temporal IoMT applications, spatial query processing plays a key role for the decision making. In our work, we focus on the range-skyline query for this kind of IoMT applications.

A range-skyline query is an extension of skyline query with a distance threshold in spatial databases. Borzsony et al.~\citep{914855} introduced skyline operator into the database systems with algorithms \emph{Block Nested Loop} (BNL) and \emph{Divide-and-Conquer} (D\&C). A great number of researchers also keep their eyes on skyline query processing from then on. Papadias et al.~\citep{Dimitris:2005:SkylineComputation} proposed a \emph{Branch-and-Bound Skyline} (BBS) method based on the best-first nearest neighbor algorithm~\citep{Hjaltason:1999:DBS:320248.320255}. Cheema et al.~\citep{Cheema:2013:SZB:2452376.2452409} proposed a \emph{safe zone} based approach and combined it with Vonronoi cells to provide a better BBS algorithm for processing skyline queries. Hose and Vlachou~\citep{Hose2012} provided comprehensive analysis of previous skyline algorithms without indexing supports, and proposed a new hybrid method with improvement. Lin et al.~\citep{6081864} also discussed both the indexing and non-indexing algorithms, and then extended their work to process probabilistic RSQ. All these works discuss the issues in a centralized data storage.

\begin{table*}[t]
	\renewcommand{\arraystretch}{1.2}
	\caption{Comparisons of Existing Works for Skyline Query}
	\label{compared_methods}
	\centering
	\small
	\begin{tabular}{lccccccc}
		\hline
		& \multicolumn{7}{c}{\textbf{Methods}} \\
		\cline{2-8}
		\textbf{Considered Issues} & BBS~\citep{Dimitris:2005:SkylineComputation} & LDSQ~\citep{4511446} & PadSkyline~\citep{10.1109/TKDE.2010.103} & EDDS~\citep{jsan5010002} & RSQ~\citep{6081864} & L-SQ~\citep{1617434} & \textbf{DRSQ \& DCRSQ}\\
		\hline
		\hline
		\textbf{Distributed Databases} & \texttimes & \texttimes & \checkmark & \checkmark & \texttimes & \checkmark & \textbf{\checkmark} \\
		\textbf{Distributed Computing} & \texttimes & \texttimes & \checkmark & \checkmark & \texttimes & \checkmark & \textbf{\checkmark} \\
		\textbf{Moving Objects} & \texttimes & \texttimes & \texttimes & \texttimes & \checkmark & \texttimes & \textbf{\checkmark} \\
		\textbf{Moving Queries} & \texttimes & \texttimes & \texttimes & \texttimes & \checkmark & \checkmark & \textbf{\checkmark} \\
		\textbf{Non-Index based} & \texttimes & \texttimes & \texttimes & \texttimes & \checkmark & \texttimes & \textbf{\checkmark} \\
		\hline
	\end{tabular}
\vspace{-10pt}
\end{table*}

To make the applications scalable and improve the performance of skyline query processing, many parallel or distributed algorithms have been proposed. Wu et al.~\citep{Wu:2006:PSQ:2117976.2117990} first attempted a progressive processing of skyline queries on a CAN-based P2P network~\citep{Ratnasamy:2001:SCN:383059.383072}. By using the query range to recursively partition the data region involved and encoding each involved sub-region dynamically, their method can progressively report skyline objects without accessing the data sites not containing potential skyline objects, thus saving computation overhead. Chen et al.~\citep{10.1109/TKDE.2010.103} proposed a parallel approach to filter data objects efficiently from distributed databases for processing constrained (or range) skyline queries. Zheng et al.~\citep{4511446} introduced a variant notion of the valid scope for skyline queries, that can save the re-computation if the next query is still inside the valid scope. Although the above existing works considered the distributed databases, they still used some high-performance servers to process skyline queries with the data objects from multiple data sources. Alternatively, we consider an infrastructure-less environment, IoMT, in this work.

Huang et al.~\citep{1617434} proposed techniques for skyline query processing in MANETs. Lightweight devices in MANETs are able to issue spatially constrained skyline queries that involve data stored on many mobile devices. Queries are forwarded through the whole MANET without routing information. However, they only considered the mobile distributed data sites over MANETs but did not consider the moving objects. Ahmed et al.~\citep{jsan5010002} proposed  an approach, Enhanced Distributed Dynamic Skyline (EDDS), to handle skyline queries over the IoMT. EDDS used disc track and sector to map the data locations. Such a way improves the performance of searching the new input data objects for computing and updating the skyline. Although EDDS considered the dynamically data input from the distributed sensor nodes, EDDS did not consider the mobile sensor nodes (moving objects). %Nakayama et al.~\citep{Nakayama:2015:RCT:2837126.2837131} proposed a method to deal with the movement of nodes and monitor the current result of the continuous range-based threshold query by utilizing a geographic routing protocol, \emph{Greedy Perimeter Stateless Routing} (GPSR)~\citep{Karp:2000:GGP:345910.345953}. However, the threshold query is simpler than the range skyline query and their objective is not developing a cooperative method for query processing.

In fact, the conventional works can be categorized into following models: (1) single data source with a centralized computing model, (2) multiple data sources with a centralized computing model, and (3) multiple data sources with a distributed/decentralized computing model. The third model is more popular in recent years. However, it is not easy to compare all the works in type (3) by simulation or experiments since the considered environments, assumptions, and requirements are quite different. To the best of our knowledge, most of works in type (3) consider the distributed computing model with multiple "powerful" computing servers for the query processing. Only very few works consider query processing over a lightweight mobile environment whose computing resource is limited. However, these few works only consider the static spatial data and do not support moving data objects. We therefore present a comparison summary of the existing methods related to skyline query and our work in Table~\ref{compared_methods}.

%This work is inspired by these prior distributed algorithms and focuses on the range-skyline queries over moving spatial data objects with non-spatial sensed attributes. The above related works just assume that all the data are well collected and do not discuss the phase of data collection. Only~\citep{Nakayama:2015:RCT:2837126.2837131} considers the moving spatial data objects in an infrastructure-less MANETs which is similar to our considered environment. This work focuses on utilizing the geographic routing protocol, GPSR, to guarantee the efficient and reliability of transmission between the sensor nodes and the query node, and thus helps the range-based threshold query. However, the considered objective, query type and underlying routing protocol in~\citep{Nakayama:2015:RCT:2837126.2837131} are quite different from our work. Therefore, we propose a reasonable distributed approach which integrates the data pruning process with the data collecting phase to monitor continuous range-based skyline queries while considering relative speeds between the data objects and query node.

\section{Preliminaries}
\label{preliminary}
In this section we give some preliminaries of the problem, including some fundamental notations and definitions. We consider a data set $S$ of sensor nodes. Each mobile sensor node $s \in S$ is associated with a spatial (i.e., location or distance) attributes and several other non-spatial attributes (e.g., temperature, trust rank, and possibility). %The Euclidean distance that we use in the spatial attribute is considered as defined in Definition~\ref{dist}.
Note that we use Euclidean distance in the spatial attribute and the non-spatial dominance relation between the mobile objects is described as Definition~\ref{nsdom}.

%\begin{definition}
%	\label{dist}
%	\textbf{(Euclidean distance)}\\
%	Given two sensor nodes $s=(x_s,y_s)$ and $s'=(x_{s'},y_{s'})$ in a plane, we denote the distance between these two nodes as $dist(s,s')=\sqrt{(x_s-x_{s'})^2+(y_s-y_{s'})^2}$.
%\end{definition}
%
\begin{definition}[\textbf{Non-spatial Dominance}]
	\label{nsdom}
	Given two sensor nodes $s$ and $s'$, if $s'$ is no worse than $s$ in all non-spatial attributes, then we say $s'$ \textit{non-spatially dominates} $s$. We say that $s'$ is a non-spatial dominator object of $s$, and $s$ is a non-spatial dominance object of $s'$. Formally, it is denoted as $s'\triangleleft s$. The set of $s$'s non-spatial dominator objects is denoted as $Dom(s)$, i.e., $s$ is dominated by any object in $Dom(s)$ on non-spatial attributes.
\end{definition}

If $s'\triangleleft s$ and $s\triangleleft s'$ are hold, it means that non-spatial attributes of $s'$ and $s$ are equivalent. In this case, the system is going to check the dominance relation between the spatial attributes of $s'$ and $s$. So the complete dominance relation can be described as
\begin{definition}
	\label{sdom}
	\textbf{(Dominance)}\\
	Given a query node $q$ and two sensor nodes $s$ and $s'$, if 1) $s'$ non-spatially dominates $s$, and 2) $dist(q,s')\leq dist(q,s)$ (i.e., $s'$ also spatially dominates $s$), then we say $s'$ dominates $s$ w.r.t. the query node $q$. Formally, it is denoted as $s'\triangleleft_q s$.
\end{definition}

Note that if $s'\triangleleft_q s$ and $s'\triangleleft_q s$, it means that both spatial and non-spatial attributes of $s'$ and $s$ are equivalent with respect to the query node $q$.

With the above definitions, point-skyline query (PSQ) can be defined as
\begin{definition}
	\label{psq}
	\textbf{(Point-Skyline Query (PSQ))}\\
	Given a data set $S$, the \textit{skyline} of a query node $q$ is a subset of $S$ in which each object (sensor node) is not dominated by any other object in $S$ w.r.t. $q$. We call this subset skyline set and denote it as $PSQ(S,q)=\{s|s \in S \wedge \forall s' \in S-\{s\}: s' \ntriangleleft_q s\}$.	
\end{definition}

In accordance with the above definitions, the range-skyline query can be defined in Definition~\ref{trsq} and such a definition comes from a global view of system.
\begin{definition}
	\label{trsq}
	\textbf{(Range-Skyline Query (RSQ))}\\
	Given a data set $S$ with a range $R$, the range-skyline query with respect to $q$ returns the skyline set of the subset of objects (sensor nodes) that locate in $R$. Formally, it is denoted as $RSQ(R,S,q)$, and $RSQ(R,S,q)=\{s \in S\wedge s$ locates in $R| \forall s'\in S-\{s\}\wedge s'$ locates in $R: s' \ntriangleleft_q s\}$.	
\end{definition}

\begin{figure}[ht]
	\begin{center}
		\includegraphics[width=0.45\textwidth,keepaspectratio]{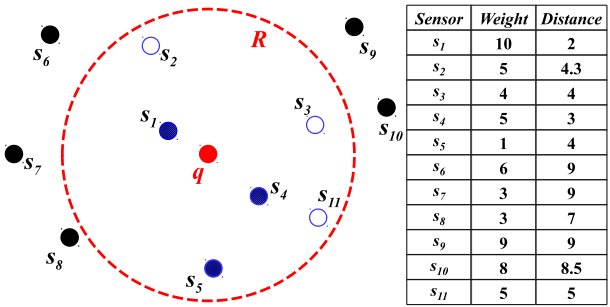}
	\end{center}
	\caption{An example of an $RSQ(R,S,q)$ where $S$ is the data set, $R$ is the range with the query node $q$ as the center, and the output is $\{s_1, s_4, s_5\}$}
	\label{fig:rsq_ex}
\end{figure}

Fig.~\ref{fig:rsq_ex} is an example of a range-skyline w.r.t. query node $q$, where the mobile user wants to search a nearby taxi around the query node with a high rank of service quality. The system firstly uses $q$'s range value, $R$, to prune the irrelevant moving objects which are out of the range. Then the system examines dominance relations between the remaining data objects. We assume that the mobile objects with smaller weights have a higher priority in this example. Since $s_1$ is the nearest neighbor of $q$ and spatially dominates all the other sensor nodes, $s_1$ must be in the resulting range skyline. Sensor node $s_4$ non-spatially dominates the other nodes in the range $R$, except $s_3$ and $s_5$. So the possible RSQ is $\{s_1,s_3,s_4,s_5\}$. However, $s_3$ is dominated by $s_5$ in the non-spatially attribute. Thus, $RSQ(R,S,q)$ will be $\{s_1, s_4, s_5\}$. It means that the system returns taxi $s_1$, $s_4$ and $s_5$ to the mobile user.

A continuous range-skyline query (CRSQ) is an extension of RSQ. CRSQ will monitor the environmental information within a given range and continuously produce the skyline for a period of time. It means that the system monitors each continuous range-skyline query $q$ within a specific range $R$ for a time period $\varDelta t=[t_0,t_{end}]$. Since each sensor node can move in the considered environment, such a phenomenon will make the answer of an RSQ change during the monitoring time $\varDelta t$. We formally define the continuous range-skyline query as below.

\begin{definition}
	\label{crsq}
	\textbf{(Continuous Range-Skyline Query (CRSQ))}\\
	Suppose that the notations are defined as above. Given a query $q$ with a query range $R$ and a time period $\varDelta t=[t_0,t_{end}]$, the \emph{continuous range-skyline query} returns a collection of range-skyline sets $RSQ_{t_i}(R,S,q)$, where $t_i \in \varDelta t$ and $i$ is the number of updated results. Formally, it is denoted as $CRSQ(R,S,q,\varDelta t)=\{RSQ_{t_i}(R,S,q)|t_i \in \varDelta t, i \in N\}$.	
\end{definition}

\begin{figure*}[h]
	\centering
	\subfigure[$RSQ_{t_0}(R,S,q)=\{s_1, s_4, s_5\}$]{
		\label{fig:crsq_ex:a} %% label for 1st subfigure
		\includegraphics[width=0.45 \textwidth]{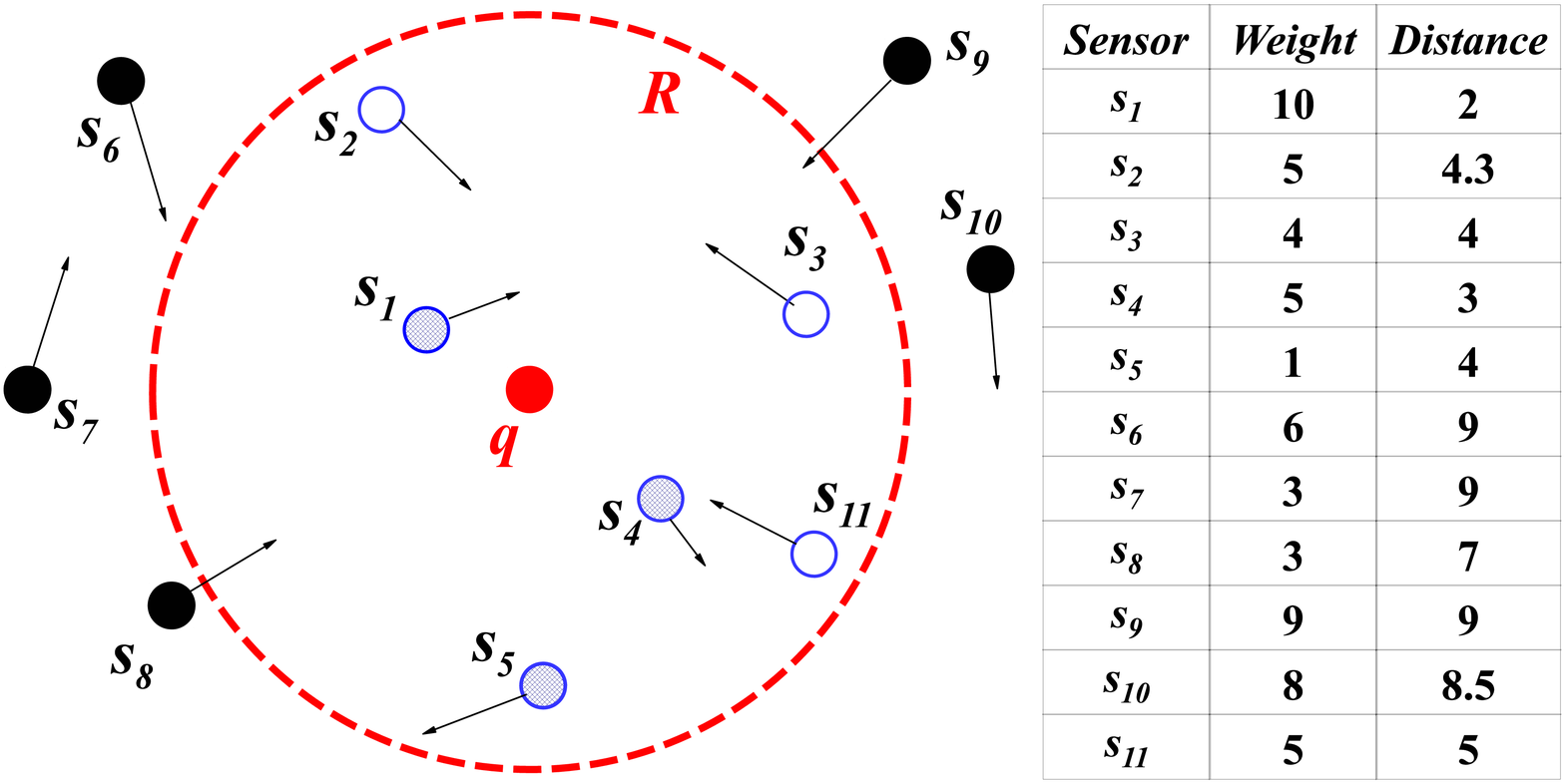}}%
	\\	
	\subfigure[$RSQ_{t_1}(R,S,q)=\{s_1, s_8, s_{11}\}$]{
		\label{fig:crsq_ex:b} %% label for 2nd subfigure
		\includegraphics[width=0.45 \textwidth]{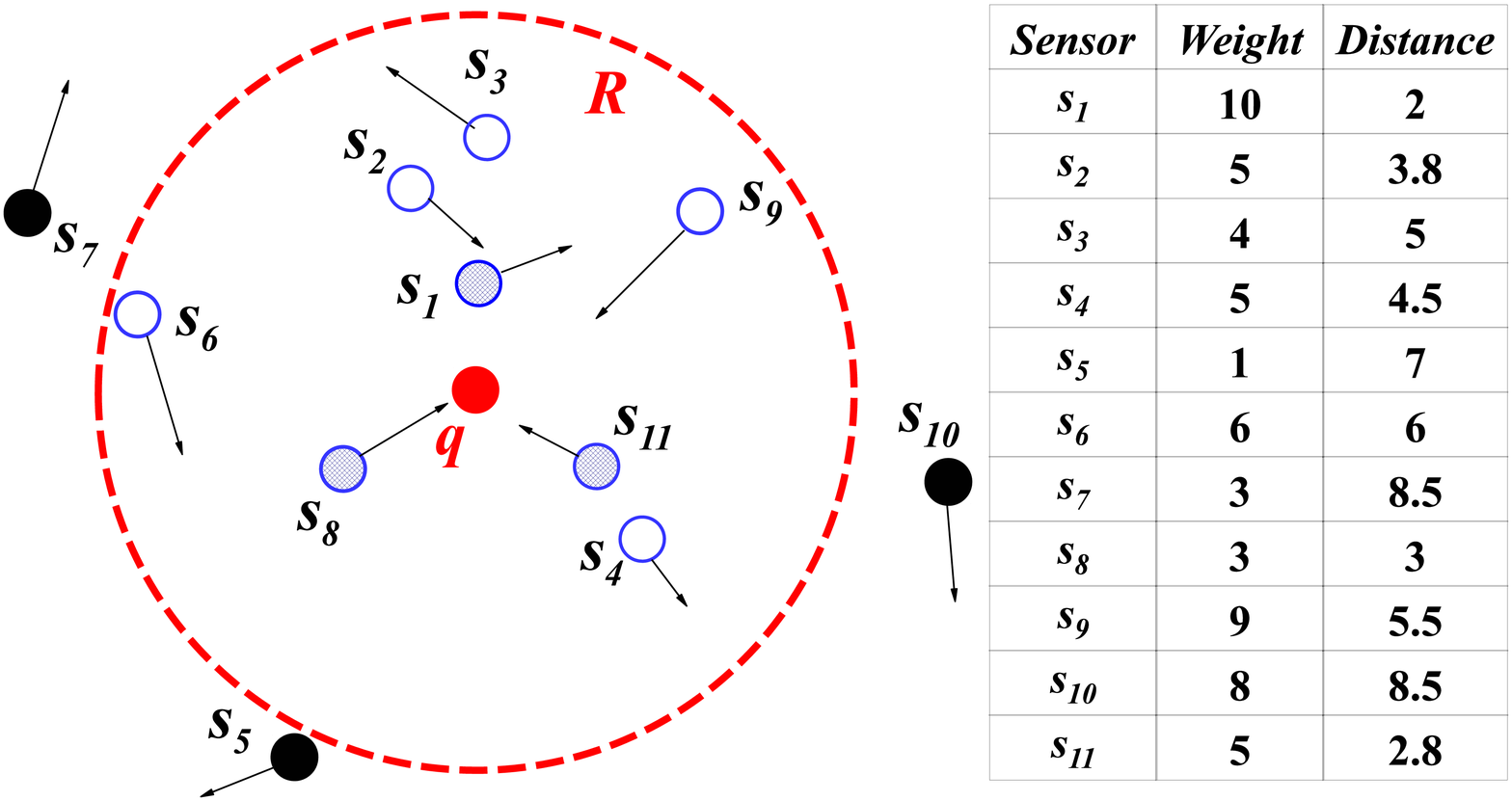}}\hspace{0.05 \textwidth}
	\subfigure[$RSQ_{t_2}(R,S,q)=\{s_8\}$]{
		\label{fig:crsq_ex:c} %% label for 2nd subfigure
		\includegraphics[width=0.45 \textwidth]{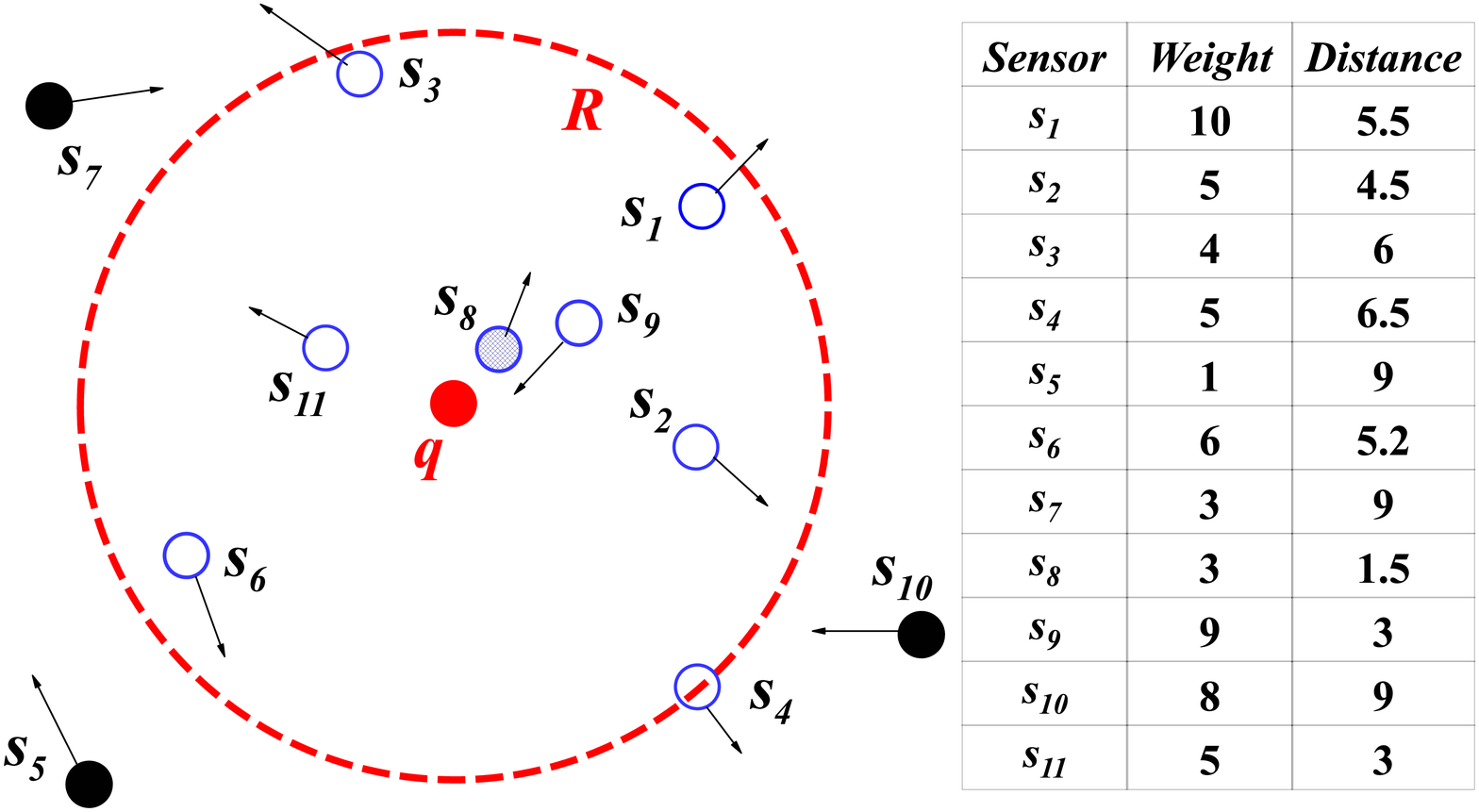}}
	\caption{An example of a $CRSQ(R,S,q,\varDelta t)$ and the output is $\{<\{s_1, s_4, s_5\},[t_0,t_1)>,<\{s_1, s_8, s_{11}\},[t_1,t_2)>,<\{s_8\},[t_2,t_{end}]>\}$ where $t_0$, $t_1$, and $t_2$ are different times during $\varDelta t=[t_0,t_{end}]$}
	\label{fig:crsq_ex} %% label for entire figure	
	\vspace{-10pt}
\end{figure*}

An example of a continuous range-skyline query $q$ is shown in Fig.~\ref{fig:crsq_ex}. In this example, the system monitors the range-skyline of $q$ for a time period $\varDelta t$ and the result may be a collection of answers that contains different RSQ answers at different time since the answer may change. The answer collection contains 3 RSQ results at time $t_0$, $t_1$, and $t_2$ and these results are respectively shown in Fig.~\ref{fig:crsq_ex:a}, Fig.~\ref{fig:crsq_ex:b}, and Fig.~\ref{fig:crsq_ex:c}. The result of $q$ is $CRSQ(R,S,q,\varDelta t)=\{RSQ_{t_0}(R,S,q),RSQ_{t_1}(R,S,q),RSQ_{t_2}(R,S,q)\}$, where $RSQ_{t_0}(R,S,q)=\{s_1, s_4, s_5\}$, $RSQ_{t_1}(R,S,q)=\{s_1, s_8, s_{11}\}$, and $RSQ_{t_2}(R,S,q)=\{s_8\}$. The final output therefore will be $\{<\{s_1, s_4, s_5\},[t_0,t_1)>,<\{s_1, s_8, s_{11}\},[t_1,t_2)>,<\{s_8\},[t_2,t_{end}]>\}$.
%

%\IEEEpubidadjcol

\section{The Distributed Continuous Range-Skyline Query Process}
\label{DCRSQ}
This section describes in detail the proposed distributed range-skyline query process over the IoMT. The proposed approach includes two parts: distributed range-skyline query process (DRSQ process) and distributed continuous range-skyline query process (DCRSQ process). The fundamental distributed approach for processing snapshot RSQs will be introduced in the DRSQ process. In second part, the DCRSQ process is extended from the DRSQ process with the consideration of node mobility to support CRSQ. Thus, the system can predict the change of RSQ result when monitoring a CRSQ during a period of time in IoMT environments.

\subsection{Distributed Range-Skyline Query}
In general, the query processing in mobile and distributed environments like IoMT based on MWSNs or MANETS, contains three steps. The first step is the local process that computes the skyline set based on local data and filters information received along with the query. The second step is query routing by which the query message can be forwarded to some of the neighboring nodes in order to retrieve their partial results. Thus, the node decides whether a neighbor can contribute to the skyline set based on available routing information. The last step is to merge the results, where the node receives the local result sets from queried neighbors and merges all the partial results by checking for dominated objects. Then, each node sends the merged result to the query node hop by hop.

\subsubsection{Description of DRSQ}
We assume that each mobile sensor node can hold a small database to store the collected sensing data and the query's information for the distributed query process. The collected sensing data set is called \emph{local data set} and all of the data are collected from the sensor node's one-hop neighbors and itself. Hence, each mobile sensor node can derive the result of \emph{local range-skyline query process} (LRSQ process) which may be the subset of range-skyline and return the result to the query node for computing the final (global) range-skyline answer. The result of LRSQ is defined in Definition~\ref{define_lrsq}.
\begin{definition}
	\label{define_lrsq}
	\textbf{(Local Range-Skyline Query)}\\
	Suppose that the notations are defined as above and a query node $q$ with a query range $R$ is given. After a mobile sensor node $s_j$ receives the query $q$, $s_j$ will return a subset of the local data set $S_{s_j}$ and each object $s$ in $S_{s_j}$ is $s_j$'s neighbor and not dominated by any other object $s'$ in $S_{s_j}$ w.r.t. $q$, where $S_{s_j}\subseteq S$. We refer to this result as a local range-skyline set and denote it as $LRSQ_{s_j}(R,S,q)=\{s$ locates in $R \wedge s \in S_{s_j} | \forall s' \in S_{s_j}-\{s\}\wedge s'$ locates in $R: s' \ntriangleleft_q s\}$.	
\end{definition}

According to Definition~\ref{define_lrsq}, the query node $q$ will receive the results of $LRSQ_{s_j}(R,S,q)$, where $s_j$ is $q$'s one-hop neighbor ($1\leq j\leq k$) and $k$ is the maximum number of neighbors. Then the query node takes the union of these results as the candidate set, $RSQ_{candidate}=\bigcup^{k}_{j=1}{LRSQ_{s_j}(R,S,q)}$. After $RSQ_{candidate}$ is derived, the query node will use Definition~\ref{nsdom} and Definition~\ref{sdom} to examine the dominance relations of all the mobile objects in $RSQ_{candidate}$ again and then save the final result in $RSQ_{distributed}$ set.
%Thus, query node $q$ can use these received LRSQ sets to derive the final answer.
Such a cooperative and distributed process is refer to \emph{distributed range-skyline query process} (DRSQ process). As a result, DRSQ can be defined as Definition~\ref{define_drsq}.

\begin{figure*}[t]
	\centering
	\subfigure[]{
		\label{fig:DRSQ-Running-example:a} %% label for 1st subfigure
		\includegraphics[width=0.31 \textwidth]{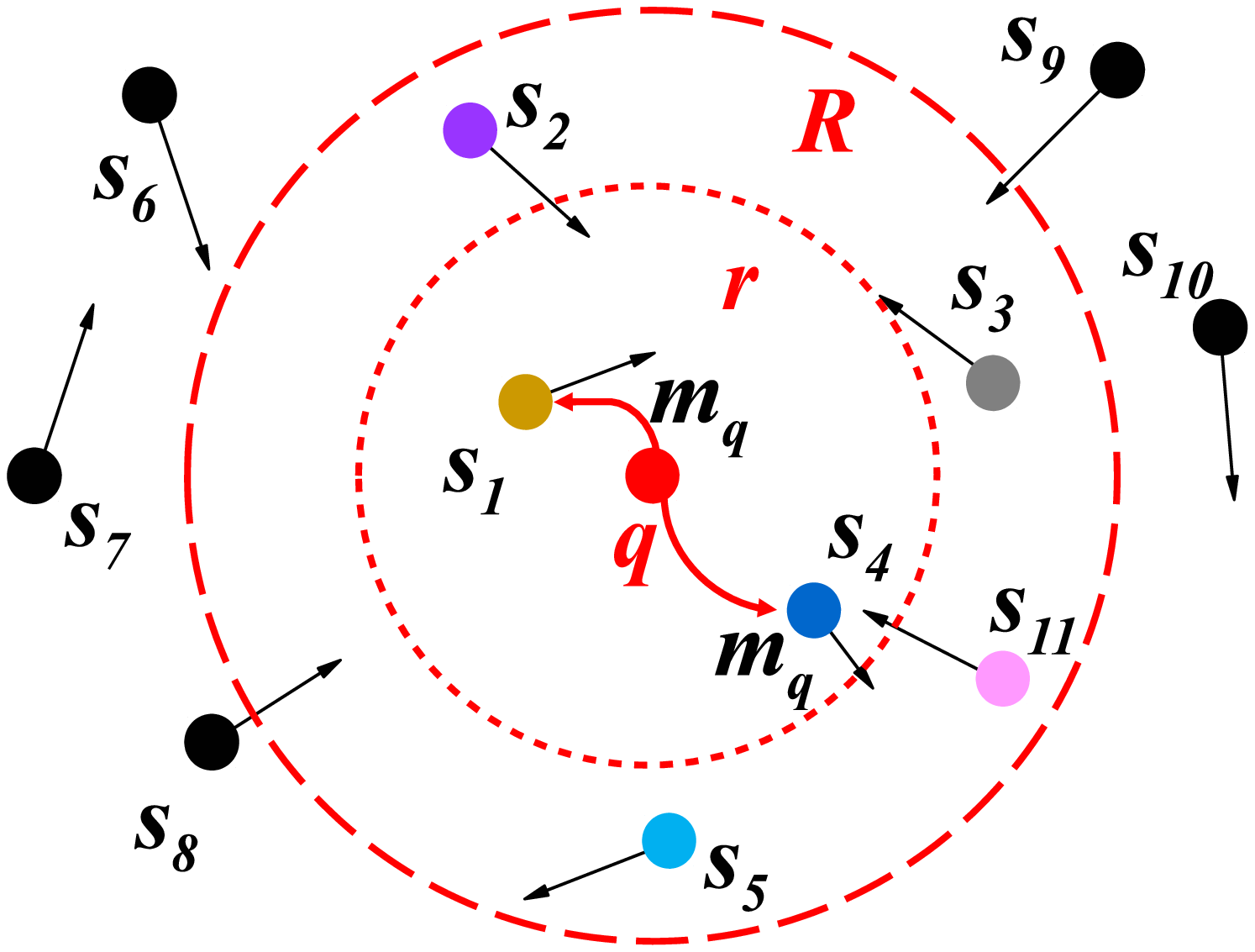}}%
	\subfigure[]{
		\label{fig:DRSQ-Running-example:b} %% label for 2nd subfigure
		\includegraphics[width=0.31 \textwidth]{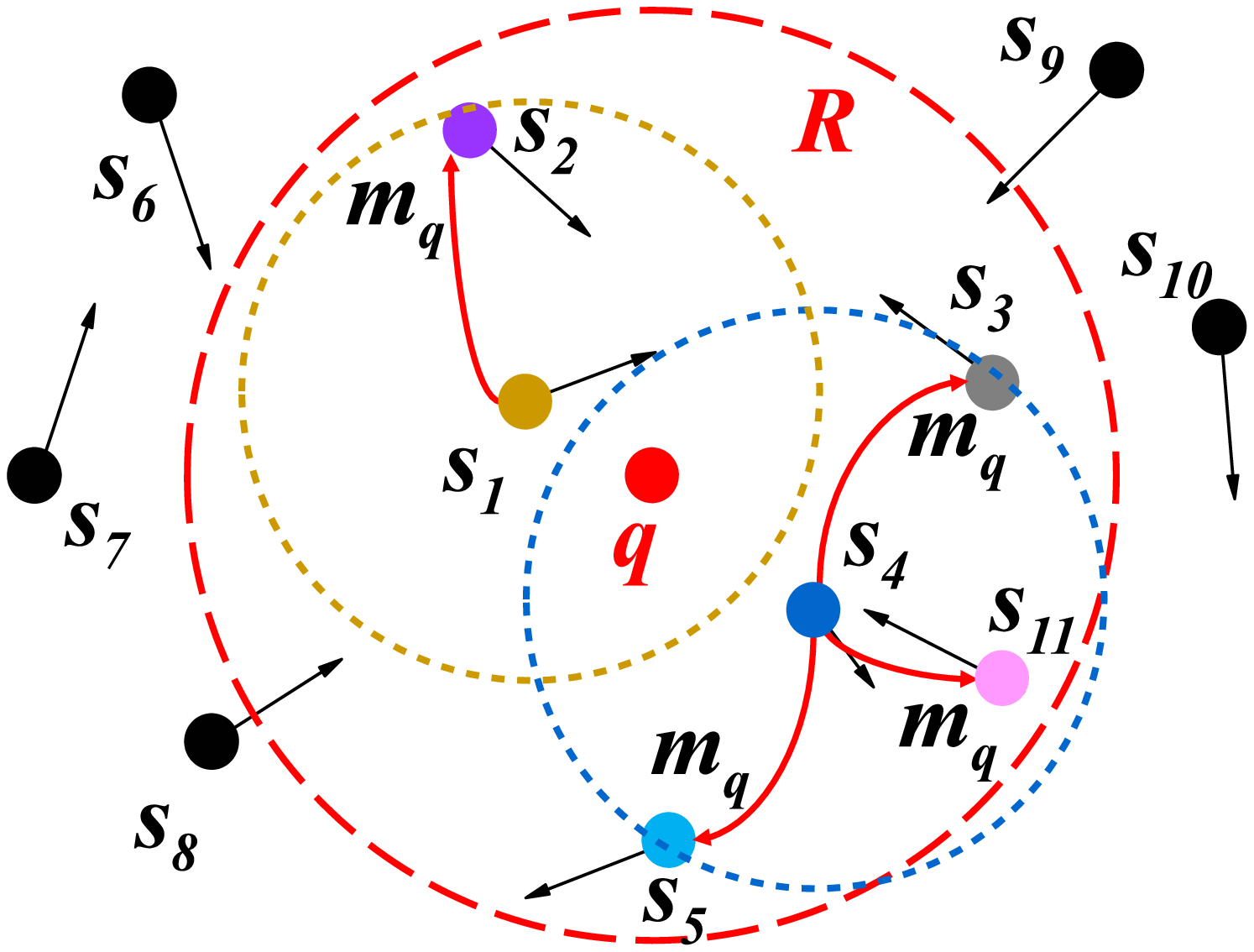}}%
	\subfigure[]{
		\label{fig:DRSQ-Running-example:c} %% label for 2nd subfigure
		\includegraphics[width=0.31 \textwidth]{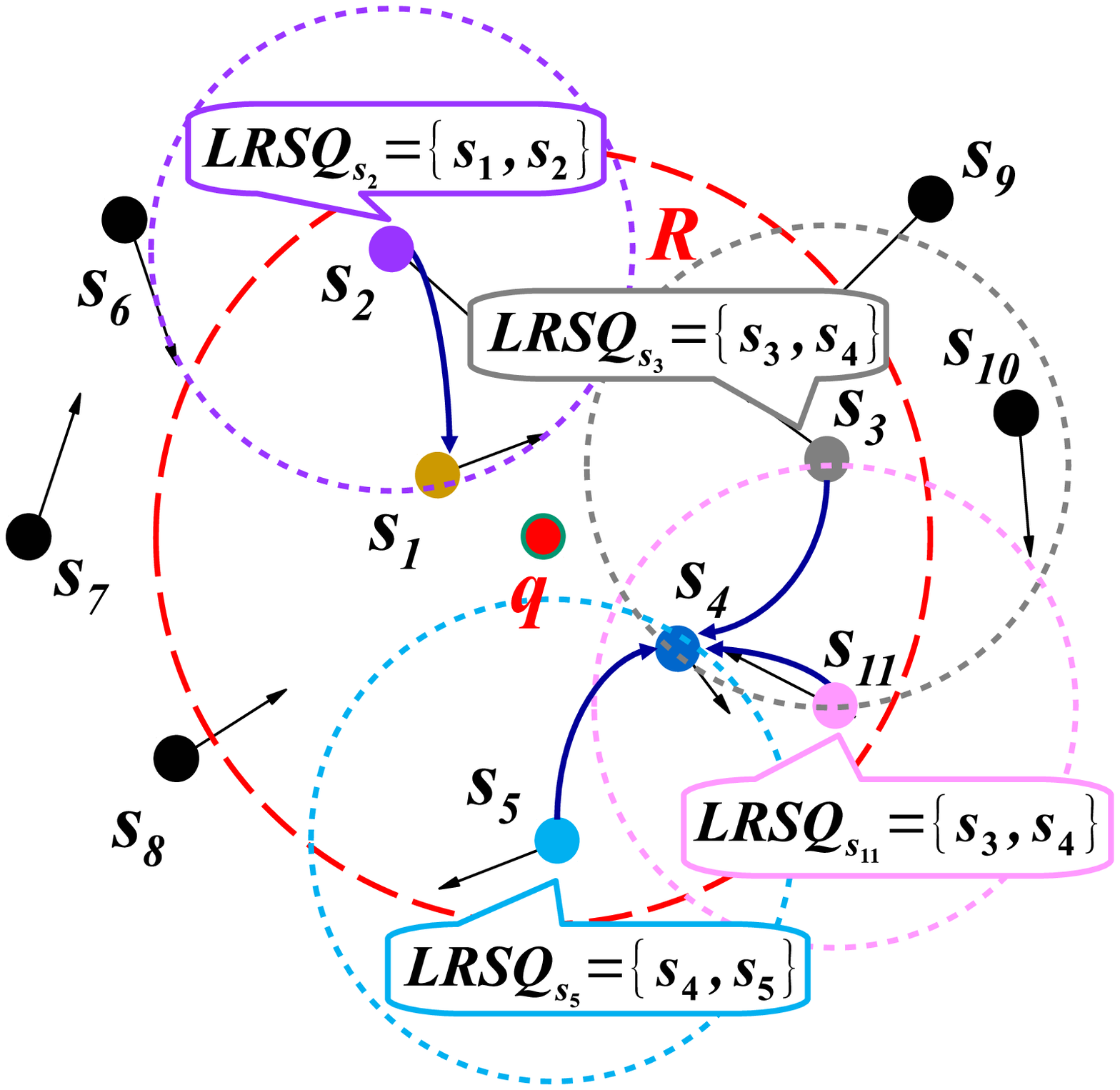}}\\
	\vspace{-15pt}
	\subfigure[]{
		\label{fig:DRSQ-Running-example:d} %% label for 1st subfigure
		\includegraphics[width=0.31 \textwidth]{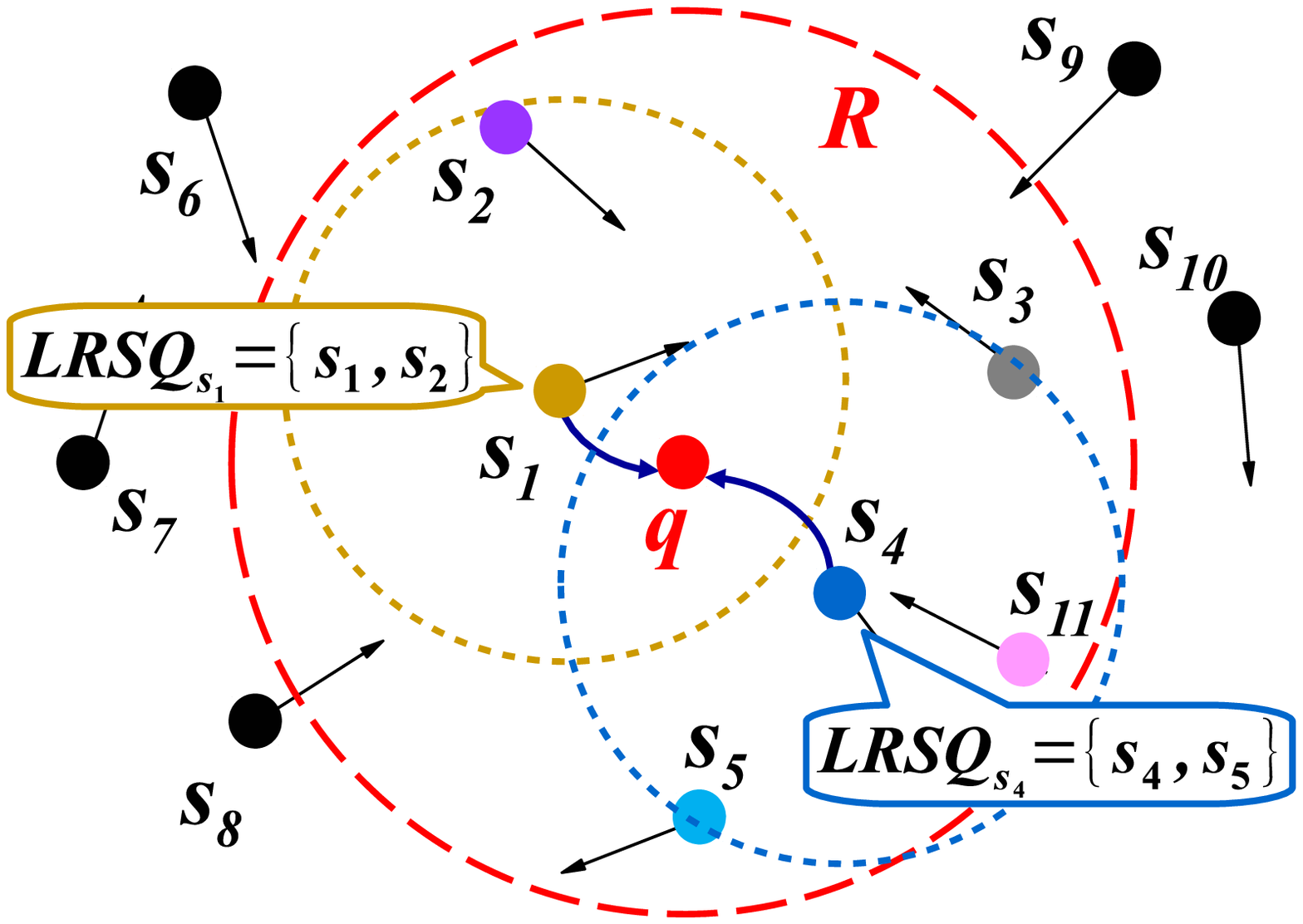}}%
	\subfigure[]{
		\label{fig:DRSQ-Running-example:e} %% label for 2nd subfigure
		\includegraphics[width=0.31 \textwidth]{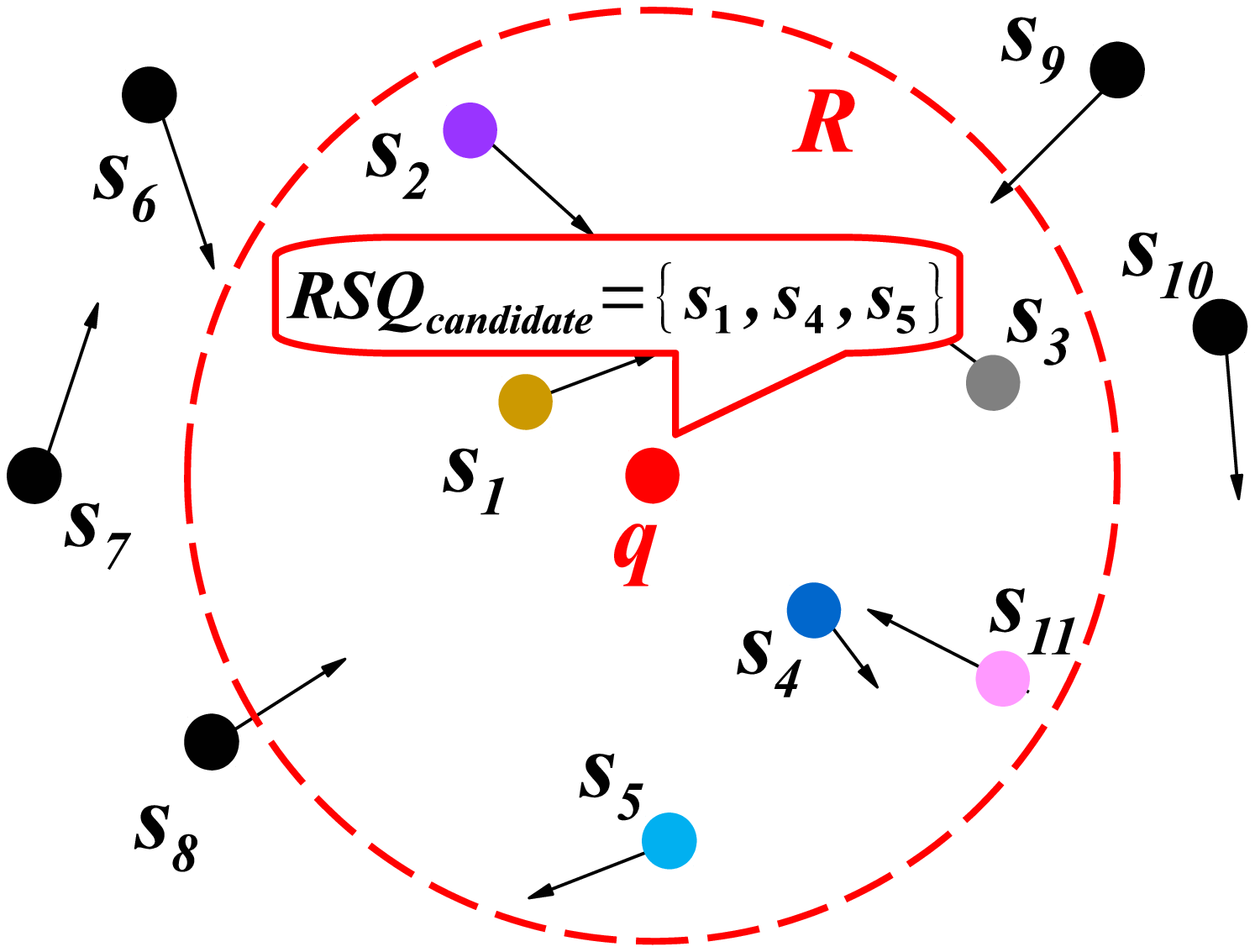}}%
	\subfigure[]{
		\label{fig:DRSQ-Running-example:f} %% label for 2nd subfigure
		\includegraphics[width=0.31 \textwidth]{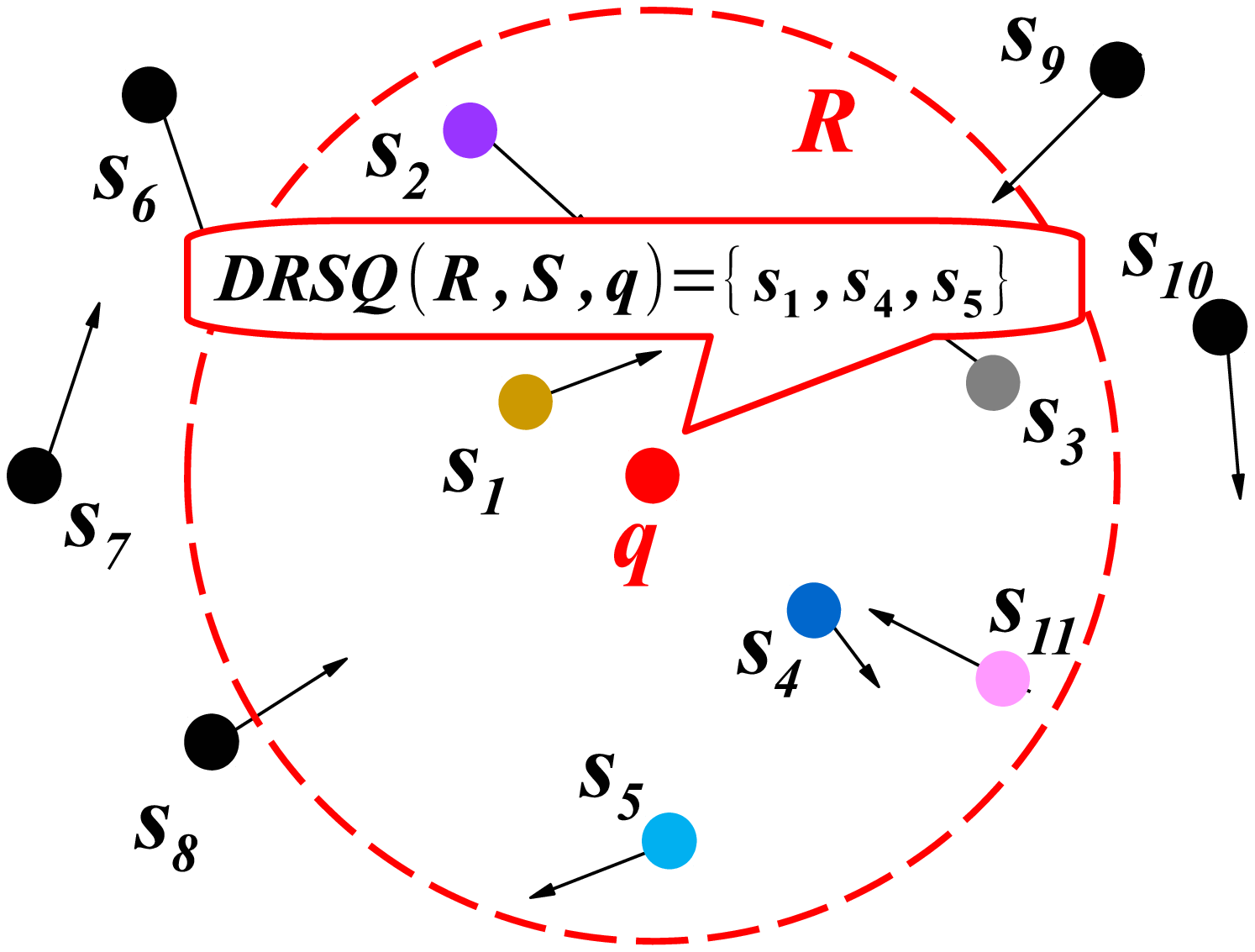}}
	\caption{A running example of DRSQ process where the data set is $S=\{s_1,...,s_{11}\}$, $m_q$ is a query message, and the $TTL$ of $m_q$ is $2$: \subref{fig:DRSQ-Running-example:a} one-hop neighbors $s_1$ and $s_4$;
		\subref{fig:DRSQ-Running-example:b} two-hop neighbors $s_2$, $s_3$, $s_5$, and $s_{11}$;
		\subref{fig:DRSQ-Running-example:c} two-hop neighbors derive and return their LRSQ results to the one-hop neighbors;
		\subref{fig:DRSQ-Running-example:d} one-hop neighbors merge all the received information, prune the irrelevant information, and then return LRSQ of themselves to $q$;
		\subref{fig:DRSQ-Running-example:e}
		$q$ unites all the received information and obtains a candidate set;
		\subref{fig:DRSQ-Running-example:e} $q$ derives the final result after checking the dominance relations between all the data objects in the candidate set.
	}
	\label{fig:DRSQ-Running-example}
	%% label for entire figure	
	\vspace{-10pt}
\end{figure*}
\begin{definition}
	\label{define_drsq}
	\textbf{(Distributed Range-Skyline Query)}\\
	Suppose the candidate set $RSQ_{candidate}$ of query node $q$ has been computed. Then, the query node $q$ uses $RSQ_{candidate}$ to derive the skyline set of the data objects in $R$ and the result of distributed range-skyline query can be denoted as $DRSQ(R,S,q)=\{s$ locates in $R\wedge s\in RSQ_{candidate}| \forall s' \in RSQ_{candidate}-\{s\}\wedge s'$ locates in $R: s' \ntriangleleft_q s\}$.	
\end{definition}
%\begin{myproof}
%	The query node $q$ will receive the results of $LRSQ_{s_j}(R,S,q)$, from its one-hop neighbors and takes the union of these results as the candidate set, $RSQ_{candidate}=LRSQ_{s_1}(R,S,q)\cup LRSQ_{s_2}(R,S,q)\cup \dots \cup LRSQ_{s_k}(R,S,q)$. After $RSQ_{candidate}$ is derived, $q$ will use Definition~\ref{nsdom} and Definition~\ref{sdom} to examine the dominance relations of all the mobile objects in $RSQ_{candidate}$ again and then save the final result in $RSQ_{distributed}$ set.
%\end{myproof}

The above distributed process for LRSQ derivation has a benefit that many irrelevant data objects are also pruned during the process. Thus, the computation overhead of the query node can be reduced.

\subsubsection{Overview of DRSQ process}
Before introducing the DRSQ processes on a query node and a sensor node with pseudo-codes in detail, we use a running example in Fig.~\ref{fig:DRSQ-Running-example} to depict the overview of DRSQ process and explain it step by step. Note that the spatial and non-spatial attributes of each sensor node (data object) are shown in Fig.~\ref{fig:rsq_ex}.

As shown in Fig.~\ref{fig:DRSQ-Running-example:a} and Fig.~\ref{fig:DRSQ-Running-example:b}, a query node spreads the query message $m_q$ ($TTL=2$) to its one hop and two-hop neighbors. Each message $m_q$ contains the information of query node, such as query range $R$, location, and speed of $q$. Fig.~\ref{fig:DRSQ-Running-example:c} then shows that two-hop neighbors of $q$, $s_2, s_3, s_5,$ and $s_{11}$, use their own local information to derive their local range-skyline results and return these local range-skyline results to $q$'s one-hop neighbors, $s_1$ and $s_4$. After $s_1$ and $s_4$ receive the local range-skyline results from the two-hop neighbors of $q$, they will merge the received local range-skyline results and do the dominance check with the information of themselves. As Fig.~\ref{fig:DRSQ-Running-example:c} shows, the local skyline candidate sets of $s_1$ and $s_4$ are $LRSQ_{s_2}=\{s_1,s_2\}$ and $LRSQ_{s_3}\cup LRSQ_{s_5}\cup LRSQ_{s_{11}}=\{s_3,s_4,s_5\}$ respectively. Sensor nodes $s_1$ and $s_4$ then check the dominance relations between all the candidate data objects and obtain their local range-skyline results, $LRSQ_{s_1}=\{s_1,s_2\}$ and $LRSQ_{s_4}=\{s_4,s_5\}$ because of $s_5\triangleleft_q s_3$ respectively, as Fig.~\ref{fig:DRSQ-Running-example:d} shows. After aggregating the local range-skyline sets, $LRSQ_{s_1}$ and $LRSQ_{s_4}$, $s_1$ and $s_4$ return them to the query node $q$ respectively.

After receiving the local range-skyline sets from $s_1$ and $s_4$, query node $q$ merges these sets to derive a candidate set $RSQ_{candidate}=LRSQ_{s_1}\cup LRSQ_{s_4}=\{s_1,s_4,s_5\}$. Fig.~\ref{fig:DRSQ-Running-example:e} presents the step of obtaining a candidate set of $q$. Finally, in Fig.~\ref{fig:DRSQ-Running-example:f}, query node $q$ checks the dominance relations between all the data objects in local candidate set and derives the final range-skyline, $DRSQ(R,S,q)$ $=\{s_1,s_4,s_5\}$.

\subsubsection{The DRSQ Process}
\label{sec_drsq_process}
In this subsection, we introduce DRSQ process with pseudo-codes. The whole DRSQ process includes two parts: LRSQ and GRSQ processes. The pseudo-codes of Algorithm~\ref{alg:lrsq} and Algorithm~\ref{alg:crsq} respectively show the LRSQ process on a mobile sensor node and the GRSQ process on the query node as well as present the ideas and frameworks of our proposed approaches.
Note that each sensor node repeatedly runs the LRSQ process in Algorithm~\ref{alg:lrsq} for a query and thus continuously receives messages from the network. When a user (mobile device) issues a query $q$, the device floods query messages to its one-hop neighbors with a maximum hop count, \emph{Time-To-Live} (TTL). After flooding the query messages, the query node starts GRSQ process (in Algorithm~\ref{alg:crsq}) to collect the local skyline sets from its one-hop neighbors and then derives the final result, $RSQ_{distributed}$, for the query.

When each one-hop neighbor of $q$ receives the query messages, it will do the operations from Line~\ref{alg:lrsq:get_query} to Line~\ref{alg:lrsq:end_get_query} of LRSQ process (Algorithm~\ref{alg:lrsq}). If the TTL value in the received query message is larger than 0, it means that the mobile sensor node is an intermediate node in the routing path of the query and the sensor node will forward the query to its neighbors at Line~\ref{alg:lrsq:forward_query}. Otherwise, the mobile sensor node is an end node in the routing path of the query. The sensor node will stop forwarding the query message, add the data objects of itself to the local skyline set $RSQ_{local}$ at Line~\ref{alg:lrsq:add_to_rsq_local}, and then start to return the $RSQ_{local}$ to the query node (at Line~\ref{alg:lrsq:return_rsq_local_result}) through the reversed routing path of the query. Note that $RSQ_{local}$ is equal to the term, $LRSQ_{s{i}}(R,S,q)$, and we may use both interchangeably afterward in this paper.
\begin{algorithm2e}[t]
	\footnotesize
	\SetAlgoLined
	\KwIn{received message $m$ and neighbor list $list_{neighbor}$}	
	\KwOut{local range-skyline $RSQ_{local}$}
	$q\leftarrow$ new query object\tcc*[r]{create a temporary empty query object}
	$s\leftarrow$ new node\tcc*[r]{create a temporary empty source node}
	$RSQ_{local}\leftarrow \emptyset$\tcc*[r]{create a local range-skyline set}
	$o\leftarrow$ this.sense()\tcc*[r]{save self's environmental data in a temporary object}
	\uIf{$m.type==$ RSQ\_TYPE\label{alg:lrsq:get_query}}{
		$q\leftarrow$ this.parse($m$, \textit{RSQ\_TYPE})\tcc*[r]{save the query information to $q$}
		$s.address\leftarrow m.source\_address$\tcc*[r]{record the previous node $s$}
		\uIf{$m.TTL > 0$}{
			this.flood($m$, $m.TTL-1$)\label{alg:lrsq:forward_query}\tcc*[r]{forward message $m$ to all the neighbors}	
			\Return \;
		}
		\ElseIf{$m.TTL == 0$}{
			\If{$o$ locates in $q$'s range}{
				add $o$ into $RSQ_{local}$\label{alg:lrsq:add_to_rsq_local}\;
			}
		}
	}\label{alg:lrsq:end_get_query}
	\ElseIf(\tcc*[f]{LRSQ}){$m.type==$ RSQ\_REPLY\_TYPE\label{alg:lrsq:rsq_local_compute}}{
		$RSQ_{neighbor}\leftarrow \emptyset$\tcc*[r]{create a temporary range-skyline set}
		\tcc{get the neighbor's local range-skyline set}
		$RSQ_{neighbor}\leftarrow$ this.parse($m$, \textit{RSQ\_REPLY\_TYPE})\;
		\tcc{check dominance relations between the recieved objects and itself}
		int $isDominated = 0$\;
		\ForEach{data object $o'$ in $RSQ_{neighbor}$\label{alg:lrsq:dominance_check}}{
			\uIf{$o \triangleleft_q o'$}{
				\textbf{continue}\;
			}
			\ElseIf{$o' \triangleleft_q o$}{
				$isDominated=1$\;
			}
			$RSQ_{local}\leftarrow RSQ_{local}\cup \{o'\}$\;					
		}\label{alg:lrsq:end_dominance_check}
		\If{$isDominated==0$}{
			$RSQ_{local}\leftarrow RSQ_{local}\cup \{o\}$\label{alg:lrsq:add_to_local_skyline}\;
		}
	}\label{alg:lrsq:end_rsq_local_compute}
	\tcc{return $m'$ to $q$ through the previous sensor node $s$ in $q$'s routing path}
	$m'\leftarrow$ this.createMessage($RSQ_{local}$, \textit{RSQ\_REPLY\_TYPE})\tcc*[r]{only in DRSQ approach}
	this.forward($m'$, $s$)\label{alg:lrsq:return_rsq_local_result}\tcc*[r]{only in DRSQ approach}
	\Return $RSQ_{local}$\;
	\caption{LRSQ process on a mobile sensor node}
	\label{alg:lrsq}
\end{algorithm2e}
\begin{algorithm2e}[t]
	\footnotesize
	\SetAlgoLined
	\KwIn{received message $m$ and neighbor list $list_{neighbor}$}
	\KwOut{distributed range-skyline $RSQ_{distributed}$}
	$RSQ_{distributed}\leftarrow \emptyset$\tcc*[r]{create a set to save the distributed range-skyline}
	$RSQ_{candidate}\leftarrow \emptyset$\tcc*[r]{create a set to save the candidate range-skyline}
	int $i=0$\;
	\Repeat{$i==list_{neighbor}.length$}{
		\label{alg:crsq:get_local_rsq}
		\If{$m.type==$ RSQ\_REPLY\_TYPE}{ \label{alg:crsq:get_local_rsq:check_msg_type}		
			$RSQ_{neighbor}\leftarrow \emptyset$\tcc*[r]{create a temporary range-skyline set}
			\tcc{obtain the neighbor's local range-skyline set}
			$RSQ_{neighbor}\leftarrow$ this.parse($m$)\;
			$RSQ_{candidate}\leftarrow RSQ_{candidate} \cup RSQ_{neighbor}$\;
			$i++$\;
		}
	}\label{alg:crsq:end_get_local_rsq}
	\tcc{check dominance relations between the recieved candidiates and itself}
	\ForEach{data object $o$ in $RSQ_{candidate}$\label{alg:crsq:dominance_check}}{
		int $isDominated=0$\;
		\ForEach{data object $o'$ in $(RSQ_{candidate}-\{o\})$}{
			\If{$o' \triangleleft_q o$}{
				$isDominated=1$\;
				\textbf{break}\;
			}
		}
		\If{$isDominated==0$}{
			$RSQ_{distributed}\leftarrow RSQ_{distributed}\cup \{o\}$\;
		}
	}\label{alg:crsq:end_dominance_check}
	\Return $RSQ_{distributed}$\;
	\caption{GRSQ process on the query node}
	\label{alg:crsq}
\end{algorithm2e}

When an intermediate sensor node receives a response message for the query, it will perform the operations from Line~\ref{alg:lrsq:rsq_local_compute} to Line~\ref{alg:lrsq:end_rsq_local_compute} in Algorithm~\ref{alg:lrsq} to compute the latest local range-skyline $RSQ_{local}$. Since the received response message contains a local range-skyline set w.r.t. the neighboring node which sent this message, the intermediate mobile sensor node will save the received local range-skyline in a temporary set $RSQ_{neighbor}$. Note that all the data objects in $RSQ_{neighbor}$ do not dominate each other, so the sensor node will check the dominance relations between each data object $o'$ in $RSQ_{neighbor}$ and the data object $o$ of itself. If data object $o'$ in $RSQ_{neighbor}$ is not dominated by data object $o$, the data object $o'$ is still a local range-skyline member. The operations of dominance relation checking are presented from Line~\ref{alg:lrsq:dominance_check} to Line~\ref{alg:lrsq:end_dominance_check} of Algorithm~\ref{alg:lrsq}. The mobile sensor node executes the operation at Line~\ref{alg:lrsq:add_to_local_skyline} if the data object $o$ of itself is not dominated by any other data objects in $RSQ_{neighbor}$. It indicates that the sensed data object $o$ of the intermediate sensor node becomes one of the local range-skyline member. After the dominance validation, the intermediate sensor node keeps forwarding the response message, including the latest local range-skyline, back to the query node. All the intermediate sensor nodes do the above operations and update the local skyline set which is saved in the response message until the response message is received by the query node.

Algorithm~\ref{alg:crsq} describes the operations executed by the query node $q$ after it floods the query messages. From Line~\ref{alg:crsq:get_local_rsq} to Line~\ref{alg:crsq:end_get_local_rsq}, the query node collects all the local range-skyline sets from its one-hop neighbors and merges them into the candidate range-skyline set $RSQ_{candidate}$. Such a union operation is done to avoid recording the same data objects multiple times. Note that the Line~\ref{alg:crsq:get_local_rsq:check_msg_type} is a operation to check whether the received message is a reply message or not. Such a check can avoid the routing loop of a query message. In the considered environment, the query node $q$ is a sensor node and may be an $i$-hop neighbor of a node $s$, so $q$ may receive the query message from $s$ if $TTL > 2i-1$, where the appropriate value of $TTL$ will be discussed in Section~\ref{analysis}. Such a scenario may only occur when the query range $R$ is much larger than the transmission range $r$.

However, $RSQ_{candidate}$ is not the final result for the query because the data objects in all the received local range-skyline sets may dominate each other. So the query node executes the operations from Line~\ref{alg:crsq:dominance_check} to Line~\ref{alg:crsq:end_dominance_check} for checking the dominance relations between all the candidate points and then saves the non-dominated data objects in a set, $RSQ_{distributed}$. Finally, the query node (the user's device) returns the final range-skyline, $RSQ_{distributed}$, to the user.

\subsection{Distributed Continuous Range-Skyline Query}
This subsection is organized as follows. We will first present some important notations and assumptions for the DCRSQ process. Second, we will introduce the proposed DCRSQ approach with a running example extended from the CRSQ example in Fig.~\ref{fig:crsq_ex}. Last, the proposed DCRSQ process will be explained in details with some definitions.

\subsubsection{System Assumptions}
In order to make the DRSQ process able to support CRSQ in the DCRSQ process, some additional and modified assumptions of the system are needed and we will describe them in detail before introducing the DCRSQ process. Details of the system assumptions are given as follows:
\begin{itemize}
	\item Each mobile node can always obtain the spatial (location and mobility) and non-spatial (sensed data) informations of its one-hop neighbors and itself with its GPS equipment. We call such collected informations as \emph{local information}.	
	\item Each mobile node has a limited buffer to store the received CRSQ queries and each stored query will be continuously processed with local information until the query is expired.
	\item Each node also collects the information of existing queries from its neighbors while generating local information of itself.
	\item The size of a network packet (message) is fixed and one packet only can store one data object.
\end{itemize}

We here show all the important notations used in this paper in Table~\ref{notations}.
\begin{table}[h]
	\caption{Important Notations}
	\label{notations}
	\centering
	\footnotesize
	\begin{tabular}{c|p{0.75\columnwidth}}
		\hline
		\textbf{Notation} & \textbf{Description} \\
		\hline
		\hline
		$\mathcal{A}$ & The sensing area\\
		$S$ & The set of all the mobile nodes\\
		$s$ & Sensor node (mobile object) \\
		$q$ & Query node\\
		$R$ & Query range \\
		$r$ & Transmission range (Sensing range) \\		
		$N$ & Number of mobile sensor nodes\\	
		$N_R$ & Number of mobile sensor nodes in the query range\\
		$N_r$ & Number of neighbors in the transmission range\\
		$\varDelta t$ & A period of time for monitoring a CRSQ\\
		$T$ & A time interval for each sensor node return its information periodically\\
		$T_{safe}$ & A predicted time period that the answer may change\\
		$t_{M_{s}}$ & The monitoring time of node $s$ w.r.t. $q$\\
		$m_q$ & Query message \\
		$m_r$ & Reply message \\
		$TTL$ & Time-To-Live (Hop count) of a message \\
		$dist$ & Euclidean distance \\
		\hline
	\end{tabular}
\end{table}

\subsubsection{Overview of DCRSQ}
To support CRSQ, the DCRSQ process should take node mobility into account because the movement of nodes may cause frequent change of the answer. We thus adapt the idea of \citep{conf/mdm/KimCT08}, \emph{safe-time}, for considering the mobility of nodes. The safe time is derived for a node to predict when a neighbor node enters and leaves the query $q$'s range. If a neighboring object leaves the range of $q$, this object will not to a point of range-skyline and thus it will not be processed on the sensor node. It means that the node can determine that processing this neighboring object is necessary or not with the safe-time information.

\begin{figure}[ht]
	\begin{center}
		\includegraphics[width=0.45\textwidth,keepaspectratio]{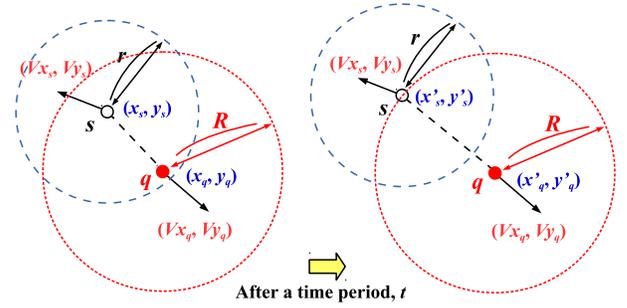}
	\end{center}
	\vspace{-10pt}
	\caption{An example for deriving the safe time $t_{\overline{qs}}$ when $dist(q,s)=R$}
	\label{fig:safe_time_ex}
\end{figure}

For deriving the precise safe time, we consider the movement of mobile nodes simultaneously and use Fig.~\ref{fig:safe_time_ex} to illustrate. Initially, $dist(q,s)$ is the distance from the query node $q$ to
sensor node $s$, where $dist(q,s)\leq R$. In the following, we present the equation to calculate the safe time $t_{|\overline{qs}|}$ when $dist(q,s)=R$. Suppose the initial location of object $q(s)$ is $(x_q,y_q)$($(x_s,y_s)$) with speed $(v_{x_q},v_{y_q})$($(v_{x_s},v_{y_s})$). Since $dist(q,s)=R$, we can have
\small{
	\begin{eqnarray}
	R^2&=&[(x_q+v_{x_q}*t_{\overline{qs}})-(x_s+v_{x_s}*t_{\overline{qs}})]^2
	\notag\\
	&+&[(y_q+v_{y_q}*t_{\overline{qs}})-(y_s+v_{y_s}*t_{\overline{qs}})]^2.
	\label{eq1}
	\end{eqnarray}
}\normalsize
After transposing each term, \eqref{eq1} can be a quadratic equation. Then, we can simply use the discriminant of quadratic equation to get two values of the safe time. We select the minimum positive value as the safe time $t_{\overline{qs}}$. Note that the above example shows the case of a leaving node. The other scenario is that a node may enter the range of query $q$. If the node is out of the $q$'s range and receives the query message in advance, the time periods of its entering and leaving can also be obtained by the same way.

\begin{figure}[ht]
	\begin{center}
		\includegraphics[width=0.32\textwidth,keepaspectratio]{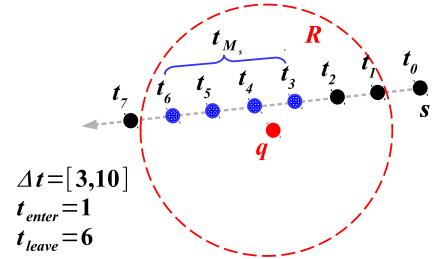}
	\end{center}
	\vspace{-10pt}
	\caption{The monitoring time ($t_{M_{s}}=[3,6]$) of node $s$ w.r.t. $q$ where $\varDelta t=[3,10]$,}
	\label{fig:safe_time_for_enter_leave_ex}
	\vspace{-10pt}
\end{figure}

For example, a query $q$ issues a CRSQ with $\varDelta t=[3,10]$ at time $t_0$. Fig~\ref{fig:safe_time_for_enter_leave_ex} shows the relative locations of $s$ and $q$ at each time step $t_i$ where $i\geq 0$. The query $q$ can use~\eqref{eq1} to obtain the safe time of node $s$, $t_{\overline{qs}}=[t_{enter},t_{leave}]=[1,6]$. So the exact monitoring time of node $s$ w.r.t $q$, $t_{M_{s}}$, is $[3,6]$, since the $q$ only concerns the results during the time $\varDelta t=[3,10]$ and the node $s$ will leave the range of $q$ after time $t_6$.

Combine the prediction of monitoring time with the LRSQ process, the continuous local range-skyline candidate sets also can be obtained. We refer such a process to \emph{continuous local range-skyline query process} (CLRSQ process). Consider the query $CRSQ(R,S,q,\varDelta t=[0,3])$ in Fig.~\ref{fig:crsq_ex}, the nodes $s_2, s_3, s_5,$ and $s_{11}$ in CLRSQ process (modified from the LRSQ process in Fig.~\ref{fig:DRSQ-Running-example}), respectively return the information of their local range-skyline candidate sets during the time $\varDelta t=[0,3]$. Node $s_2$ returns $CLRSQ_{s_2}=\{<\{s_1,s_2\},[t_0,t_1)>,<\{s_1,s_2\},[t_1,t_2)>,<\{s_2\},[t_2,t_3]>\}$ back to the intermediate node $s_1$. Nodes $s_3$, $s_5$, and $s_{11}$ respectively return $CLRSQ_{s_3}=\{<\{s_3,s_4\},[t_0,t_1)>,<\{s_3,s_{11}\},[t_1,t_2)>,<\{s_3,s_{11}\},[t_2,t_3]>\}$, $CLRSQ_{s_5}=\{<\{s_4,s_5\},[t_0,t_1)>,<\{s_4\},[t_1,t_2)>,<\{s_4\},[t_2,t_3]>\}$, and $CLRSQ_{s_{11}}=\{<\{s_3,s_4\},[t_0,t_1)>,<\{s_3,s_{11}\},[t_1,t_2)>,<\{s_3,s_{11}\},[t_2,t_3]>\}$ to the intermediate node $s_4$. In such a case, node $s_1$ will use the received $CLRSQ_{s_2}$ and the local information of itself to derive the $CLRSQ_{s_1}=\{<\{s_1,s_2\},[t_0,t_1)>,<\{s_1,s_2\},[t_1,t_2)>,<\{s_2\},[t_2,t_3]>\}$; and node $s_4$ will use the received $CLRSQ_{s_3}$, $CLRSQ_{s_5}$, $CLRSQ_{s_{11}}$, and the local information of itself to calculate the $CLRSQ_{s_4}=\{<\{s_4,s_5\},[t_0,t_1)>,<\{s_3,s_{11}\},[t_1,t_2)>,<\{s_3,s_{11}\},[t_2,t_3]>\}$. Finally, with received $CLRSQ_{s_1}$ and $CLRSQ_{s_4}$, the query node $q$ can obtain a predicted final result of CRSQ, $DCRSQ(R,S,q,\varDelta t)=\{<\{s_1,s_4,s_5\},[t_0,t_1)>,<\{s_1,s_3,s_{11}\},[t_1,t_2)>,<\{s_3,s_{11}\},[t_2,t_3]>\}$, at time $t_0$.

Unfortunately, the predicted result may not be correct. In the above example, $DCRSQ(R,S,q,\varDelta t)$ is not equal to $CRSQ(R,S,q,\varDelta t)$ since the query node $q$ cannot obtain the information of node $s_8$ at time $t_0$. So the result needs to be updated continuously. In the DCRSQ process, three cases may happen if a node enters the range of $q$. First, if $s_8$ received a query message from $q$, $s_8$ would return its local range-skyline while entering the range of $q$. It means that at least one node locates in the range of $q$ at time $t_0$ and it is the intermediate (relay) node between $s_8$ and $q$. In such a case, the intermediate node has the information of $s_8$ and uses that to derive the predicted CLRSQ result. Then the predicted DCRSQ result should be a correct answer, $DCRSQ(R,S,q,\varDelta t)=\{<\{s_1,s_4,s_5\},[t_0,t_1)>,<\{s_1,s_8,s_{11}\},[t_1,t_2)>,<\{s_8\},[t_2,t_3]>\}$. However, the mentioned example is not in this case since there is no intermediate node between $s_8$ and $q$. It is the second case that $s_8$ does not receive any information of $q$ at time $t_0$. According the assumptions of the DCRSQ process, $s_8$ will obtain the information of $q$ from its neighbors when $s_8$ enters the range of $q$ after time $t_1$. Hence, the RSQ results at time $t_1$ and $t_2$, $<\{s_1,s_3,s_{11}\},[t_1,t_2)>$ and $<\{s_3,s_{11}\},[t_2,t_3]>$, will be updated to $<\{s_1,s_8,s_{11}\},[t_1,t_2)>$ and $<\{s_8\},[t_2,t_3]>$, since $s_8 \triangleleft_q \{s_3,s_{11}\}$. The last case is that $s_8$ cannot successfully obtain the information of $q$ when it enters the range of $q$. Such a case will be recognized as an incorrect result and it only occurs when the mobile environment is too sparse.

\subsubsection{Description of DCRSQ}
According to Definition~\ref{crsq}, the system will process the CRSQ for a period of time $\varDelta t$ and derive the collection of possible answers. However, such a definition comes from the global and centralized view of system. In the previous subsection, the distributed method for processing RSQ has been introduced with Definition~\ref{define_lrsq} and Definition~\ref{define_drsq}. In the DCRSQ process, we use a mechanism to make each node able to predict the change of LRSQ answer with the node mobility. With above definitions and examples, the formal descriptions of CLRSQ and can be defined as Definition~\ref{define_clrsq} and Definition~\ref{define_dcrsq} respectively.
\begin{definition}
	\label{define_clrsq}
	\textbf{(Continuous Local Range-Skyline Query)}\\
	Suppose that the notations are defined as above and a query $CRSQ(R,S,q,\varDelta t)$ is issued by the query node $q$. After a mobile sensor node $s_j$ receives the query message from $q$, $s_j$ will return a collection of local range-skyline sets $LRSQ_{s_j}(R,S,q,t_i)$, where $t_i \in \varDelta t$ and $i$ is the number of local answer change. Formally, it is denoted as $CLRSQ_{s_j}(R,S,q,\varDelta t)=\{LRSQ_{s_j}(R,S,q,t_i)|t_i \in \varDelta t, i\in N\}$.
\end{definition}
According to Definition~\ref{define_clrsq}, the query node $q$ will receive the results of $CLRSQ_{s_j}(R,S,q,\varDelta t)$, where $s_j$ is one-hop neighbor of $q$, $1\leq j\leq k$, and $k$ is the maximum number of neighbors. Then the query node $q$ will take the union of received results, which are the local range-skyline sets for time $t_i$, as $RSQ_{candidate}(R,S,q,t_i)=\bigcup^{k}_{j=1}LRSQ_{s_j}(R,S,q,t_i)$. Thus, the candidate collection can be denoted as $CRSQ_{candidate}=\{<RSQ_{candidate}(R,S,q,t_0),[t_0,t_1)>,<RSQ_{candidate}(R,S,q,t_1),[t_1,t_2)>,\dots,<RSQ_{candidate}(R,S,q,t_i),[t_i,t_{end}]>\}$. After deriving $CRSQ_{candidate}$, $q$ will check the dominance relations of all objects in each $RSQ_{candidate}(R,S,q,t_i)$ set again and then obtain the final result $DRSQ(R,S,q,t_i)$ at each time $t_i$. We call such a process \emph{distributed continuous range-skyline query process} (DCRSQ process) and the definition is given in Definition~\ref{define_dcrsq}.

\begin{definition}
	\label{define_dcrsq}
	\textbf{(Distributed Continuous Range-Skyline Query)}\\
	Suppose the candidate collection $CRSQ_{candidate}$ of query node $q$ has been computed. Query node $q$ uses $CRSQ_{candidate}$ to derive a collection of the range-skyline sets for different time $t_i$ and we use $DCRSQ(R,S,q,\varDelta t)$ to represent such a collection of continuous range-skyline sets, where $DCRSQ(R,S,q,\varDelta t)=\{DRSQ(R,S,q,t_i)|t_i \in \varDelta t, i\in N\}$.
\end{definition}

Note that the fundamental process of DCRSQ is similar to DRSQ process mentioned in section~\ref{sec_drsq_process}. The main difference is that each mobile node $s_j$ generates a continuous local range-skyline set $CLRSQ_{s_j}(R,S,q,\varDelta t)$ with the safe-time information of candidate nodes. Thus the DCRSQ process can provide sufficient information to the query node $q$ for deriving, predicting, and updating the answer as time continuously goes on. Algorithm~\ref{alg:dcrsq} gives the high-level description of DCRSQ process. To implement Line 12 and Line 13 of Algorithm~\ref{alg:dcrsq}, we use the idea of \emph{sliding window}, which is already a widely used design in many domains. Since it is out of the scope of this paper, we will not address it. In addition, if the $\varDelta t$ is the specific time of the query issuing, $[t_i,t_i]$, $t_i\geq 0$, the query will be a snapshot RSQ and the DCRSQ will do the same process as the DRSQ process does.
\begin{algorithm2e}[t]
	\footnotesize
	\SetAlgoLined
	\KwIn{received message $m$ and neighbor list $list_{neighbor}$}
	$RSQ_{distributed}\leftarrow \emptyset$\tcc*[r]{create a set to save the distributed range-skyline}
	$RSQ_{local}\leftarrow \emptyset$\tcc*[r]{create a set to save the latest local range-skyline}
	$RSQ_{current\_local}\leftarrow \emptyset$\tcc*[r]{create a set to save the previous local range-skyline}
	$list_{safe\_time}\leftarrow \emptyset$\tcc*[r]{create a list to record the safe time of neighbors}
	\uIf{$this.nodeType==$ QUERY\_NODE}{
		\Repeat{$m$.isExpired()}{		
			\tcc{call the Algorithm~\ref{alg:crsq} to update the final distributed range-skyline}
			$RSQ_{distributed}\leftarrow$GRSQ($m$, $list_{neighbor}$)\;
		}\label{alg:dcrsq:end_get_local_rsq}
	}
	\ElseIf{$this.nodeType==$SENSOR\_NODE}{ \label{alg:dcrsq:get_local_rsq:check_msg_type}	
		\If{$m.type==$ RSQ\_REPLY\_TYPE}{	
			\tcc{derive the safe time of each neighbor}
			$list_{safe\_time}\leftarrow$ UpdateSafeTime($list_{neighbor}$)\;
			\tcc{call the Algorithm~\ref{alg:lrsq} to update the local range-skyline}
			$RSQ_{local}\leftarrow$ LRSQ($m$, $list_{neighbor}$)\;			
			\tcc{update the local range-skyline with the safe time values of neighbors}
			$RSQ_{local}\leftarrow$ SafeTimeCheck($RSQ_{local}$, $list_{safe\_time}$)\;
			\tcc{return the update message when the local range-skyline changes}
			\If{$RSQ_{current\_local}$ is not equal to $RSQ_{local}$}{
				$s\leftarrow$new node\;
				$s.address\leftarrow m.source\_address$\;
				$m'\leftarrow$ this.createMessage($RSQ_{local}$, \textit{RSQ\_REPLY\_TYPE})\;
				this.forward($m'$, $s$)\;	
			}	
		}		
	}
	\Return $RSQ_{distributed}$\;
	\caption{DCRSQ process on a mobile node (both query and sensor node)}
	\label{alg:dcrsq}
\end{algorithm2e}

\section{Cost Analysis and Discussion}
\label{analysis}
Suppose that $N$ mobile data objects are distributed independently and uniformly in the sensing area, $\mathcal{A}$, and each data object has $d$ attributes. If all the attributes of each object are in a uniform distribution, the skyline search problem can be treated as the problem of finding the maxima~\citep{Bentley:1978:ANM:322092.322095} in an $N\times d$ matrix. Hence, the expected size of skyline will be $n_{sky}=O((\ln N)^{d-1})$. In the considered environment, the query node does not need to process all the mobile sensor nodes (or data objects) for the range-skyline query and thus the expected size of range-skyline will be $n_{range-sky}=O((\ln N_{R})^{d-1})\leq O((\ln N)^{d-1})$, where $N_{R}$ is the number of mobile sensor nodes in the query range $R$ and $0\leq N_{R}\leq N$. Note that the value of $N_{R}$ is influenced by the value of $N$, sensing area $|\mathcal{A}|$ and the query range $R$ and $N_{R}=\lfloor\frac{\pi R^2N}{|\mathcal{A}|}\rfloor$. According to the above notations, the average number of data objects in the transmission range of a mobile sensor node will be $N_r=\lfloor\frac{\pi r^2N}{|\mathcal{A}|}\rfloor$, where $r$ is the transmission range of a mobile sensor node. If $N_r \leq 1$, the density of mobile sensor nodes is too sparse and thus it is too hard to route messages. In such a case, none of the conventional centralized and proposed approaches can perform well in the CRSQ processing. Hence, we only discuss the case, $N_r>1$, in this work. Note that we do not discuss the case here $N_{R}\leq1$ since none of mobile sensor nodes can serve this query.

In the considered IoMT, the mobile sensor nodes in the query range have to return information to the query node in hop-by-hop manner. The possibility distribution function of each hop in a multi-hop wireless environment has been discussed in~\citep{4277081} and we use that to obtain the possibility $P_i$ of the $i$th hop transmission. To obtain sufficient information for deriving the accurate result of a RSQ, the system must guarantee  that more than $N_R$ neighboring nodes of the query node can receive the query message. Then we can denote such an expected network cost for spreading the query message as
\begin{eqnarray}\label{eq2:expected_query_speard}
E[q_{spread}] = \sum_{i=1}^{k}N_r^i\prod_{j=1}^{i}P_j,
\end{eqnarray}
where $N_r^i$ is the average number of $i$th-hop neighbors with respect to the query node $q$.
Since all sensor nodes in the query range should be notified with the query messages from $q$, we can find a minimum value of $k\in \mathbb{N}$ that $E[q_{spread}]\geq N_R$.
Hence, the expected hop count $TTL_q$ can be derived by~\eqref{eq2:expected_query_speard} and $TTL_q=k$.

The process of data collection in the centralized approach is straightforward and each of the mobile sensor node which receives the query message will return the information of itself to the query node. Since the $i$th-hop neighbor needs to return an $i$-hop response message to the query node, the network cost of the reply messages for the $i$th-hop neighbors will be $N_r^i\times i\prod_{j=1}^{i}P_j$ without the cooperative pruning. Hence, the expected network cost for returning messages in the centralized approach can be denoted as
\begin{equation}\label{eq3:centralized_response_cost}
E_{centralized}[q_{response}] = \sum_{i=1}^{k}N_r^i\times i\prod_{j=1}^{i}P_j,
\end{equation}
where $k=TTL_q$ is determined by \eqref{eq2:expected_query_speard} with the constraint $E[q_{spread}]\geq N_R$.
In summary, the total network cost of the centralized approach for a RSQ, $q$, can be denoted as
\begin{equation}\label{eq4:centralized_cost}
E_{centralized}[q] = E[q_{spread}] + E_{centralized}[q_{response}].
\end{equation}

In the proposed approach, DRSQ process, each node derives the local range-skyline and the expected size of result is $O(\ln N_r)^{d-1}$. The reason is that DRSQ process combines the information filtering into the data collection, thus reducing a large number of irrelevant response messages. Hence, the network cost for replying the information can be denoted as
\begin{equation}\label{eq4:drsq_response_cost}
E_{DRSQ}[q_{response}] = \sum_{i=1}^k N_r^i (\ln N_r^i)^{d-1}P_{k-i},
\end{equation}
and $E_{DRSQ}[q_{response}]< E_{centralized}[q_{response}]$ in normal cases. So the total network cost of DRSQ process can be estimated as
\begin{equation}\label{eq5:drsq_cost}
E_{DRSQ}[q] = E[q_{spread}] + E_{DRSQ}[q_{response}].
\end{equation}

For monitoring a CRSQ query in the centralized approach, the query node has to spread the query message periodically during the time period $\varDelta t$ and each neighboring node also has to periodically return the information of itself. So the network cost of the centralized approach can be denoted as
\begin{equation}\label{eq6:crsq_centralized_cost}
E_{centralized}^{CRSQ}[q] = \dfrac{|\varDelta t|}{T}\times E_{centralized}[q],
\end{equation}
where $T$ is the time interval that each mobile sensor node periodically reports the updated information to the query node and the default value of $T$ is 1 second. In DCRSQ process, the query node does not have to periodically spread query messages since each mobile sensor node can buffered the information of the query. So the network cost is mainly influenced by the frequency of the answer changes and it can be derived by
\begin{equation}\label{eq7:crsq_dcrsq_cost}
E_{DCRSQ}^{CRSQ}[q] = E[q_{spread}] + \dfrac{|\varDelta t|}{\overline{T_{safe}}}\times E_{DRSQ}[q_{response}],
\end{equation}
where $\overline{T_{safe}}$ is the average safe-time that the result needs to be updated.

\section{Simulation Results}
\label{simulation}
All of the simulations are implemented as custom programs using C++ and executed on a Windows 7 system with an Intel i5-4460 3.20GHz CPU and 8GB memory. In all the simulation scenarios, the mobile sensor nodes are distributed uniformly and the results are reported with the average of 200 executions. The used mobility model is Random Way Point (RWP) and the network routing protocol is AODV~\citep{749281}. Since none of existing works provides distributed RSQ process over IoMT environments, we thereby use a centralized method~\citep{Dimitris:2005:SkylineComputation} as the compared centralized approach and it is executed on the query node for calculating the query results. In the centralized approach, the query node directly uses the flooding scheme to spread query messages and then collects information from moving data objects.

The proposed approach, DCRSQ process, can support (snapshot) range-skyline query and continuous range-skyline query. If $\varDelta t=[t_0,t_{end}]=0$, where $t_0=t_{end}$, the DCRSQ process will perform the DRSQ process for deriving the results of the (snapshot) RSQ at time $t_0$. We thus organize the simulation section as two scenarios. In the first scenario, we discuss the performance of DRSQ process in terms of \emph{response time} and \emph{number of messages}. The response time is the period of time from issuing a RSQ to obtaining the result during the DRSQ process. The number of messages represents the necessary network cost on data collection. 

In the second scenario, the performance of DCRSQ process is discussed in terms of \emph{number of accessed objects} and \emph{number of messages}. Additionally, the correctness of DCRSQ result is discussed in terms of \emph{precision} and \emph{recall}. In both scenarios, the following important factors are discussed: \emph{density (number of sensor nodes)}, \emph{number of queries}, \emph{query range}, and \emph{transmission range}. For validating the DCRSQ process, an additional factor, \emph{node speed}, is also in the discussion.

\subsection{Scenario I: Performance of DRSQ Process}
In the first scenario, we discuss the performance of DRSQ process. There are $100$ mobile sensor nodes in a $400m \times 400m$ square sensing area. The default transmission range is $75m$ and the node speed is $5 m/s$. Initially, mobile sensor nodes and queries are placed randomly in the area. The basic simulation settings for the first scenario are shown in TABLE~\ref{simulation_parameters_1} and, we execute the simulation 200 times to get the average results and the $95\%$ confidence intervals under each scenario.
\begin{table}[ht]
	\caption{Simulation Parameters for Scenario I}
	\label{simulation_parameters_1}
	\centering
	\footnotesize
	\begin{tabular}{p{3.7cm}cc}
		\hline
		\textbf{Parameter} & \textbf{Default Value} & \textbf{Range (type)}\\
		\hline
		\hline
		Sensing Area ($m \times m$) & 400 $\times$ 400 & --\\
		Number of Sensor Nodes & 100 & 50, 100, 150 ,200\\
		Number of Queries & 1 & 1, 2, 3, 4, 5\\
		Query Range, $R$ ($m$) & 80 & 60, 80, 100, 120\\
		Transmission Range, $r$ ($m$) & 75 & 50, 75, 100, 125\\
		Node Speed ($m/s$) & 2 & 1, 2, 3, 4, 5\\
		$TTL$ of Messages (centralized approach) & 5 & --\\
		Bandwidth ($Mb/s$)  & 2 & --\\
		\hline
	\end{tabular}
\end{table}

To the best of our knowledge, none of existing works proposed a method for processing range-skyline queries in such an environment, where the databases, CPUs, and memory are fully-distributed. We hence compare the proposed approach, DRSQ process, with a centralized approach which is a baseline. Note that the centralized approach does not use a powerful server. In the centralized approach, we assume that the query node is a sink node and can process the query with received information. The other mobile nodes only just forward the query and response messages without processing the local range-skyline.

\subsubsection{DRSQ: Density}
\begin{figure}[t]
	\centering
	\subfigure[Response Time]{
		\label{fig:sensor_num:time} %% label for 1st subfigure
		\includegraphics[width=0.245 \textwidth]{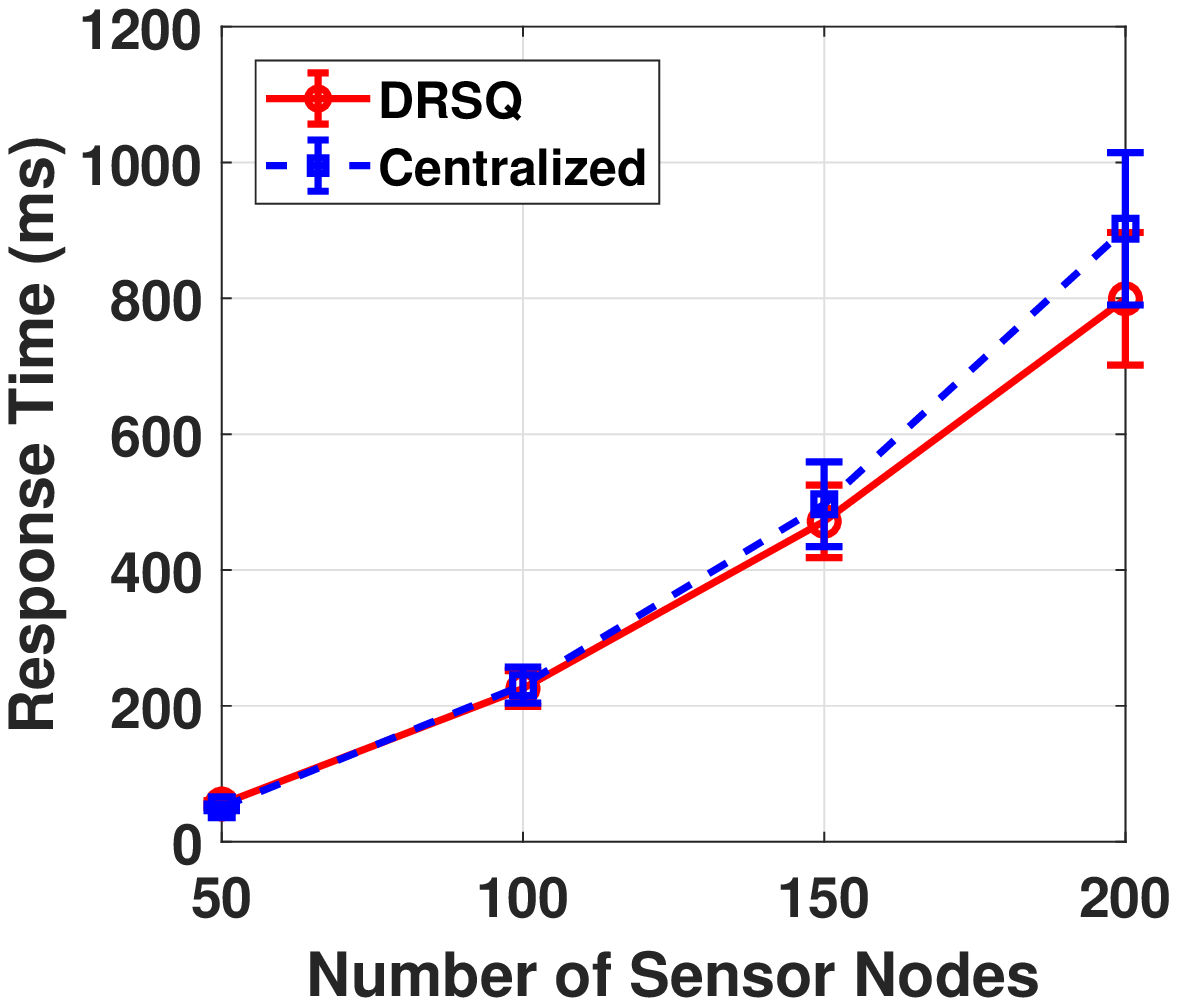}}%\hspace{.5in}
	\subfigure[Number of Messages]{
		\label{fig:sensor_num:messages} %% label for 2nd subfigure
		\includegraphics[width=0.245 \textwidth]{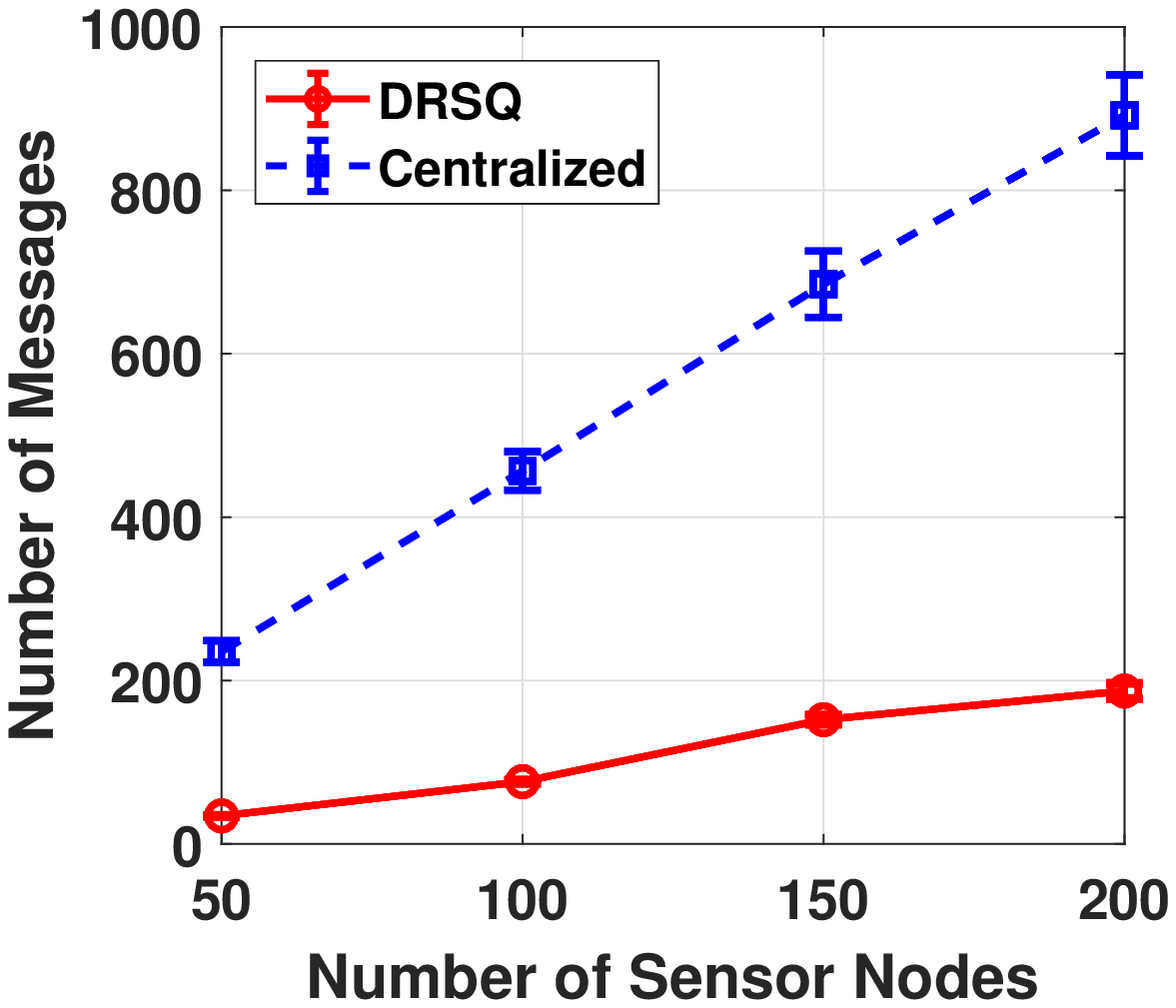}}
	\caption{Impact of the number of sensor nodes on \subref{fig:sensor_num:time} response time and
		\subref{fig:sensor_num:messages} number of messages
	}
	\label{fig:sensor_num} %% label for entire figure	
	\vspace{-10pt}
\end{figure}

Fig.~\ref{fig:sensor_num:time} shows that the response time of our proposed distributed approach, DRSQ process, is $10\%$ better than the centralized approach when the density becomes more dense (the number of sensor nodes increases). Although both centralized approach and DRSQ process need to collect the information from the other sensor nodes, DRSQ process spends less time on data collection. There are two reasons. One is that DRSQ process only needs to access the sensor nodes around the query range. Conversely, the centralized approach asks all the sensor nodes in the considered environment for data collection. The other reason is that the query node using the centralized approach needs to process many data objects and the computation overhead is thus heavy.

Fig.~\ref{fig:sensor_num:messages} presents that DRSQ process is almost $75\%$ better than the centralized approach in term of number of messages. In the DRSQ process, each mobile sensor node collects the information of its neighbors and derives a local RSQ result before sending a response message to the query node. Such a process can effectively prune a lot of irrelevant information from data objects (sensor nodes) and thus cost less number of messages on returning the local range-skyline. On the contrary, the centralized approach just floods query messages and collects the information from all the neighboring mobile sensor nodes to derive the range-skyline. It thus wastes more network cost on data collection.
\begin{figure}[t]
	\centering
	\subfigure[Response Time]{
		\label{fig:query_num:time} %% label for 1st subfigure
		\includegraphics[width=0.245 \textwidth]{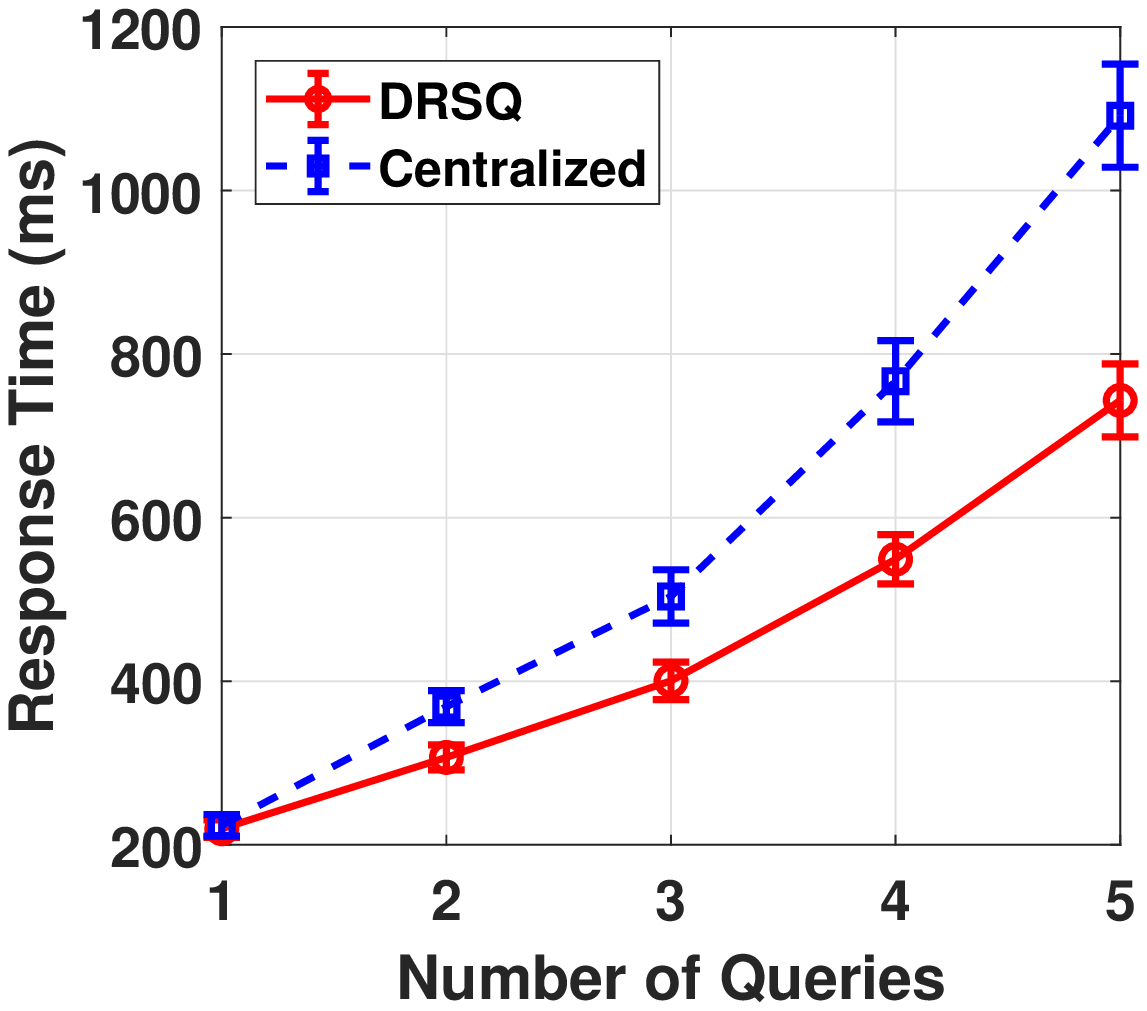}}%\hspace{.5in}
	\subfigure[Number of Messages]{
		\label{fig:query_num:messages} %% label for 2nd subfigure
		\includegraphics[width=0.245 \textwidth]{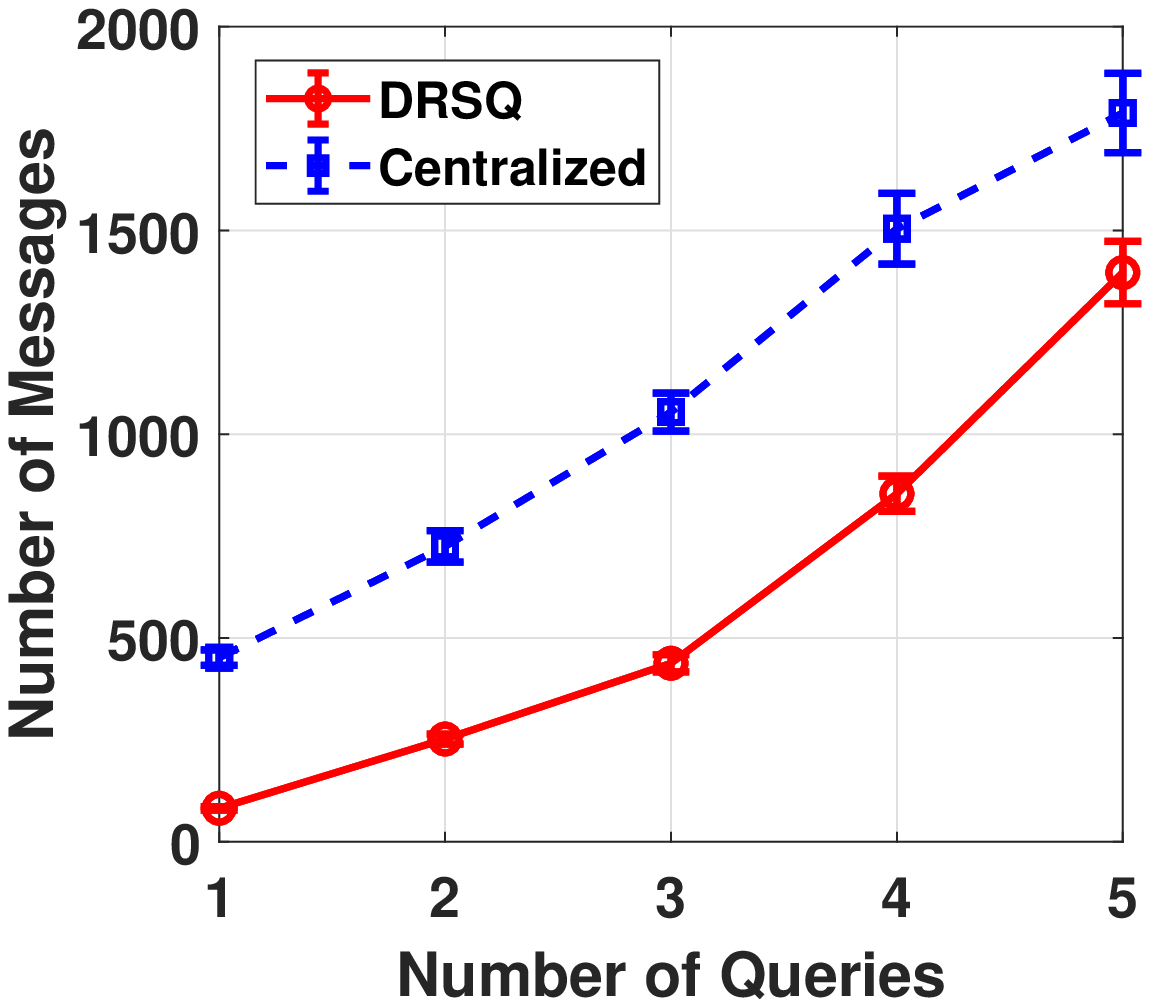}}
	\caption{Impact of the number of queries on \subref{fig:query_num:time} response time and
		\subref{fig:query_num:messages} number of messages
	}
	\label{fig:query_num} %% label for entire figure	
	\vspace{-10pt}
\end{figure}
\begin{figure}[t]
	\centering
	\subfigure[Response Time]{
		\label{fig:query_range:time} %% label for 1st subfigure
		\includegraphics[width=0.245 \textwidth]{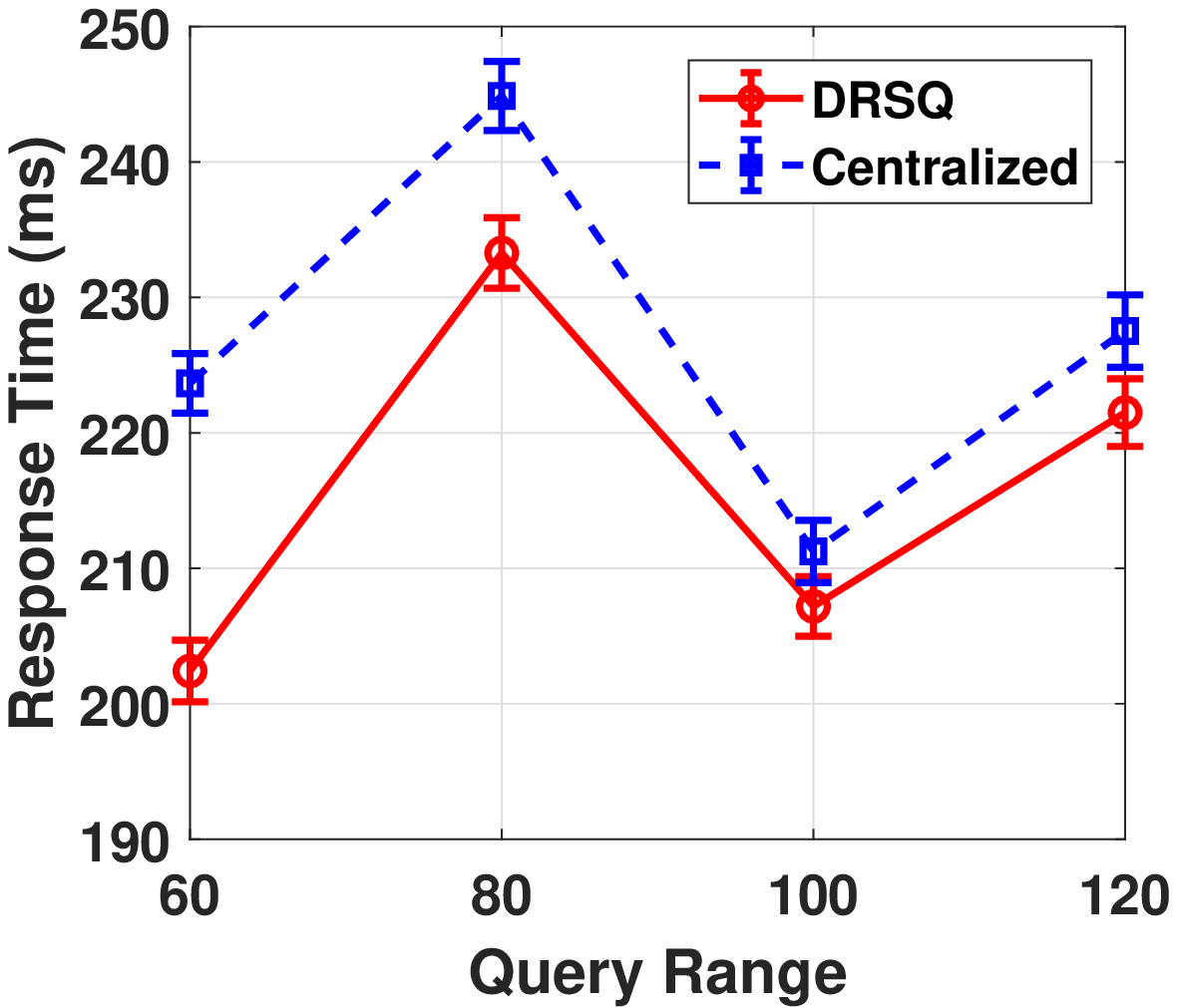}}%\hspace{.5in}
	\subfigure[Number of Messages]{
		\label{fig:query_range:messages} %% label for 2nd subfigure
		\includegraphics[width=0.245 \textwidth]{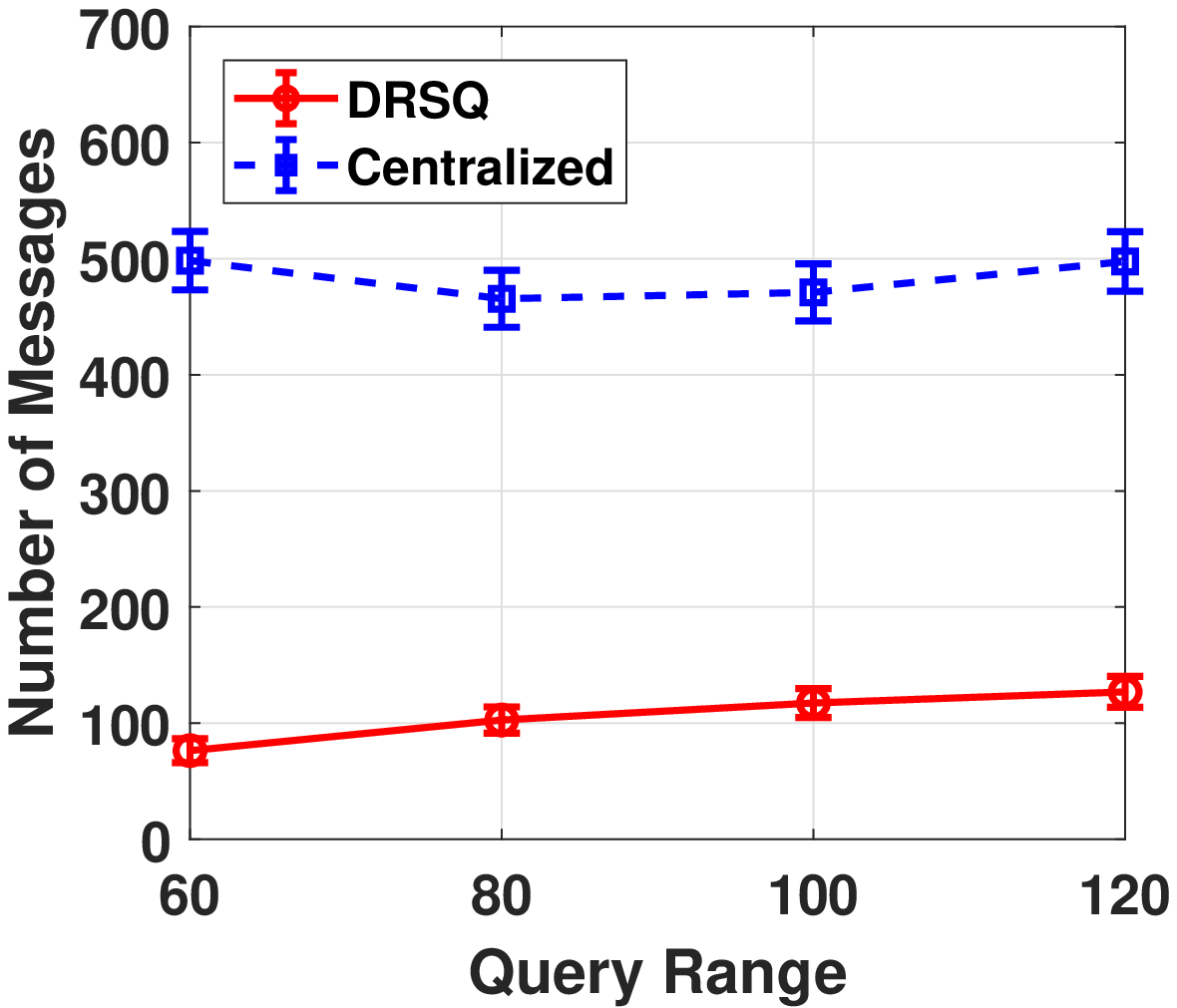}}
	\caption{Impact of query range on \subref{fig:query_range:time} response time and
		\subref{fig:query_range:messages} number of messages
	}
	\label{fig:query_range} %% label for entire figure	
	\vspace{-10pt}
\end{figure}
\begin{figure}[t]
	\centering
	\subfigure[Response Time]{
		\label{fig:radio_range:time} %% label for 1st subfigure
		\includegraphics[width=0.245 \textwidth]{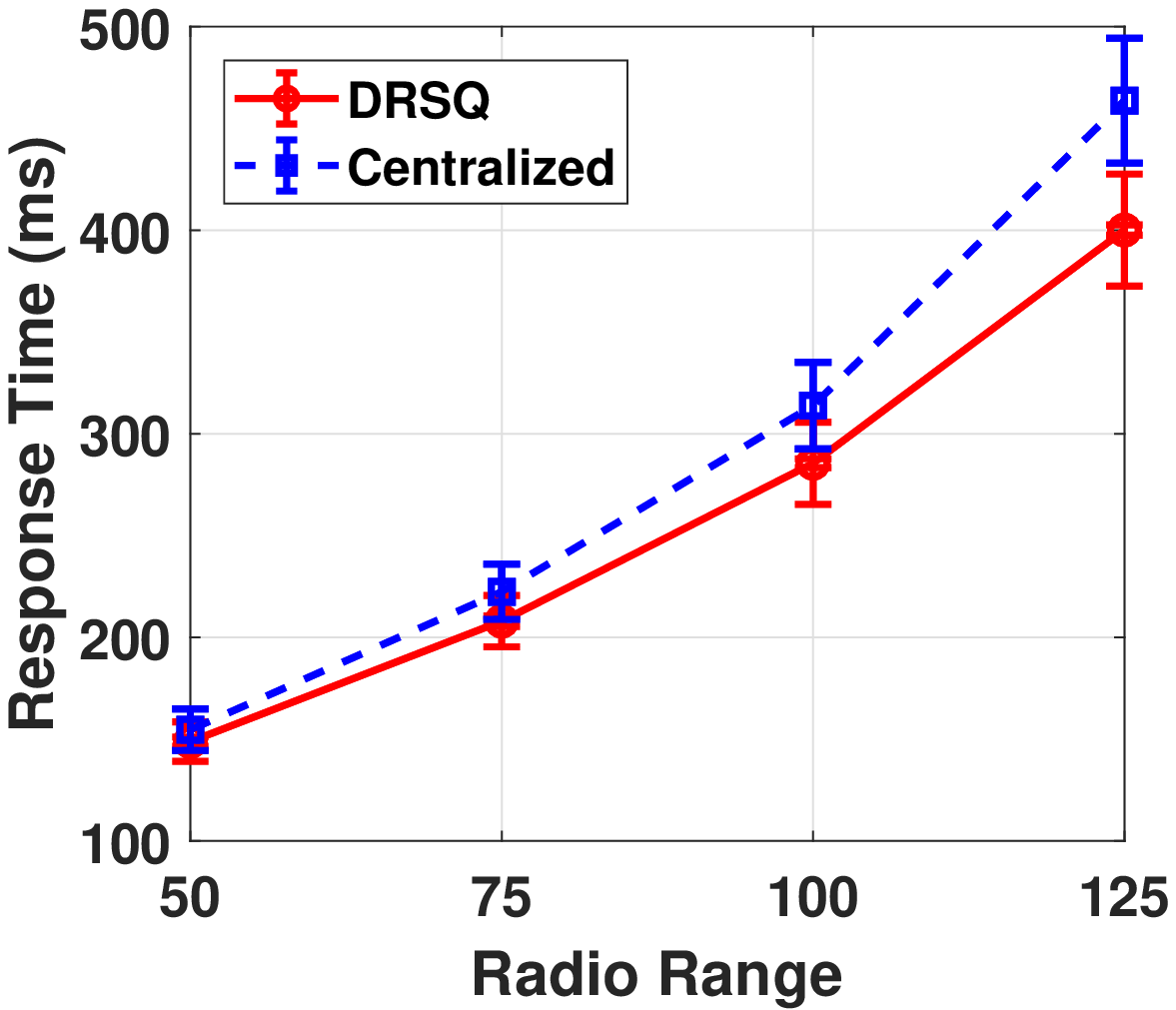}}%\hspace{.5in}
	\subfigure[Number of Messages]{
		\label{fig:radio_range:messages} %% label for 2nd subfigure
		\includegraphics[width=0.245 \textwidth]{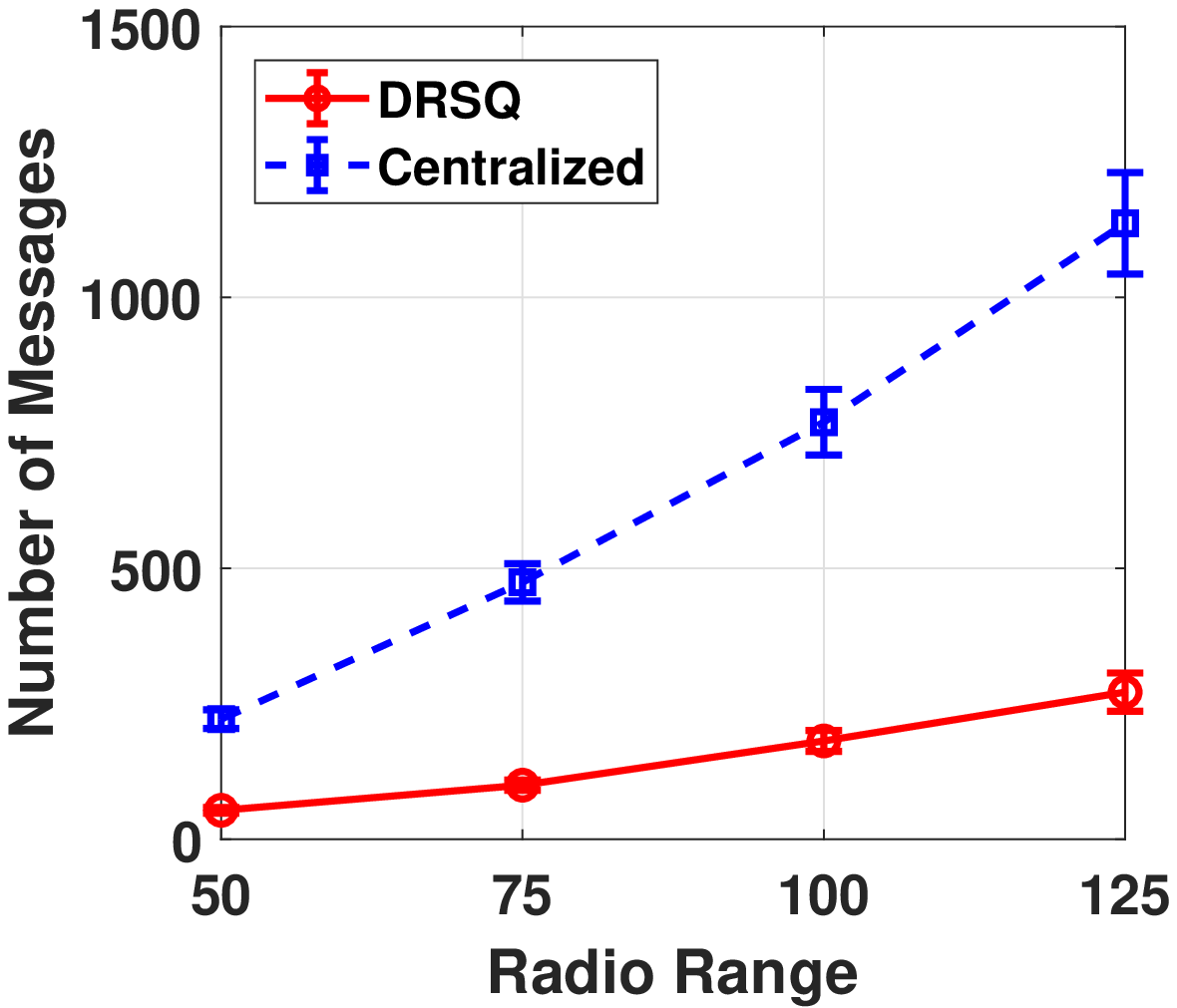}}
	\caption{Impact of transmission range on \subref{fig:radio_range:time} response time and
		\subref{fig:radio_range:messages} number of messages
	}
	\label{fig:radio_range} %% label for entire figure
	\vspace{-10pt}
\end{figure}

\subsubsection{DRSQ: Number of Queries}
In this subsection, we discuss the impact of the number of queries. The number of queries indicates the maximum number of queries concurrently processed in the system. Fig.~\ref{fig:query_num:time} shows that the response time of DRSQ process is much better than the centralized approach as the number of queries increases. When the number of queries is $5$, the DRSQ process performs almost $30\%$ faster than the centralized approach does. The DRSQ process only needs to access the sensor nodes which are around the query range. In contrast, the centralized approach floods query messages to asks all the sensor nodes in the considered environment for data collection. As the number of queries increases, a large amount of flooding messages harms the network routing performance and thus increases the response time. In addition, the query node using the centralized approach needs to process more information from data objects, so the overhead of RSQ computing on the query node becomes heavy. This can be verified by the experimental results shown in Fig.~\ref{fig:query_num:messages}. As the result indicates, the DRSQ process costs $20\%$ to $50\%$ fewer number of messages than the centralized approach does since it can avoid irrelevant data objects during data collection, and thus reducing a large amount of duplicated messages for data transmissions.

\subsubsection{DRSQ: Query Range}
Different values on the query range also influence the performance of query processing. Fig.~\ref{fig:query_range:time} shows that the DRSQ process outperforms the centralized approach by $2\%$ to $10\%$ in term of response time with all different range values. When the query range $R$ is smaller than the transmission range $r$, the probability that the whole query range falls in the transmission range is high and it thus is easier for the query node to obtain the RSQ result by collecting sufficient information from its one-hop neighbors. When the query range $R$ is larger than the transmission range $r$, the DRSQ process only needs to access the sensor nodes which are around the query range instead of accessing all the sensor nodes as the centralized approach does. So, the DRSQ process can outperform the centralized approach on response time. Fig.~\ref{fig:query_range:messages} shows that DRSQ saves almost $80\%$ transmission cost in comparison with the centralized approach. The reason is that the DRSQ process can effectively prune the irrelevant information from data objects during data (local skyline) collection.

\subsubsection{DRSQ: Transmission Range}
The last important impact is the transmission range $r$ of a sensor node. As Fig.~\ref{fig:radio_range:time} indicates, the distributed approach outperforms the centralized approach by $5\%$ to $10\%$ with different transmission ranges in terms of the response time. Unlike the dramatic increasing response time of the centralized approach, the response time of DRSQ process increases more gently. The centralized approach has to do the dominance checks after it receives a large amount of information from the neighboring sensor nodes. So, it needs more computation overhead. Instead, the DRSQ process can avoid the irrelevant data objects in a distributed way during the information collection. The query node only processes the one-hop neighbors' local range-skyline sets whose sizes are much smaller than the sizes of the data sets in the centralized approach on the query node. Effectively pruning irrelevant data objects in a distributed way also reduces a lot of required messages for returning the local range-skyline sets to the query node and this trend is demonstrated in Fig.~\ref{fig:radio_range:messages}. DRSQ can save $60\%$ to $70\%$ transmission cost on the data collection.

\subsection{Scenario II: Performance of DCRSQ Process}
In the second scenario, we present the performance results of DCRSQ process. The duration of each query $\varDelta t$ is $10$ seconds and the total duration of the simulation is 60 seconds. Initially, the mobile sensor nodes and query nodes are placed randomly in a $500m \times 500m$ square area. For each simulation set, we execute the simulation 200 times to get the average results and the $95\%$ confidence intervals.
% redundant sentence 2018-01-25-liu
% The time period of each run is 60 seconds and the $\varDelta t$ of each query is 10 seconds. 
%
The $t_0$ of each query's $\varDelta t$ is randomly generated from second 1 to 50. The other important settings are shown in Table~\ref{simulation_parameters_2}.

The system will continuously return results for a CRSQ query within the time period $\varDelta t$, so it is difficult to measure the response time precisely. Instead, we observe the number of accessed objects (collected data objects) on each query node. If the number of accessed objects on the query node is small, it means that the efficiency of DCRSQ process is better since a large number of irrelevant data objects are skipped during message routing. For the DCRSQ process in a mobile environment, the node speed is one of the important factors. If the node speed becomes fast, it may lead the answer changing more frequently and the overhead of processing CRSQ queries also becomes heavier. We thus discuss the impact of node speed on the performance of DCRSQ process. In addition, we use a server to check the correctness of results, generated by the DCRSQ process and the centralized approaches respectively in terms of precision and recall. Note that the server has a global knowledge of all the data objects and always generates the correct answer for a query.
\begin{table}[ht]
	\caption{Simulation Parameters for Scenario II}
	\label{simulation_parameters_2}
	\centering
	\footnotesize
	\begin{tabular}{p{3.7cm}cc}
		\hline
		\textbf{Parameter} & \textbf{Default Value} & \textbf{Range (type)}\\
		\hline
		\hline
		Sensing Area ($m \times m$) & 500 $\times$ 500 & --\\
		Simulation time (seconds) & 60 & --\\
		Number of Sensor Nodes & 60 & 30, 60, 90, 120\\
		Number of Queries & 1 & 1 to 10\\
		Query Range, $R$ ($m$) & 100 & 50, 100, 150, 200\\
		$\varDelta t$ of a Query (seconds) & 10 & --\\
		Transmission Range, $r$ ($m$) & 75 & 50, 75, 100, 125\\
		Maximum of Node Speed ($m/s$) & 10 & 5, 10, 15, 20, 25, 30\\
		$TTL$ of Messages (centralized approach) & 5 & --\\
		Bandwidth ($Mb/s$)  & 2 & --\\
		\hline
	\end{tabular}
\end{table}

\subsubsection{DCRSQ: Density}
\begin{figure*}[t]
	\centering
	\subfigure[Number of Accessed Objects]{
		\label{fig:dcrsq:sensor_num:access} %% label for 1st subfigure
		\includegraphics[width=0.245 \textwidth]{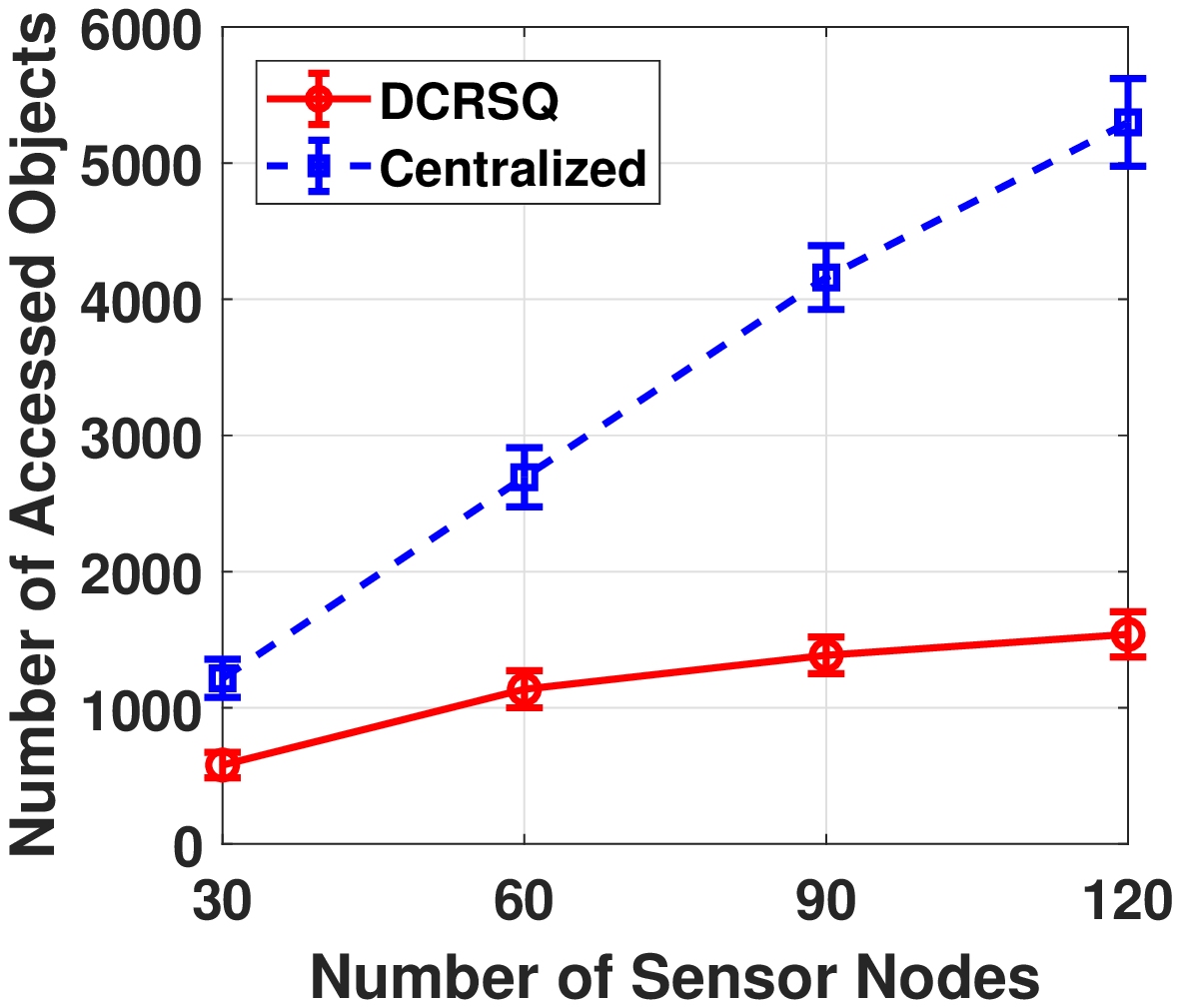}}%\hspace{.5in}
	\subfigure[Number of Messages]{
		\label{fig:dcrsq:sensor_num:messages} %% label for 2nd subfigure
		\includegraphics[width=0.245 \textwidth]{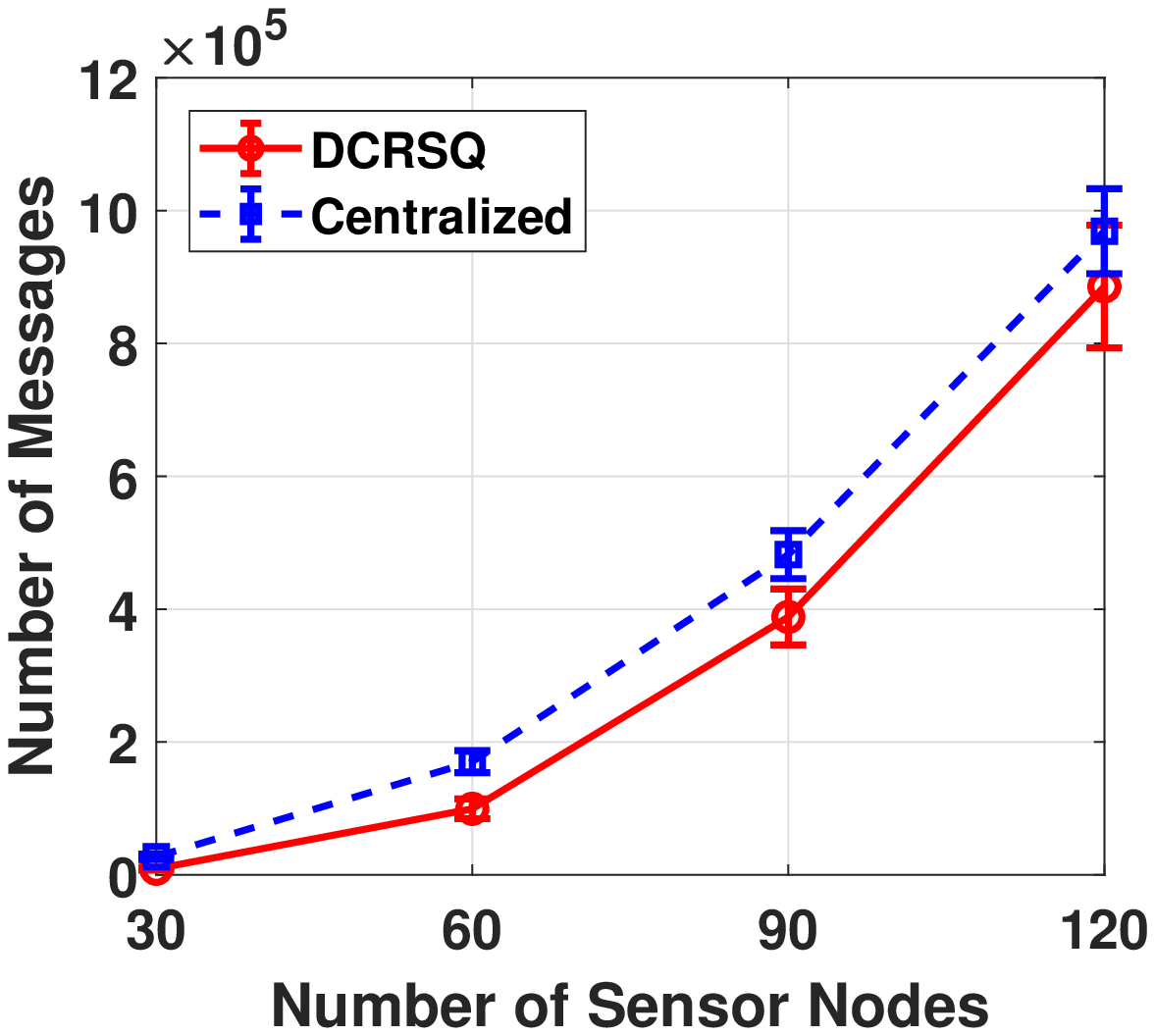}}%\hspace{.5in}
	\subfigure[Precision]{
		\label{fig:dcrsq:sensor_num:precision} %% label for 2nd subfigure
		\includegraphics[width=0.245 \textwidth]{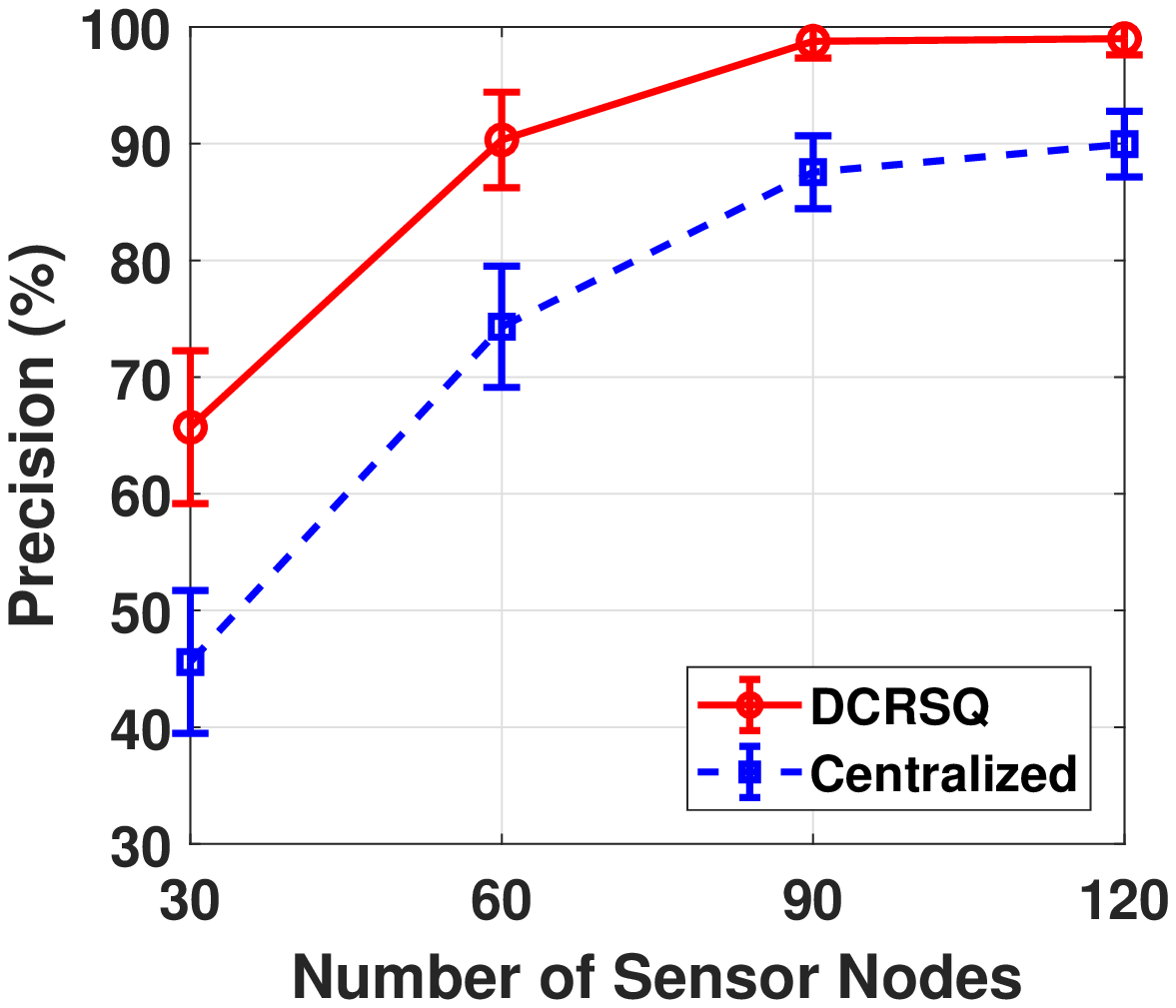}}%\hspace{.5in}
	\subfigure[Recall]{
		\label{fig:dcrsq:sensor_num:recall} %% label for 2nd subfigure
		\includegraphics[width=0.245 \textwidth]{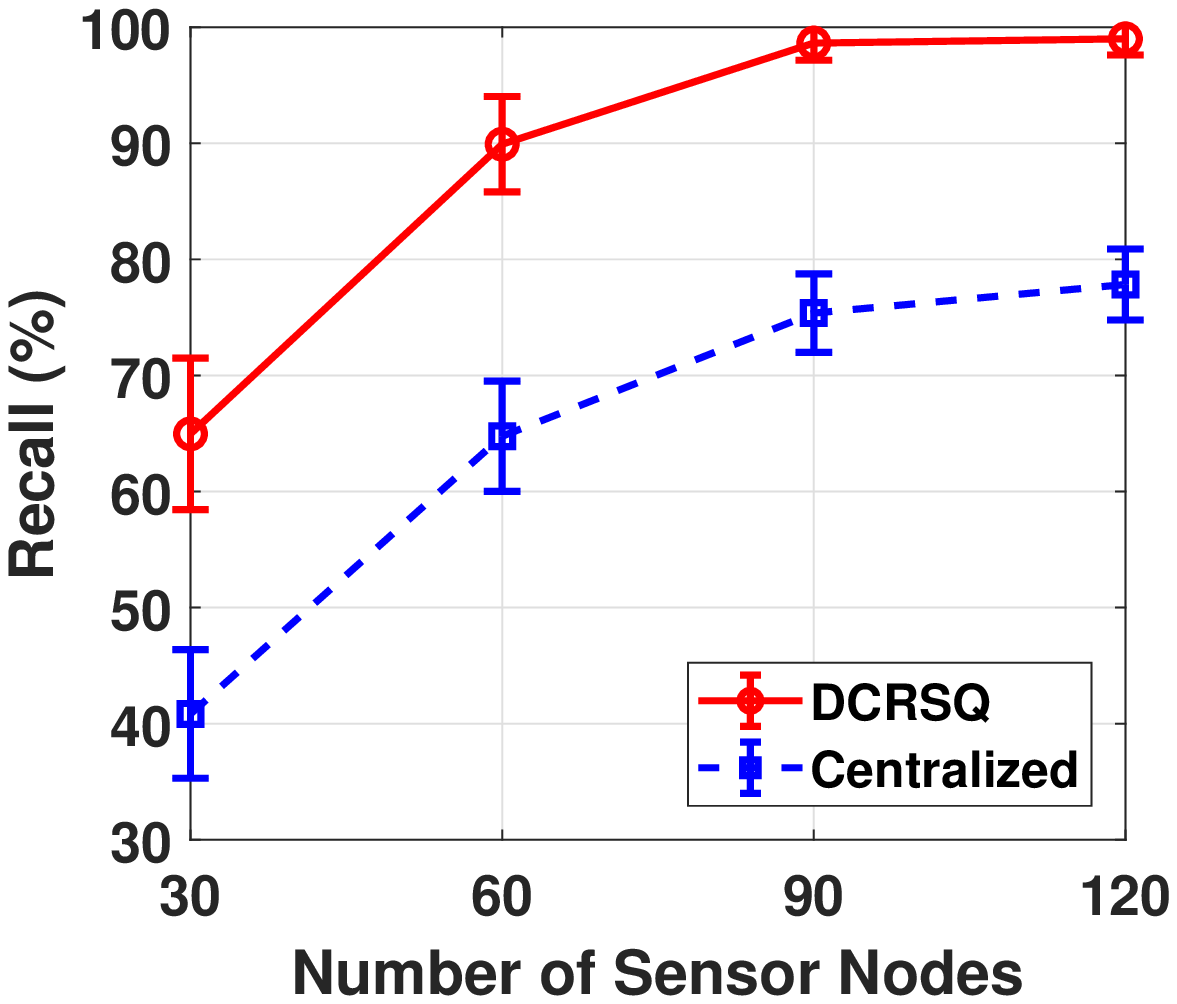}}
	\caption{Impact of the number of sensor nodes on \subref{fig:dcrsq:sensor_num:access} number of accessed objects, \subref{fig:dcrsq:sensor_num:messages} number of messages, \subref{fig:dcrsq:sensor_num:precision} precision, and \subref{fig:dcrsq:sensor_num:recall} recall
	}
	\label{fig:dcrsq:sensor_num} %% label for entire figure
	\vspace{-10pt}
\end{figure*}

Fig.~\ref{fig:dcrsq:sensor_num:access} shows that DCRSQ process is better than the centralized approach in terms of number of accessed objects. When the number of mobile sensors increases, the total number of accessed objects in the proposed method remains constant that is no more 1100 nodes. In contrast, using the centralized way, a query point needs about 2 to 5.5 times more nodes to derive the final range-skyline. This is due to no process of discarding irrelevant moving objects during data collection (local range-skyline processing) in the centralized approach. In summary, DCRSQ can save $50\%$ to $80\%$ computational cost in average for a query.

Similarly, Fig.~\ref{fig:dcrsq:sensor_num:messages} shows that DCRSQ process is better than the centralized approach in terms of number of messages. In DCRSQ process, each mobile sensor node collects the information of its neighbors and derives a local RSQ result before sending a response message to the query node. Such a process can prune a lot of moving objects which will not be the candidates and thus cost less number of messages on returning local range-skyline. On the other hand, the centralized approach just floods query messages and collects the information of all neighboring mobile sensor nodes for deriving the final range-skyline. So, the centralized approach wastes $10\%$ to $20\%$ more network cost on data collection.

For the accuracy, each mobile sensor node in the centralized approach does not consider the prediction location of the neighbor nodes. Each sensor node only forwards the collected information to the query point. The result of final range-skyline may be inaccurate, so we compare the results of DCRSQ process and centralized approach with the answer in a server to measure the precision and recall. Fig.~\ref{fig:dcrsq:sensor_num:precision} and Fig.~\ref{fig:dcrsq:sensor_num:recall} show that both precision and recall of the centralized approach are worse than DCRSQ process by $10\%$ to $20\%$. Moreover, precision and recall of DCRSQ process are almost $100\%$ correct when the number of sensor nodes is large.
\begin{figure*}[t]
	\centering
	\subfigure[Number of Accessed Objects]{
		\label{fig:dcrsq:query_num:access} %% label for 1st subfigure
		\includegraphics[width=0.245 \textwidth]{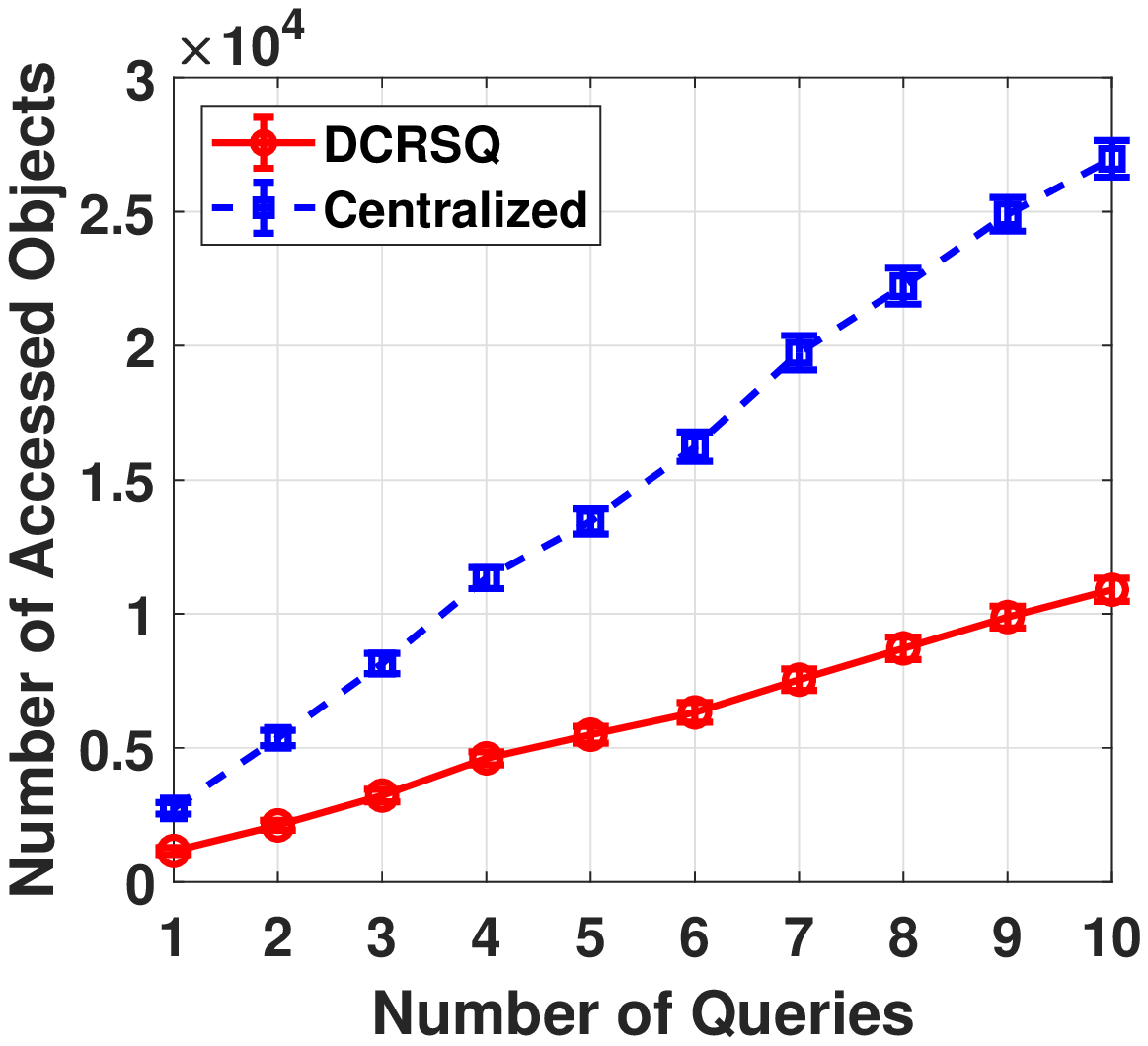}}%\hspace{.5in}
	\subfigure[Number of Messages]{
		\label{fig:dcrsq:query_num:messages} %% label for 2nd subfigure
		\includegraphics[width=0.245 \textwidth]{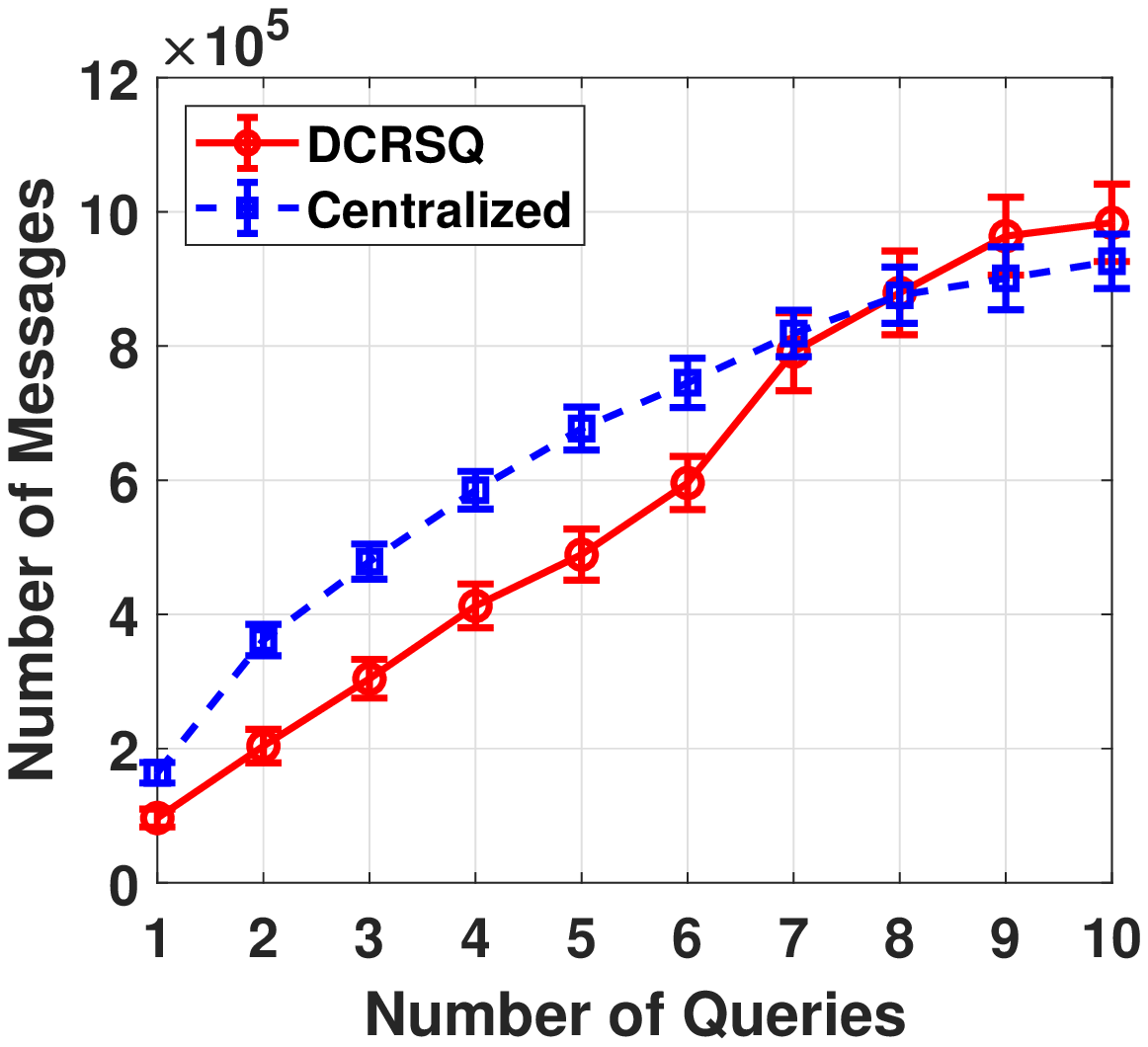}}%\hspace{.5in}
	\subfigure[Precision]{
		\label{fig:dcrsq:query_num:precision} %% label for 2nd subfigure
		\includegraphics[width=0.245 \textwidth]{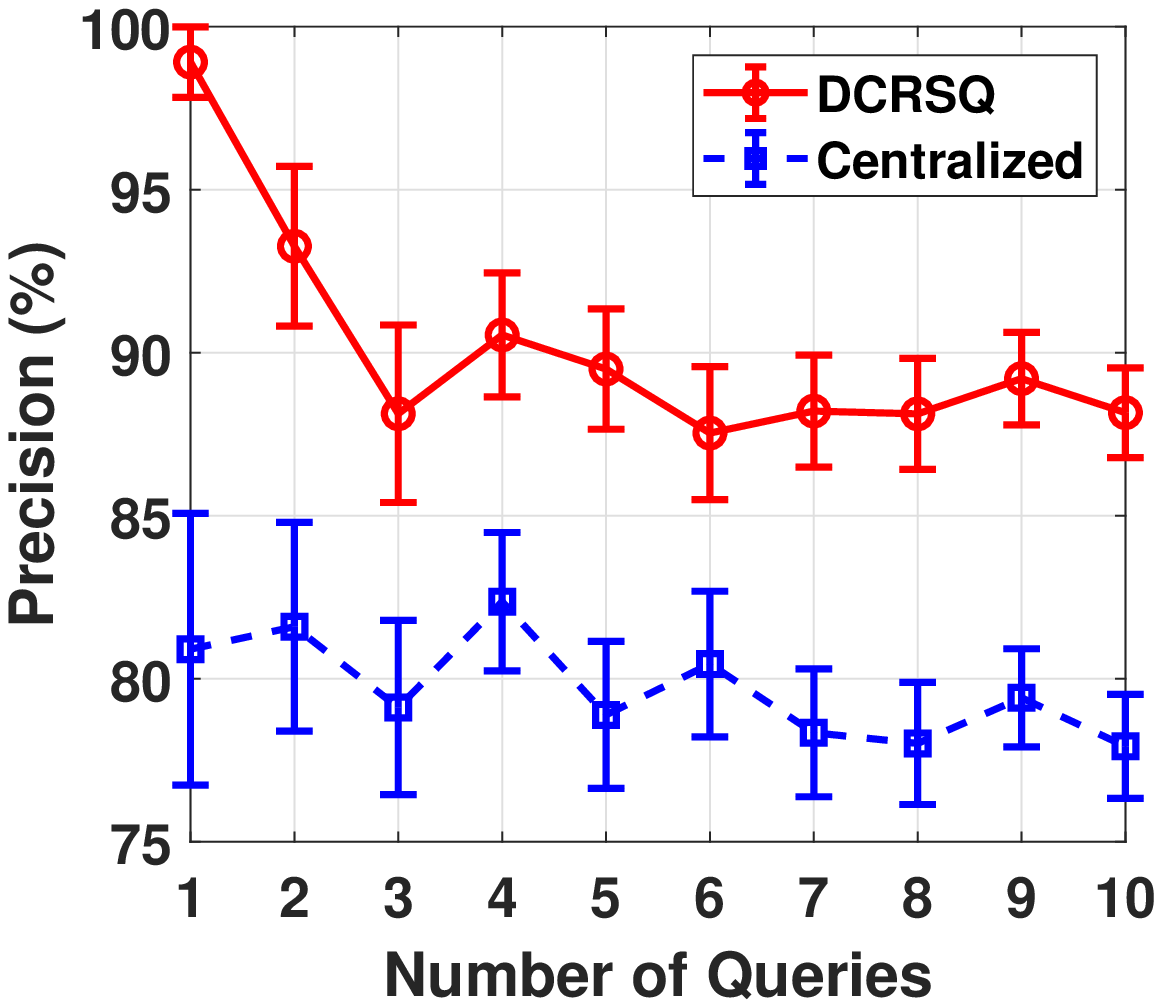}}%\hspace{.5in}
	\subfigure[Recall]{
		\label{fig:dcrsq:query_num:recall} %% label for 2nd subfigure
		\includegraphics[width=0.245 \textwidth]{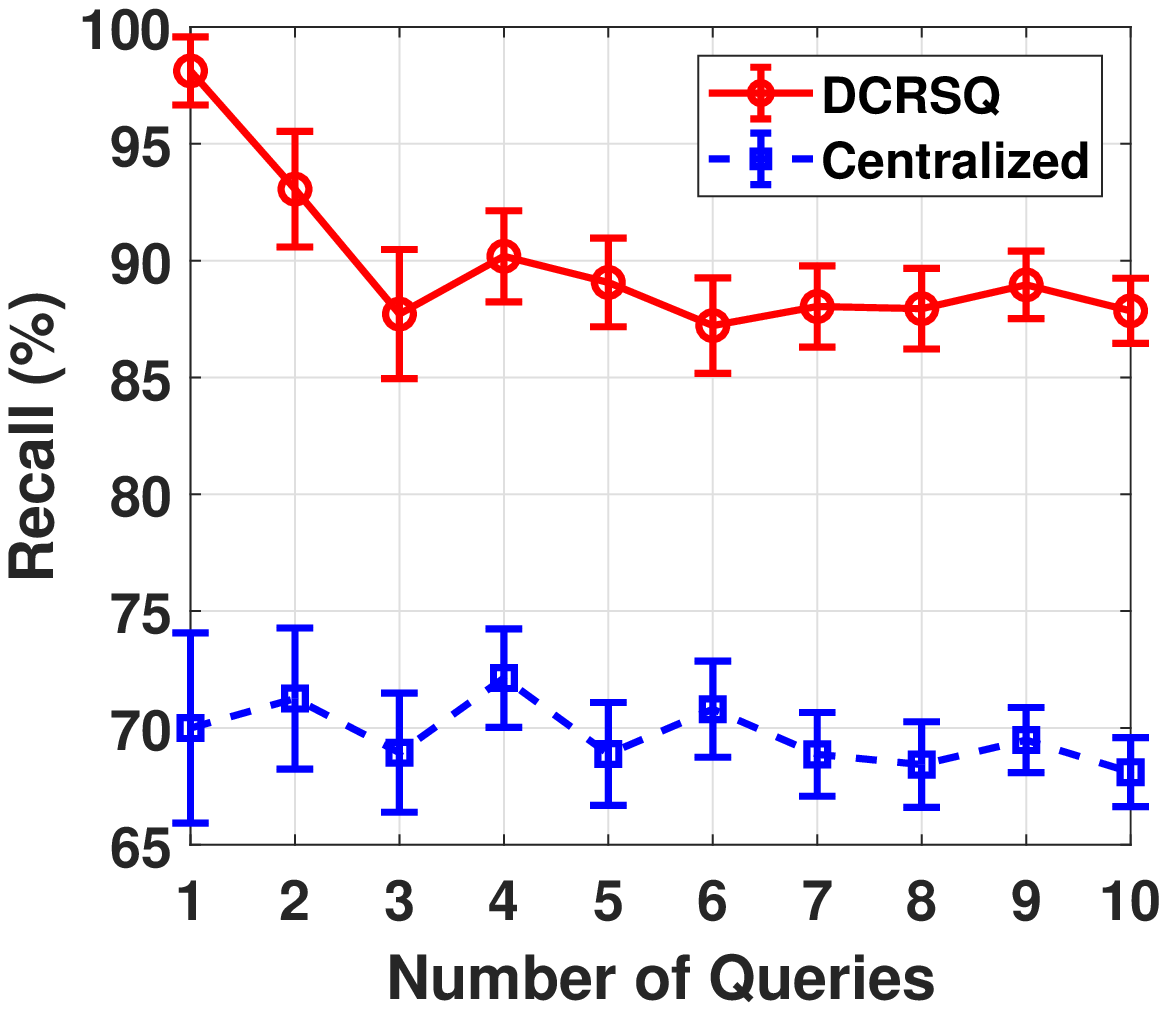}}
	\caption{Impact of the number of queries on \subref{fig:dcrsq:query_num:access} number of accessed objects, \subref{fig:dcrsq:query_num:messages} number of messages, \subref{fig:dcrsq:query_num:precision} precision, and \subref{fig:dcrsq:query_num:recall} recall
	}
	\label{fig:dcrsq:query_num} %% label for entire figure
	\vspace{-10pt}
\end{figure*}

\subsubsection{DCRSQ: Number of Queries}
In this simulation experiment, we are interested in evaluating the performance of different number of queries issued simultaneously. We set the number of queries from 1 to 10, which means that there are at most 10 queries within the time period $\varDelta t$ in the simulation. Fig.~\ref{fig:dcrsq:query_num:access} and Fig.~\ref{fig:dcrsq:query_num:messages} show that as the number of queries increases, the number of messages and the number of accessed objects grow up respectively for both approaches. However, DCRSQ process only needs to access $50\%$ to $70\%$ amount of data objects in the derivation comparing to the centralized approach. DCRSQ outperforms the centralized approach and saves about $30\%$ on network cost when the number of queries is smaller then 7. If the number of queries exceeds 7, DCRSQ will cost more network messages. The reason is that the neighboring nodes cooperatively process the local result of CRSQ and some of them store duplicated local results (objects). That is, the query node may receive many duplicated reply messages.

Fig.~\ref{fig:dcrsq:query_num:precision} and Fig.~\ref{fig:dcrsq:query_num:recall} show that DCRSQ process is better than the centralized approach in terms of precision and recall. As the number of queries increases, DCRSQ process still can achieve almost $90\%$ correctness and outperforms the centralized approach by $12\%$ to $25\%$ in terms of precision and recall, respectively. The reason is that DCRSQ process does not compute the irrelevant data objects anymore since they have already been filtered by the neighboring mobile nodes. Thus, a query point only checks the dominance objects that can guarantee to be in the final range-skyline. The other reason is that each mobile sensor node in the centralized way just gathers the information of its neighbors and sends the information back to the query node for deriving final range-skyline. Therefore, there is a possibility to compute a large number of irrelevant data objects for the query node. Thus, it makes the precision and recall of the centralized approach worse than the DCRSQ process. Although DCRSQ still has many redundant transmissions we mentioned above (in Fig.~\ref{fig:dcrsq:query_num:messages}), such duplicate local results significantly recover transmission failures and thus increase the precision and recall of the query result.

\subsubsection{DCRSQ: Query Range}
\begin{figure*}[t]
	\centering
	\subfigure[Number of Accessed Objects]{
		\label{fig:dcrsq:query_range:access} %% label for 1st subfigure
		\includegraphics[width=0.245 \textwidth]{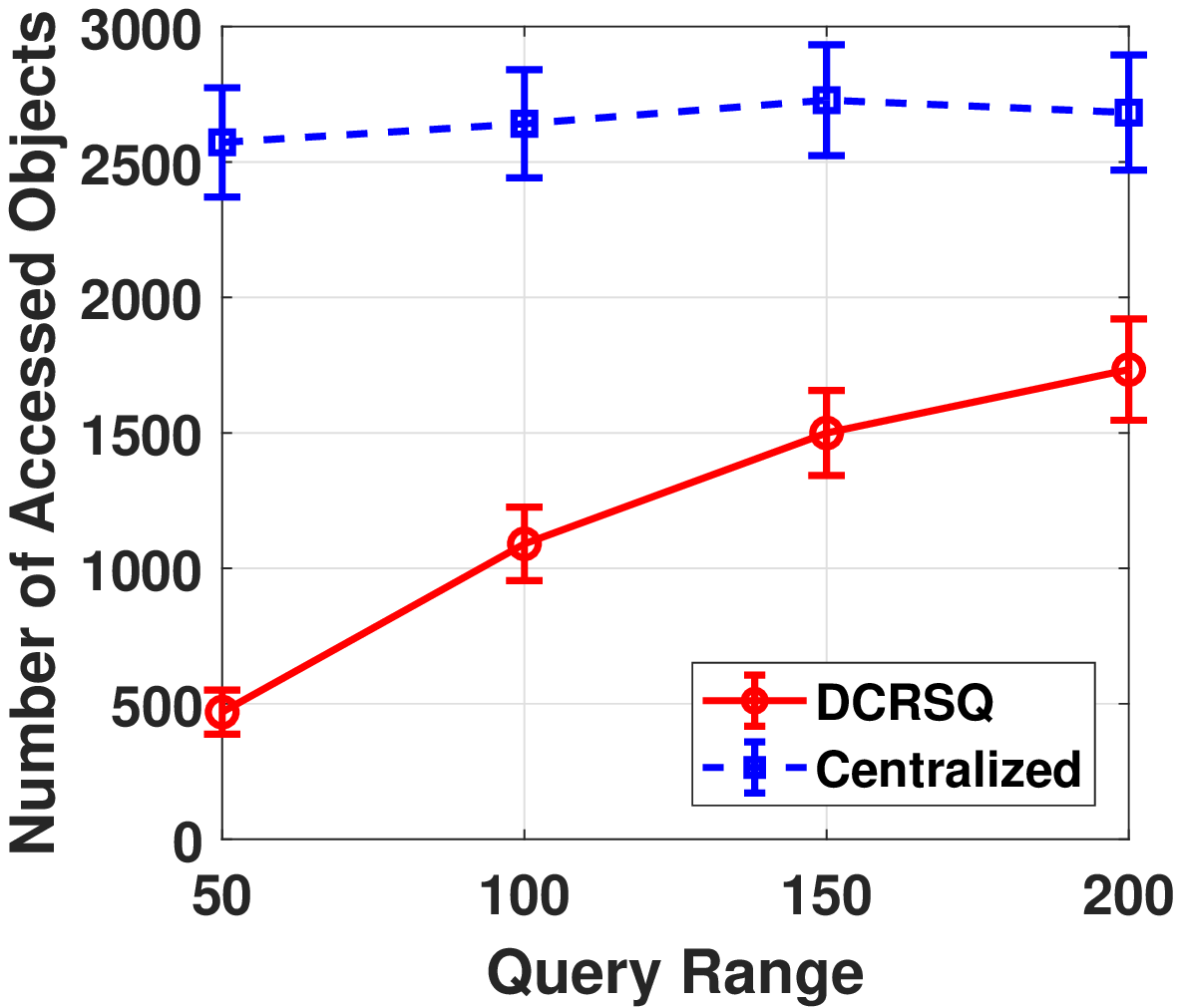}}%\hspace{.5in}
	\subfigure[Number of Messages]{
		\label{fig:dcrsq:query_range:messages} %% label for 2nd subfigure
		\includegraphics[width=0.245 \textwidth]{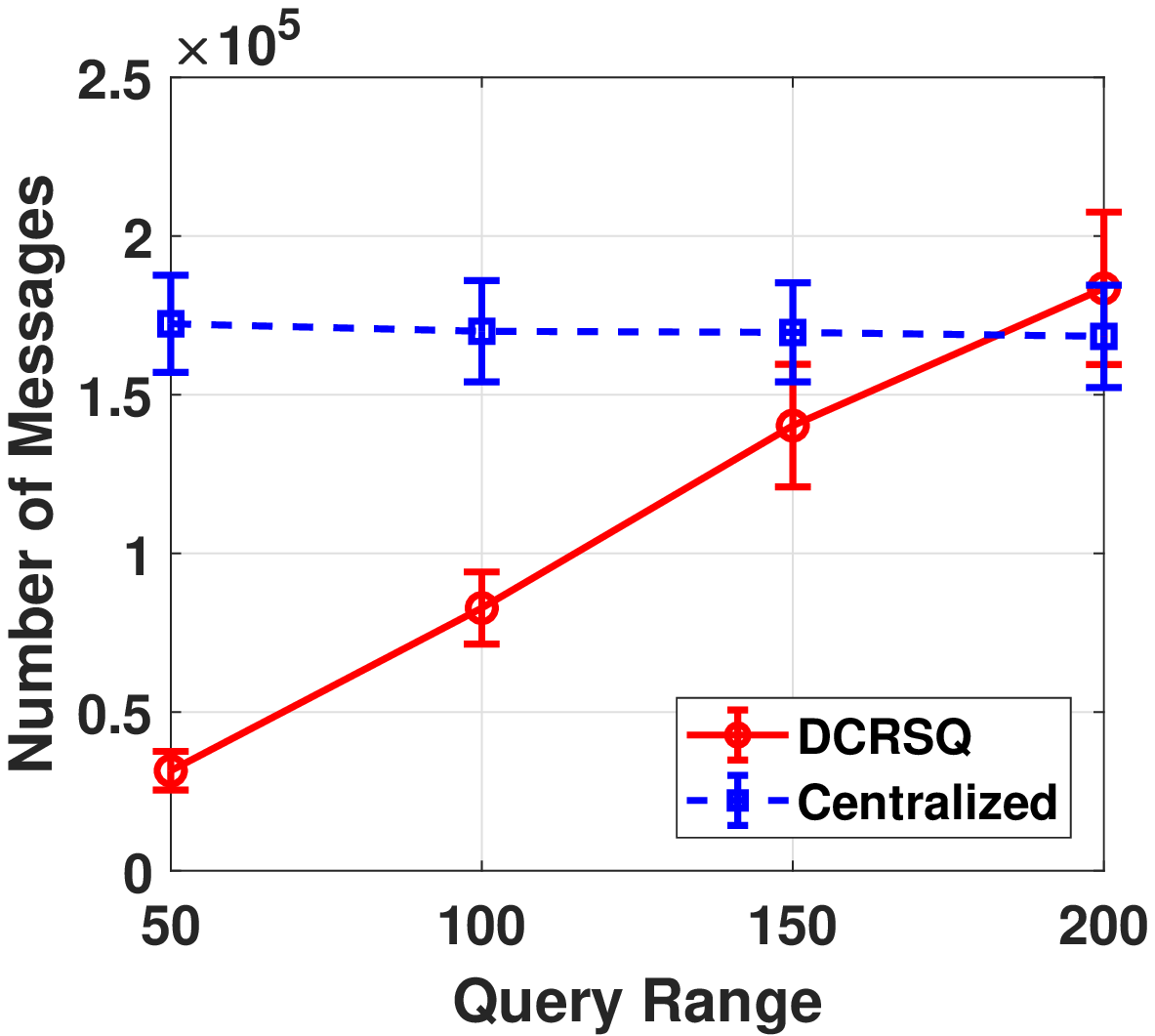}}%\hspace{.5in}
	\subfigure[Precision]{
		\label{fig:dcrsq:query_range:precision} %% label for 2nd subfigure
		\includegraphics[width=0.245 \textwidth]{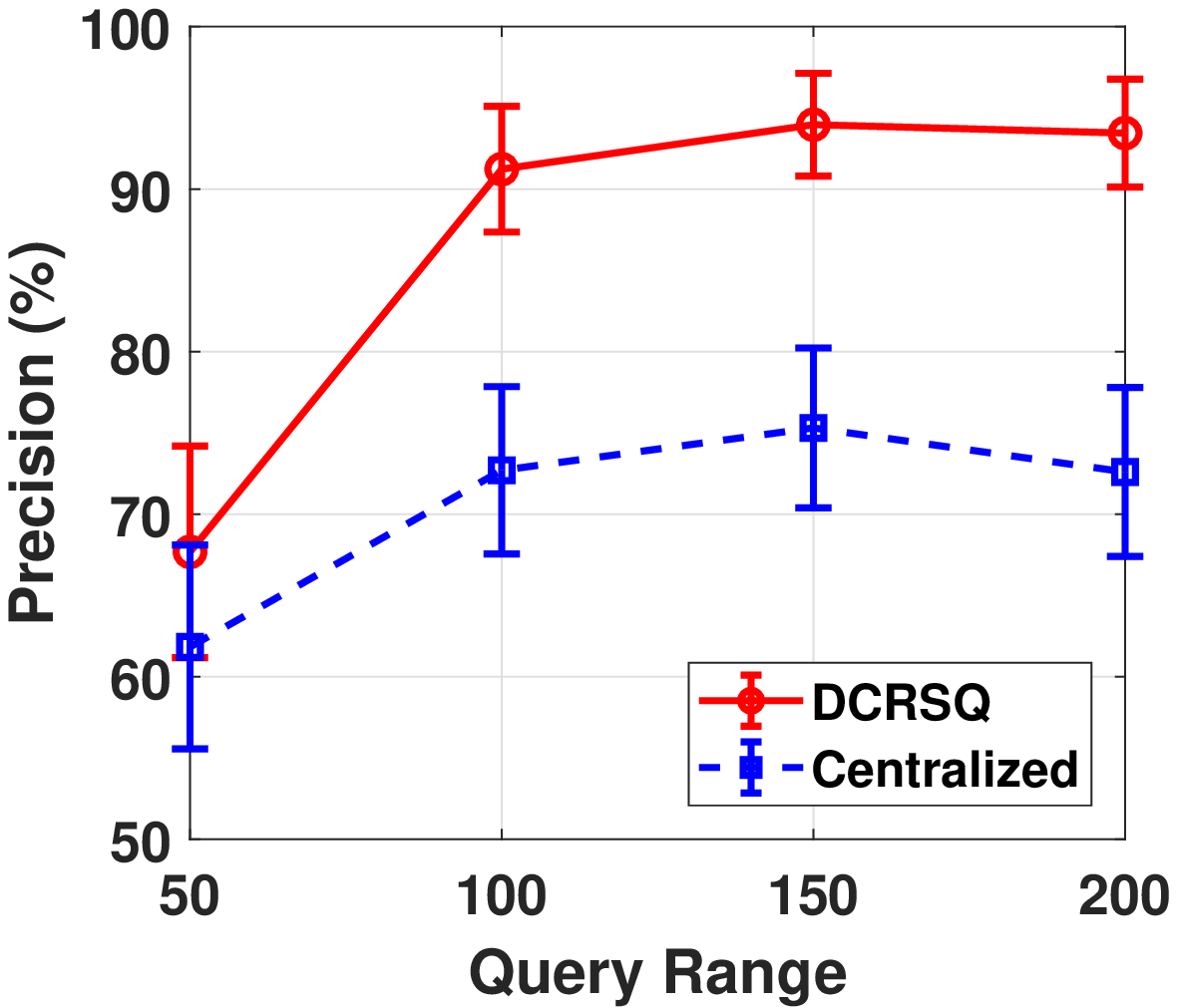}}%\hspace{.5in}
	\subfigure[Recall]{
		\label{fig:dcrsq:query_range:recall} %% label for 2nd subfigure
		\includegraphics[width=0.245 \textwidth]{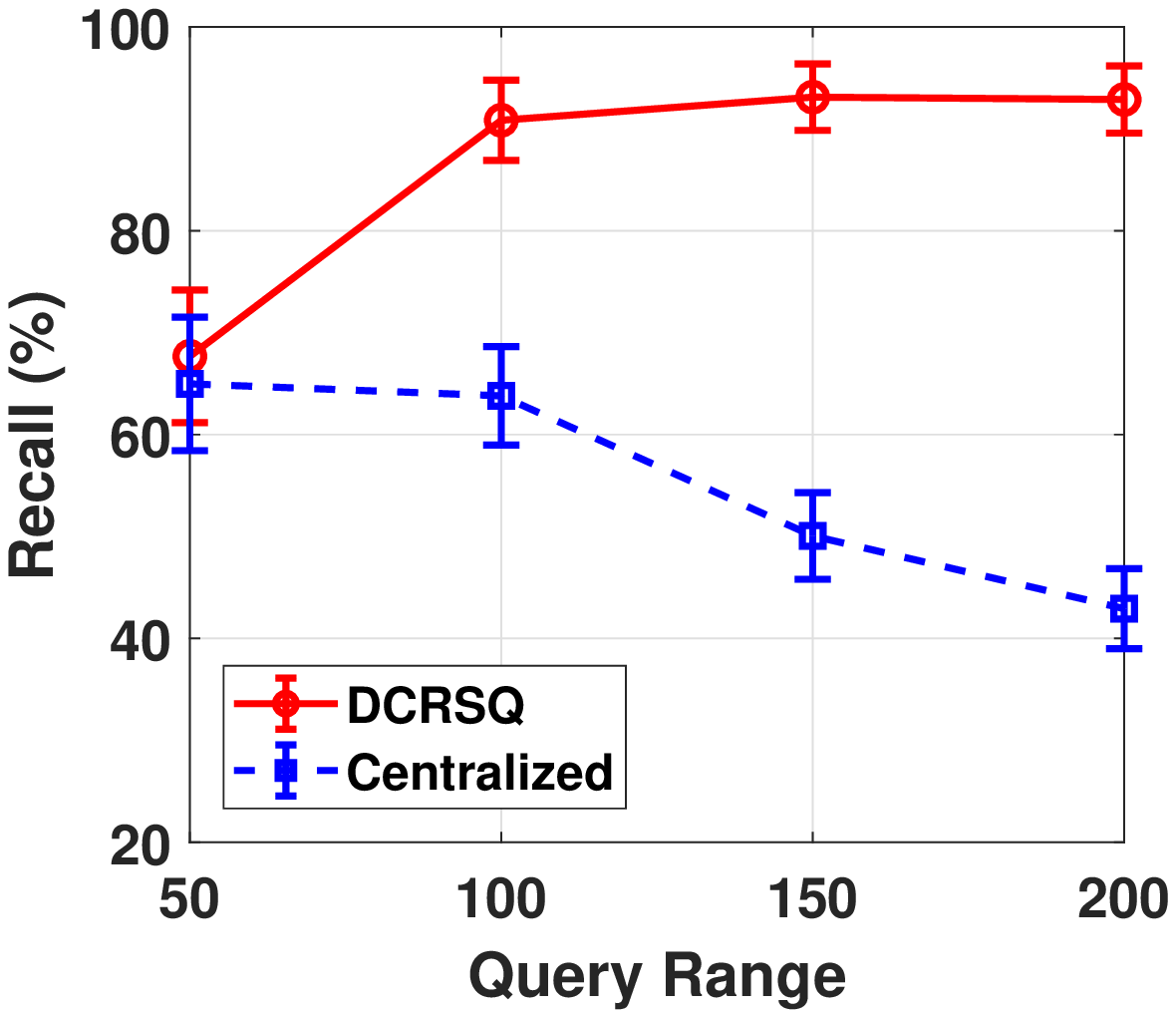}}
	\caption{Impact of query range on \subref{fig:dcrsq:query_range:access} number of accessed objects, \subref{fig:dcrsq:query_range:messages} number of messages, \subref{fig:dcrsq:query_range:precision} precision, and \subref{fig:dcrsq:query_range:recall} recall
	}
	\label{fig:dcrsq:query_range} %% label for entire figure
	\vspace{-10pt}
\end{figure*}

%explain why we discuss query range
In this simulation set, we discuss the results of varying the query range. In Fig.~\ref{fig:dcrsq:query_range:access}, the number of accessed objects in the centralized approach remains around 3000 nodes for the final range-skyline in query point. However, the number of accessed node increases slightly when the query range becomes larger. It means that the query node in DCRSQ can save almost $30\%$ to $80\%$ computational cost in comparison with the centralized approach. As shown in Fig.~\ref{fig:dcrsq:query_range:messages}, when the query range increases, the number of messages in the centralized approach is always about $1.7\times 10^5$ because it always floods messages to the whole sensing area. DCRSQ process needs much less network cost than the centralized approach when the query range is smaller than 150 meters and only performs slightly worse when the query range is 200 meters. The possible reason is that the DCRSQ still costs too much network messages on exchanging information between the irrelevant nodes which are very far away from the query node.
%add some explaination for the result

As for the accuracy shown in Fig.~\ref{fig:dcrsq:query_range:precision} and Fig.~\ref{fig:dcrsq:query_range:recall}, the trends in centralized approach, in comparison with the previous two measurements, are different. With the wider range of a query, the percentage of recall drops rapidly down about 40$\%$ in the centralized approach. Recall that all the neighboring sensor nodes have to report their information to the query node continuously for CRSQ queries in the centralized approach. As the query range becomes larger, more neighboring sensor nodes need to continuously report their information without data pruning. In such a scenario, the query node will be a bottleneck of the system and thus many messages may be dropped. Due to the above reason, the precision and recall of the centralized approach decrease significantly. On the contrary, the DCRSQ process already computes the range-skyline locally and the local range-skyline candidate sets are continuously sent back to the query node for deriving the final range-skyline. Such a way can reduce large amount of network cost and avoid the bottleneck problem on the query node. Thus, the accuracy of our approach can achieve over $92\%$ better.
\begin{figure*}[t]
	\centering
	\subfigure[Number of Accessed Objects]{
		\label{fig:dcrsq:trans_range:access} %% label for 1st subfigure
		\includegraphics[width=0.245 \textwidth]{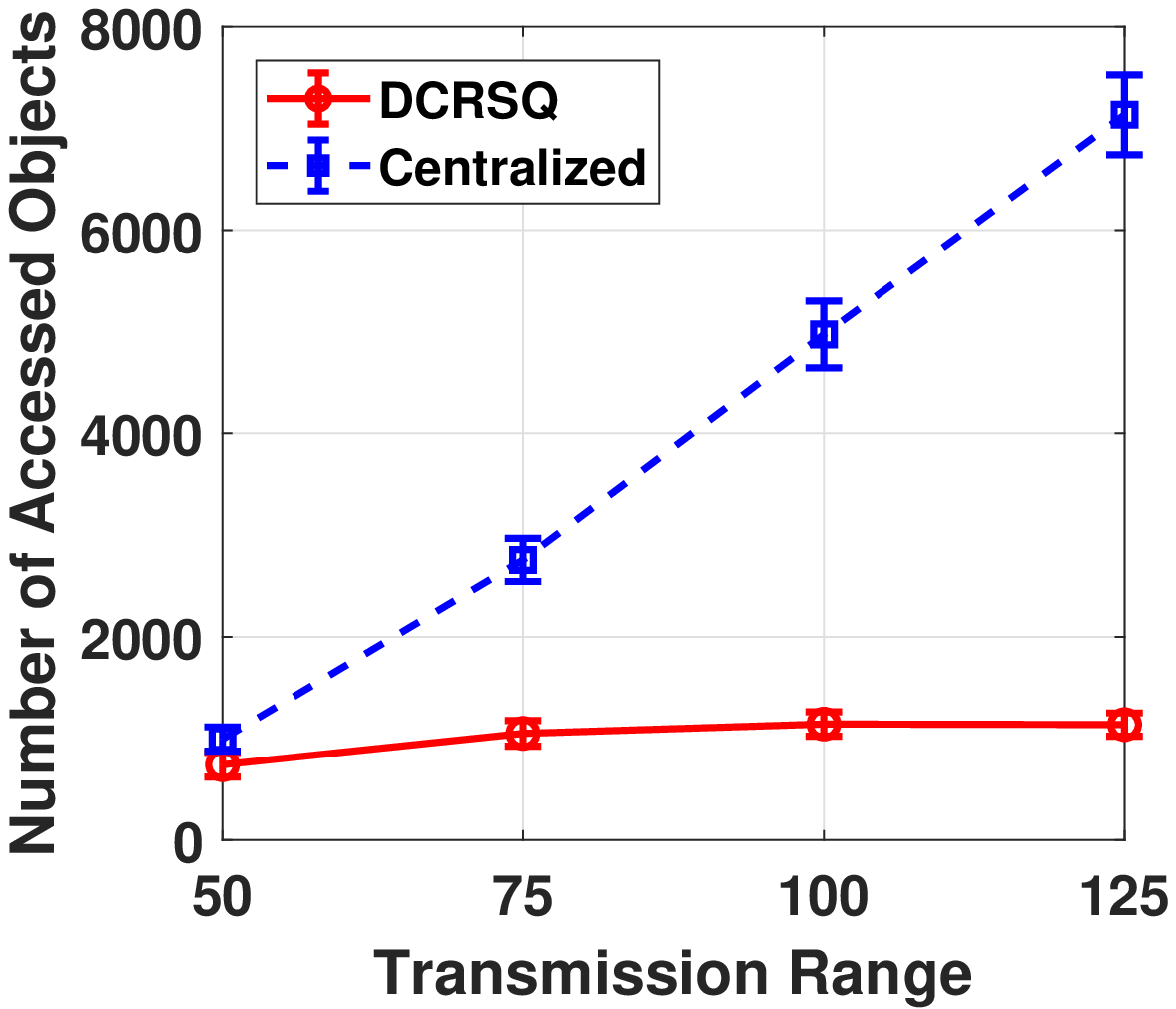}}%\hspace{.5in}
	\subfigure[Number of Messages]{
		\label{fig:dcrsq:trans_range:messages} %% label for 2nd subfigure
		\includegraphics[width=0.245 \textwidth]{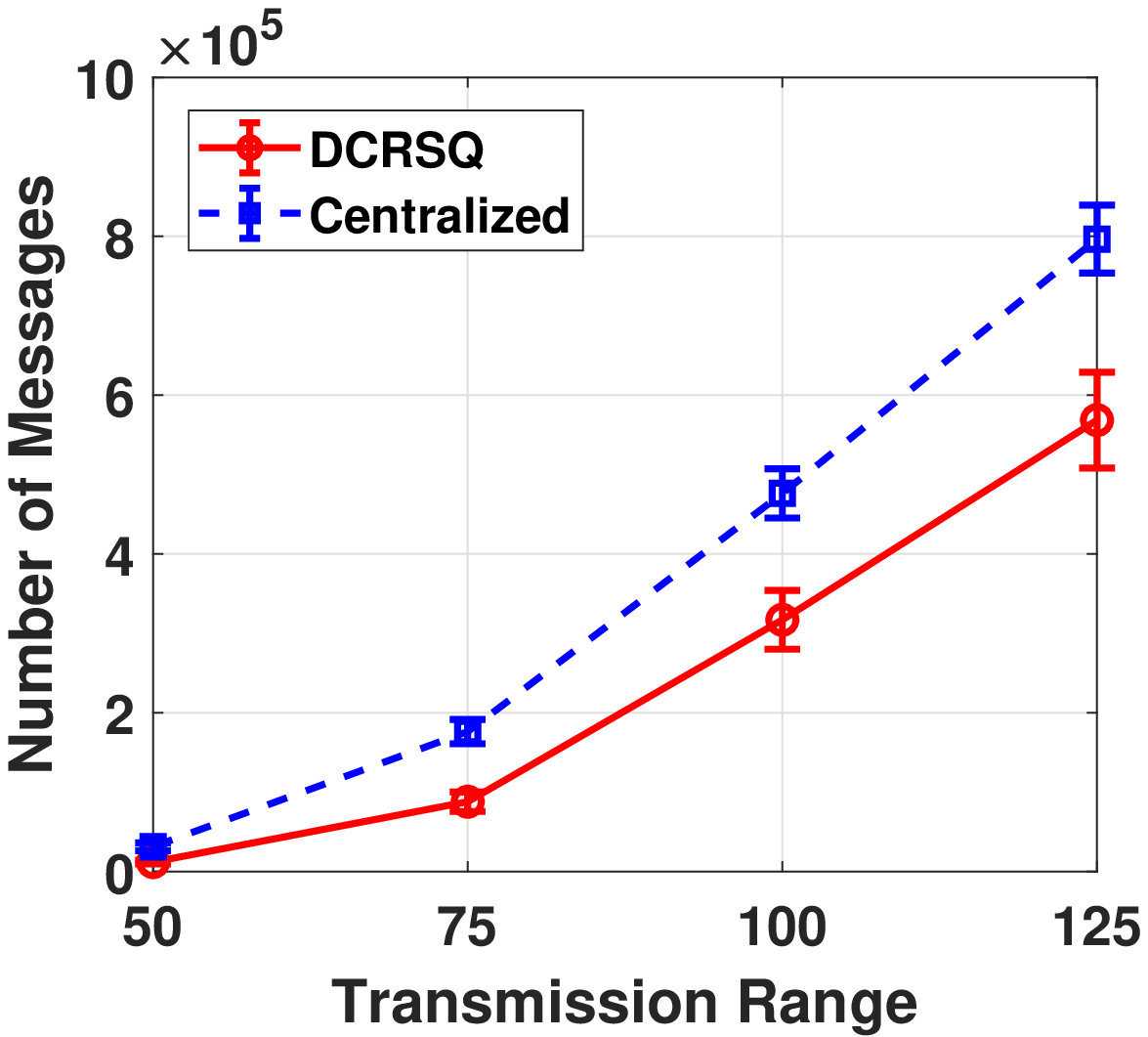}}%\hspace{.5in}
	\subfigure[Precision]{
		\label{fig:dcrsq:trans_range:precision} %% label for 2nd subfigure
		\includegraphics[width=0.245 \textwidth]{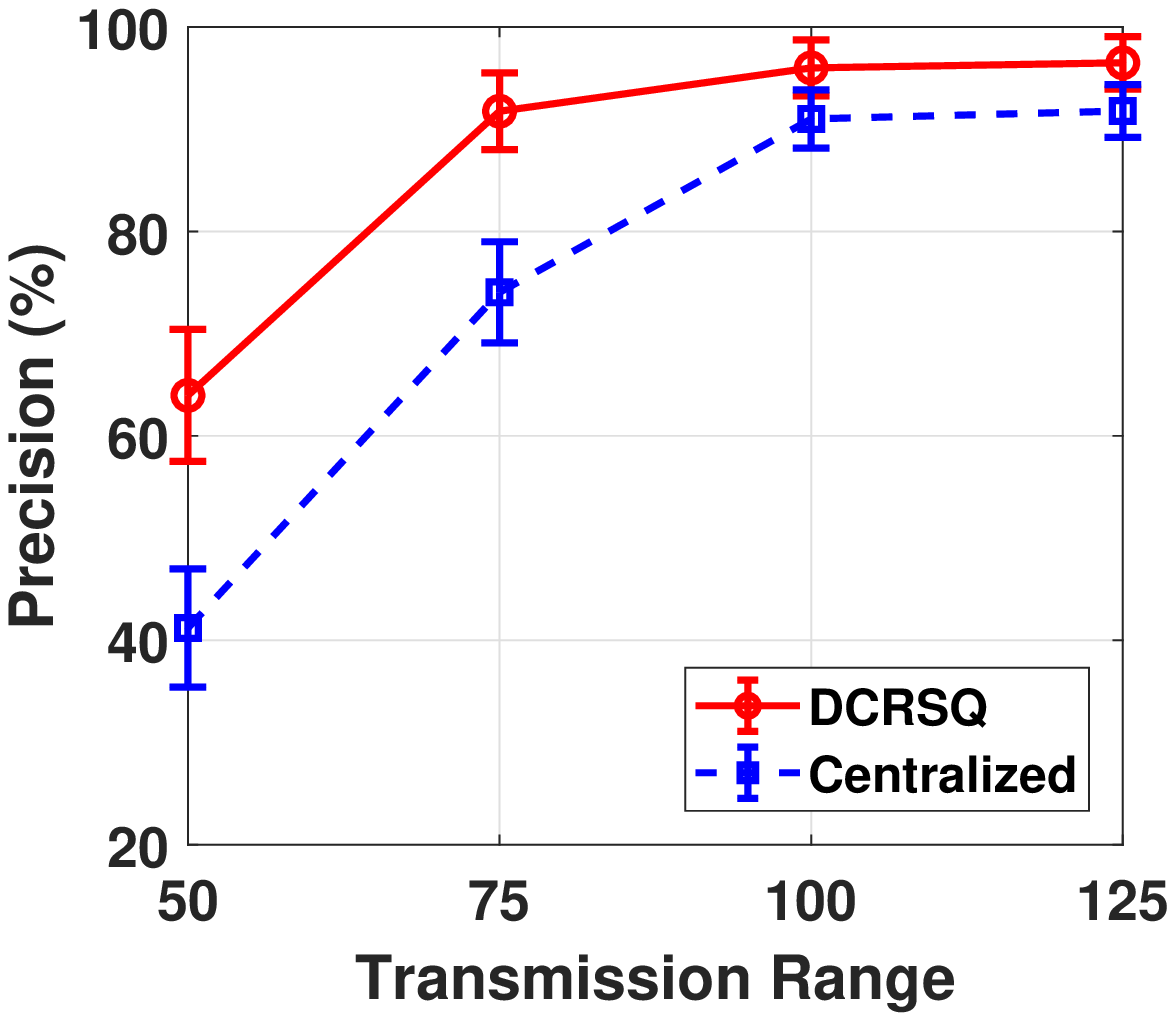}}%\hspace{.5in}
	\subfigure[Recall]{
		\label{fig:dcrsq:trans_range:recall} %% label for 2nd subfigure
		\includegraphics[width=0.245 \textwidth]{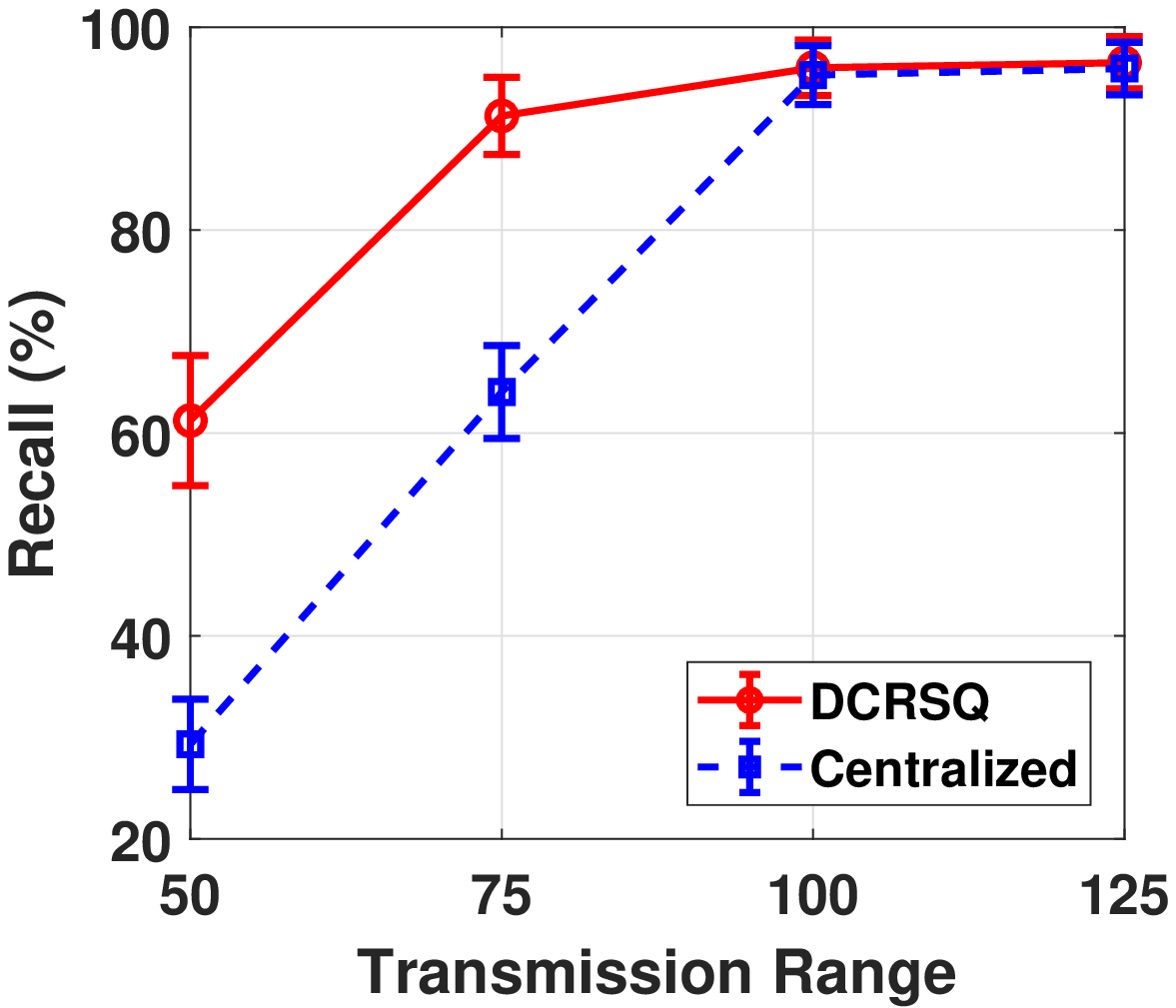}}
	\caption{Impact of transmission range on \subref{fig:dcrsq:trans_range:access} number of accessed objects, \subref{fig:dcrsq:trans_range:messages} number of messages, \subref{fig:dcrsq:trans_range:precision} precision, and \subref{fig:dcrsq:trans_range:recall} recall
	}
	\label{fig:dcrsq:trans_range} %% label for entire figure
	\vspace{-10pt}
\end{figure*}
\begin{figure*}[t]
	\centering
	\subfigure[Number of Accessed Objects]{
		\label{fig:dcrsq:node_speed:access} %% label for 1st subfigure
		\includegraphics[width=0.245 \textwidth]{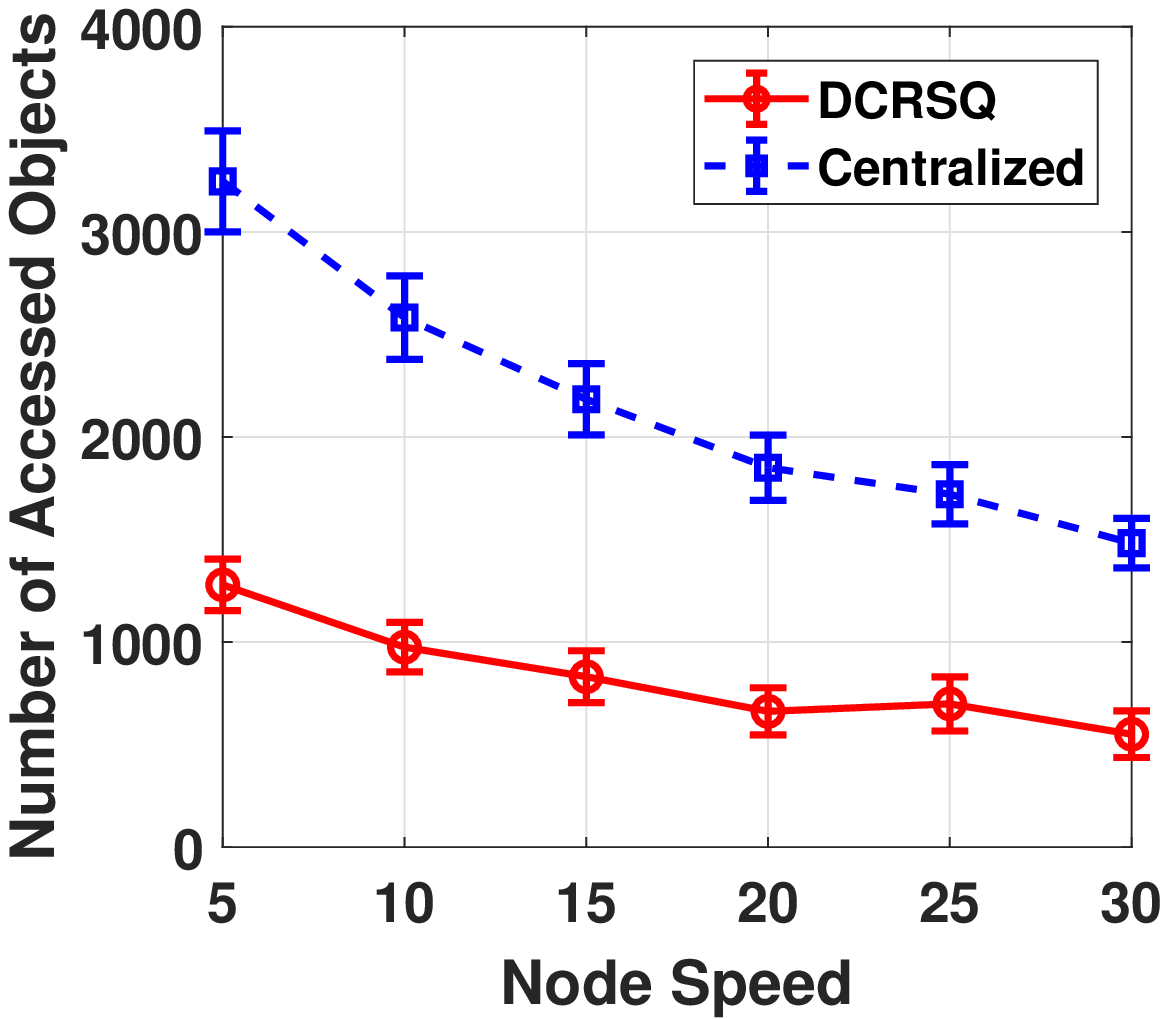}}%\hspace{.5in}
	\subfigure[Number of Messages]{
		\label{fig:dcrsq:node_speed:messages} %% label for 2nd subfigure
		\includegraphics[width=0.245 \textwidth]{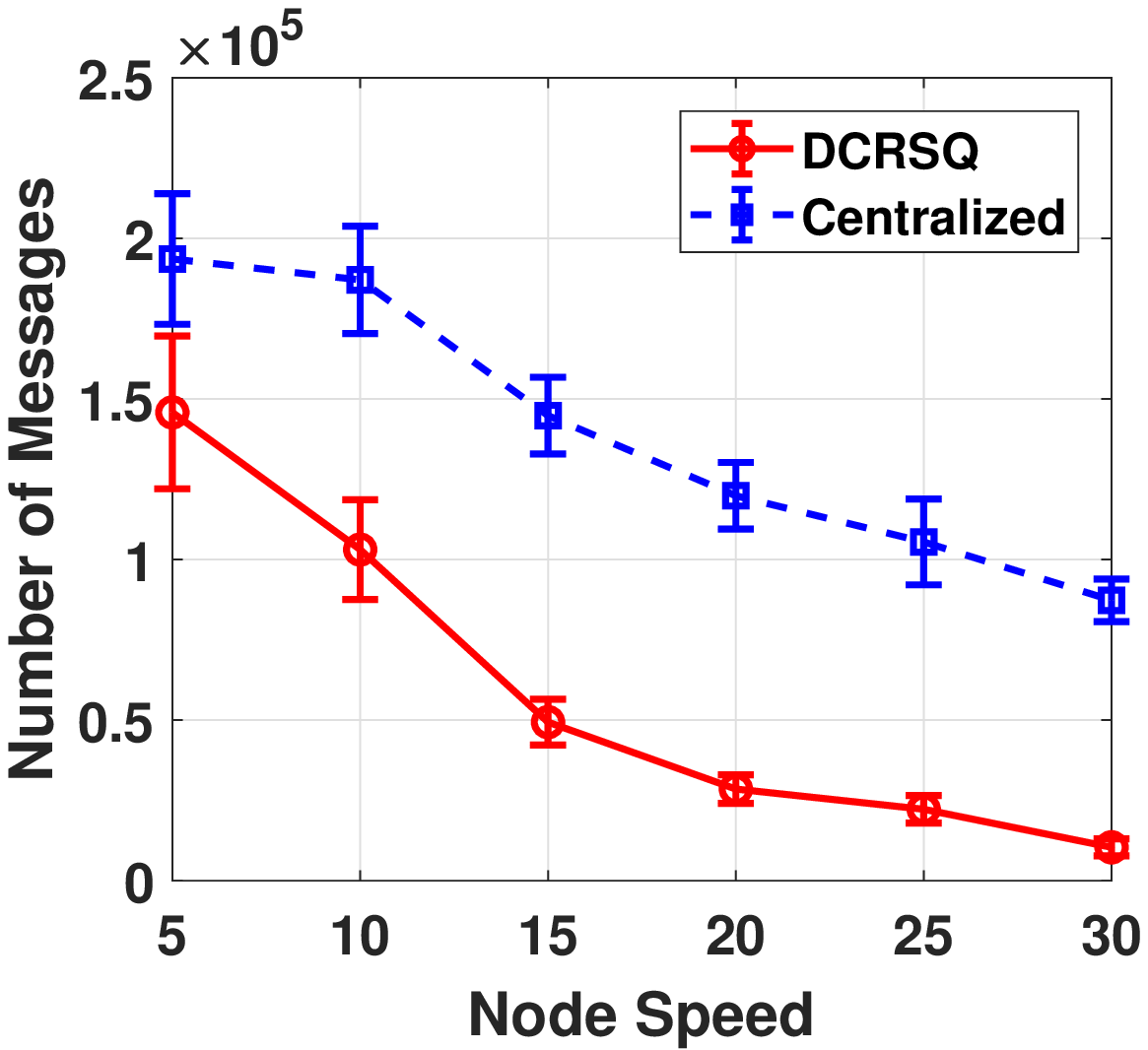}}%\hspace{.5in}
	\subfigure[Precision]{
		\label{fig:dcrsq:node_speed:precision} %% label for 2nd subfigure
		\includegraphics[width=0.245 \textwidth]{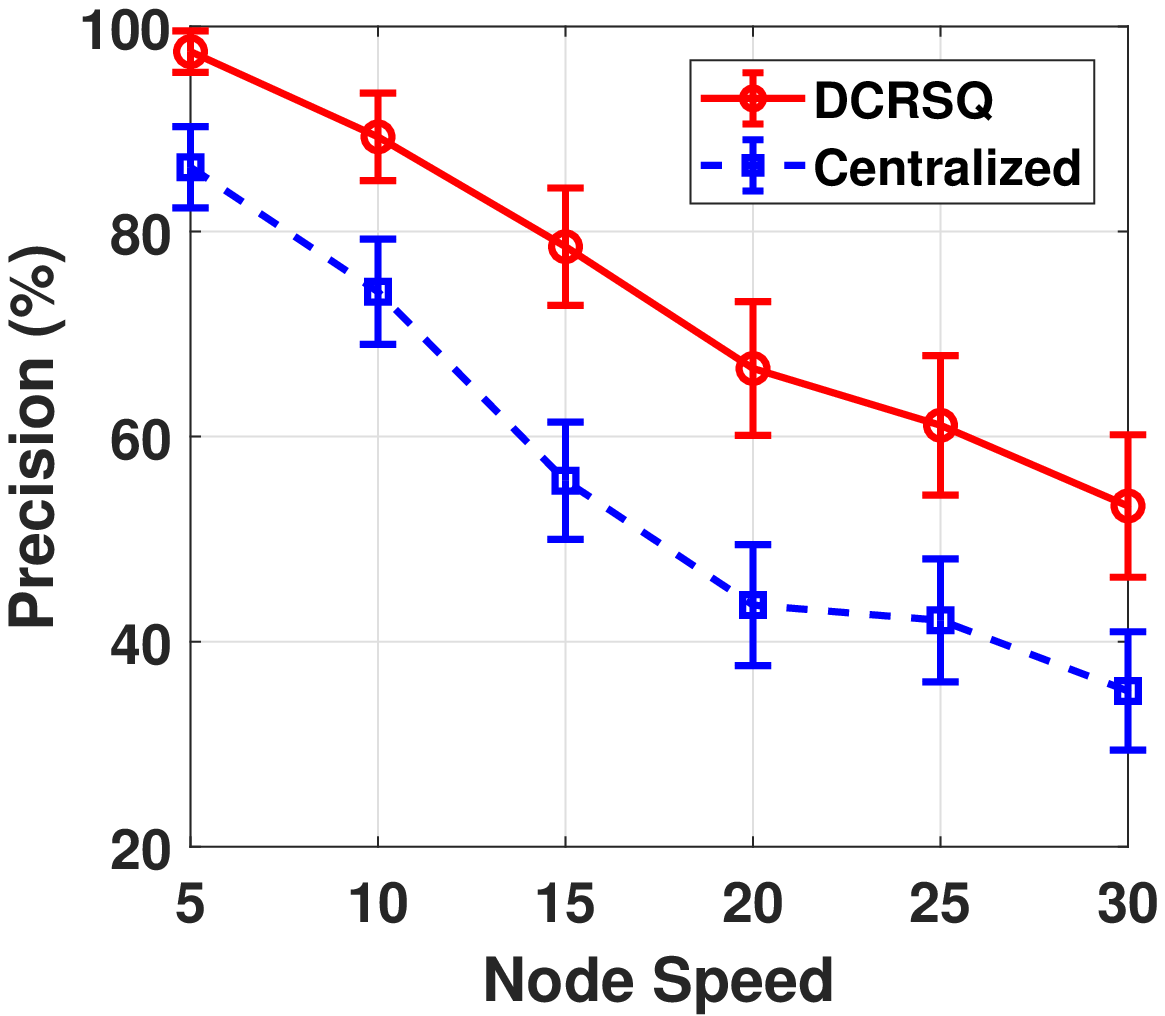}}%\hspace{.5in}
	\subfigure[Recall]{
		\label{fig:dcrsq:node_speed:recall} %% label for 2nd subfigure
		\includegraphics[width=0.245 \textwidth]{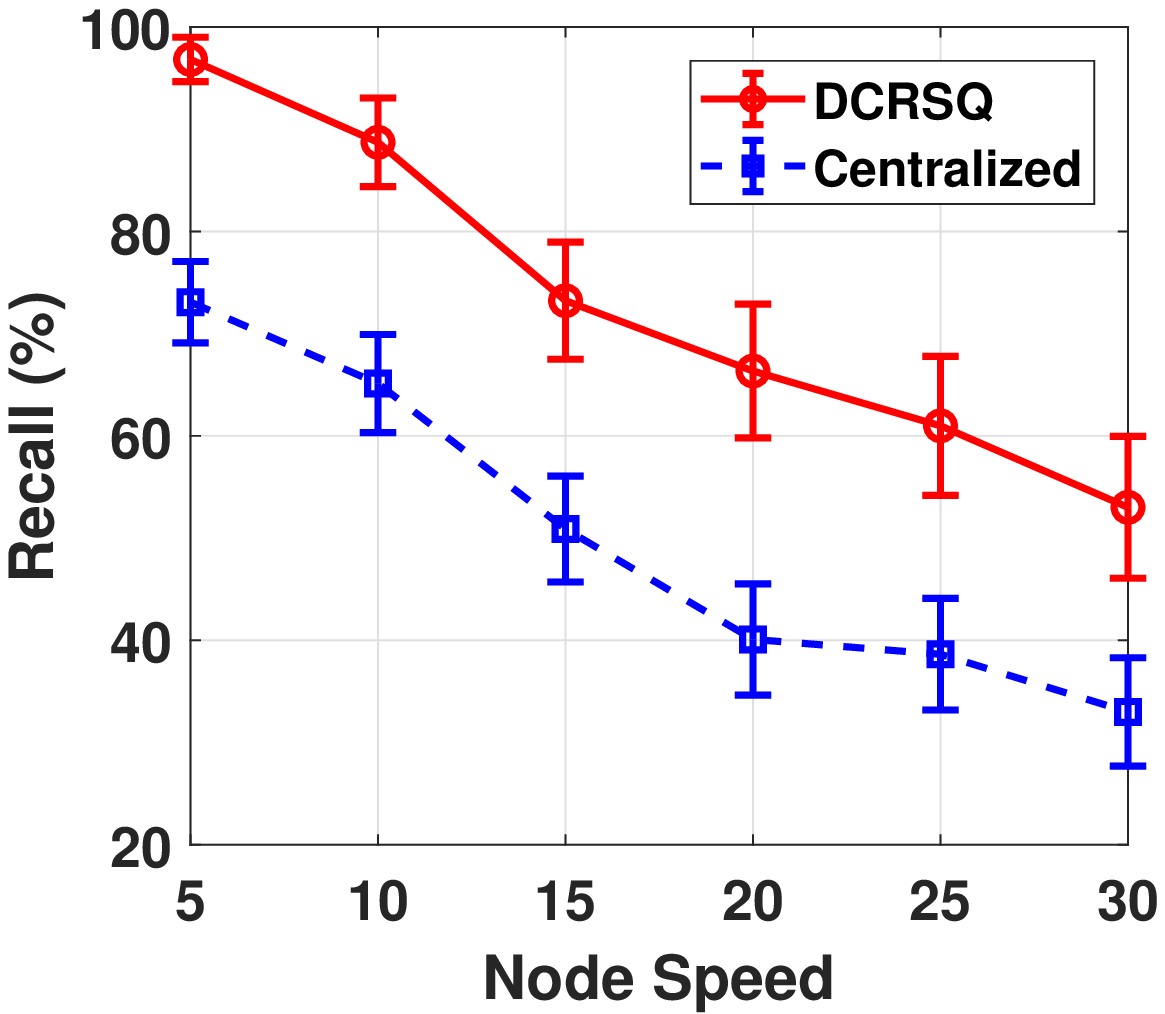}}
	\caption{Impact of node speed on \subref{fig:dcrsq:node_speed:access} number of accessed objects, \subref{fig:dcrsq:node_speed:messages} number of messages, \subref{fig:dcrsq:node_speed:precision} precision, and \subref{fig:dcrsq:node_speed:recall} recall
	}
	\label{fig:dcrsq:node_speed} %% label for entire figure
	\vspace{-10pt}
\end{figure*}

\subsubsection{DCRSQ: Transmission Range}
Transmission range (or sensing range) of each object is also an important impact factor. We therefore measure the performance on different values of transmission range. As shown in Fig.~\ref{fig:dcrsq:trans_range:access}, the number of accessed objects in  DCRSQ process is much less than the centralized approach by up to $80\%$ if the transmission range is $125$ meters. Filtering unnecessary data objects in a distributed way can reduce a lot of required messages for returning the local range-skyline sets to the query point. Fig.~\ref{fig:dcrsq:trans_range:messages} shows that DCRSQ process outperforms the centralized approach by up to $25\%$ with different settings of transmission range in terms of the number of messages. Unlike the dramatic increasing of the number of messages in the centralized approach, the number of messages in DCRSQ process increases more slowly. The reason is that each mobile node in the centralized approach brings a large number of data objects from neighboring sensor nodes before forwarding them back to the query node.

Fig.~\ref{fig:dcrsq:trans_range:precision} and Fig.~\ref{fig:dcrsq:trans_range:recall} show the accuracy of both approaches. For the centralized approach, the precision and recall are under $50\%$ when $r = 50$, and jump to more than $90\%$ if $r$ becomes larger than $100$ meters. The reason is that each mobile sensor node may not successfully transmit information to others while the transmission range becomes too small. In contrast, DCRSQ process can achieve $98\%$ precision and recall since DCRSQ process uses safe time to predict the locations of its neighbors and thus provides more accurate results in the final range-skyline sets.

\subsubsection{DCRSQ: Node Speed}
The last simulation experiment investigates the effect of mobile sensor node's speed. We vary the maximum value of node speed from $5$ to $30$ $m/s$ in the simulation. Fig.~\ref{fig:dcrsq:node_speed} indicates that all the trends are gradually decreasing and our proposed method, DCRSQ process, always has a better performance than the centralized approach. Fig.~\ref{fig:dcrsq:node_speed:access} and Fig.~\ref{fig:dcrsq:node_speed:messages} show that DCRSQ outperforms the centralized approach in terms of the number of accessed objects and the number of messages. Note that each mobile sensor node in DCRSQ process knows when the neighboring nodes enter and leave the query range in advance. It thus can save almost $20\%$ to $90\%$ message cost on query processing and reduce $60\%$ amount of accessed objects for deriving the final skyline result on the query node.

In addition, when each mobile node moves faster, the neighbors will change more frequently and wireless connections between mobile sensor nodes become more unstable. Hence, in high-speed scenarios, it is more difficult for the query node to collect sufficient information to derive the accurate result. Fig.~\ref{fig:dcrsq:node_speed:precision} and Fig.~\ref{fig:dcrsq:node_speed:recall} show that both DCRSQ and the centralized approach has the similar trends of performance on precision and recall. Comparing to the centralized approach, DCRSQ has $20\%$ improvement in average on the precision and recall.

\section{Conclusion}
\label{conclusion}
In IoMT, very few works discuss the RSQ and CRSQ queries while simultaneously considering the following constraints: distributed computing units and databases on different mobile sensor nodes databases, moving data objects, and the mobile query. We hence propose a Distributed Continuous Range-Skyline Query process (DCRSQ process), for driving the results of RSQ and CRSQ queries. The main idea of the proposed DCRSQ process is to predict the appropriate time, safe-time, that the answer of a query changes. We apply such a prediction to each mobile sensor node and query node, so each mobile sensor node can compute the local result more precisely and then provide the local result to the query node for computing the final result. Instead of processing the information of all the neighboring mobile sensor nodes on the query node, the proposed distributed and cooperative approach with safe-time prediction can effectively reduce the computation overhead of the query node. The performance of DCRSQ process is also validated by the extensive simulation experiments. In some scenarios, the performance of DCRSQ process is almost 80\% better than the performance of the centralized approach in terms of the number of accessed objects. The DCRSQ process saves more than 15\% network cost in terms of the number of messages in general. In most scenarios, the DCRSQ process outperforms the centralized approach by more than 10\% to 25\% accuracy (precision and recall). 

In this work, we propose a prototype of distributed query process for CRSQ and simply use 2 dimensional data objects (distance and sensing value) to valid the performance in the simulation. For each sensor node, it needs to buffer the received query information and then help the data filtering and local computation until the query time is expired. Hence, there is one possible future research direction to find the relation between the minimum requirement (CPU and memory/storage) of each sensor and a parameterized (report/sense rate, number of sensor nodes and number of queries, etc.) IoT environment. Another possible future research work is to implement the distributed multi-criteria decision services on different modern open source IoT constrained platform~\citep{Mosquitto} or simulators~\citep{7496514}. Such a way can help the research and open source communities for evaluating. In the future, we are going to develop distributed approaches for monitoring different spatial queries in practical drone-assisted IoMT applications~\citep{8642333}.

\appendices

%\section{}

\ifCLASSOPTIONcompsoc
\section*{Acknowledgments}
\else
\section*{Acknowledgment}
\fi

This research is partially supported by Ministry of Science and Technology under the Grant MOST 107-2221-E-027-099-MY2 and MOST 108-2634-F-009-006- through Pervasive Artificial Intelligence Research (PAIR) Labs, Taiwan.

\ifCLASSOPTIONcaptionsoff
\newpage
\fi



%\bibliographystyle{abbrv}
\bibliographystyle{ieeetr}
\bibpreamble
\renewcommand{\bibfont}{\footnotesize}
\bibliography{reference}

\begin{thebibliography}{10}

\bibitem{SeokJin:2006:IndexRangeQueryBroadcast}
S.~Im, M.~Song, S.-W.~K. Jongwan~Kim, C.-S. Hwang, and S.~Lee, ``Cell-based
  distributed index for range query processing in wireless data broadcast
  systems,'' in {\em Knowledge-Based Intelligent Information and Engineering
  Systems Lecture Notes in Computer Science}, vol.~4251, pp.~1139--1146,
  Springer Berlin Heidelberg, 2006.

\bibitem{Jianting:2004:MultiDimensionRangeQueryBroadcast}
J.~Zhang and L.~Gruenwald, ``Optimizing data placement over wireless broadcast
  channel for multi-dimensional range query processing,'' in {\em Mobile Data
  Management}, pp.~256--265, IEEE, 2004.

\bibitem{Li2015}
C.~Li, Y.~Gu, J.~Qi, R.~Zhang, and G.~Yu, ``A safe region based approach to
  moving $k$nn queries in obstructed space,'' {\em Knowledge and Information
  Systems}, vol.~45, no.~2, pp.~417--451, 2015.

\bibitem{Liu:2008:kNNIndexTreeMultiDimension}
C.-M. Liu and S.-Y. Fu, ``Effective protocols for knn search on broadcast
  multi-dimensional index trees,'' {\em Information Systems}, vol.~33,
  pp.~18--35, 2008.

\bibitem{6081864}
X.~Lin, J.~Xu, and H.~Hu, ``Range-based skyline queries in mobile
  environments,'' {\em IEEE Transactions on Knowledge and Data Engineering},
  vol.~25, pp.~835--849, April 2013.

\bibitem{Rahul:2012:ARQ:2424321.2424406}
S.~Rahul and R.~Janardan, ``Algorithms for range-skyline queries,'' in {\em
  Proceedings of the 20th International Conference on Advances in Geographic
  Information Systems}, SIGSPATIAL '12, (New York, NY, USA), pp.~526--529, ACM,
  2012.

\bibitem{Dimitris:2005:SkylineComputation}
D.~Papadias, Y.~Tao, G.~Fu, and B.~Seeger, ``Progressive skyline computation in
  database systems,'' in {\em ACM Transactions on Database Systems}, TODS, (New
  York, NY, USA), pp.~41--82, ACM, 2005.

\bibitem{Tian:2007:CMS:1254850.1254861}
L.~Tian, L.~Wang, P.~Zou, Y.~Jia, and A.~Li, ``Continuous monitoring of skyline
  query over highly dynamic moving objects,'' in {\em Proceedings of the 6th
  ACM International Workshop on Data Engineering for Wireless and Mobile
  Access}, MobiDE '07, (New York, NY, USA), pp.~59--66, ACM, 2007.

\bibitem{7903653}
L.~{Ang}, K.~P. {Seng}, A.~M. {Zungeru}, and G.~K. {Ijemaru}, ``Big sensor data
  systems for smart cities,'' {\em IEEE Internet of Things Journal}, vol.~4,
  pp.~1259--1271, Oct 2017.

\bibitem{7945539}
X.~{Cheng}, L.~{Fang}, L.~{Yang}, and S.~{Cui}, ``Mobile big data: The fuel for
  data-driven wireless,'' {\em IEEE Internet of Things Journal}, vol.~4,
  pp.~1489--1516, Oct 2017.

\bibitem{5594641}
F.~{Calabrese}, M.~{Colonna}, P.~{Lovisolo}, D.~{Parata}, and C.~{Ratti},
  ``Real-time urban monitoring using cell phones: A case study in rome,'' {\em
  IEEE Transactions on Intelligent Transportation Systems}, vol.~12,
  pp.~141--151, March 2011.

\bibitem{HORITA201584}
F.~E. Horita, J.~{a}o Porto~de Albuquerque, L.~C. Degrossi, E.~M. Mendiondo,
  and J.~Ueyama, ``Development of a spatial decision support system for flood
  risk management in brazil that combines volunteered geographic information
  with wireless sensor networks,'' {\em Computers \& Geosciences}, vol.~80,
  pp.~84--94, 2015.

\bibitem{914855}
S.~Borzsony, D.~Kossmann, and K.~Stocker, ``The skyline operator,'' in {\em
  Proceedings of 17th International Conference on Data Engineering},
  pp.~421--430, 2001.

\bibitem{Hjaltason:1999:DBS:320248.320255}
G.~R. Hjaltason and H.~Samet, ``Distance browsing in spatial databases,'' {\em
  ACM Trans. Database Syst.}, vol.~24, pp.~265--318, June 1999.

\bibitem{Cheema:2013:SZB:2452376.2452409}
M.~A. Cheema, X.~Lin, W.~Zhang, and Y.~Zhang, ``A safe zone based approach for
  monitoring moving skyline queries,'' in {\em Proceedings of the 16th
  International Conference on Extending Database Technology}, EDBT '13, (New
  York, NY, USA), pp.~275--286, ACM, 2013.

\bibitem{Hose2012}
K.~Hose and A.~Vlachou, ``A survey of skyline processing in highly distributed
  environments,'' {\em The VLDB Journal}, vol.~21, no.~3, pp.~359--384, 2012.

\bibitem{4511446}
B.~Zheng, K.~C.~K. Lee, and W.~C. Lee, ``Location-dependent skyline query,'' in
  {\em The Ninth International Conference on Mobile Data Management (mdm
  2008)}, pp.~148--155, April 2008.

\bibitem{10.1109/TKDE.2010.103}
L.~Chen, B.~Cui, and H.~Lu, ``Constrained skyline query processing against
  distributed data sites,'' {\em IEEE Transactions on Knowledge and Data
  Engineering}, vol.~23, no.~2, pp.~204--217, 2011.

\bibitem{jsan5010002}
K.~Ahmed, N.~S. Nafi, and M.~A. Gregory, ``Enhanced distributed dynamic skyline
  query for wireless sensor networks,'' {\em Journal of Sensor and Actuator
  Networks}, vol.~5, no.~1, 2016.

\bibitem{1617434}
Z.~Huang, C.~S. Jensen, H.~Lu, and B.~C. Ooi, ``Skyline queries against mobile
  lightweight devices in manets,'' in {\em 22nd International Conference on
  Data Engineering (ICDE'06)}, pp.~66--66, April 2006.

\bibitem{Wu:2006:PSQ:2117976.2117990}
P.~Wu, C.~Zhang, Y.~Feng, B.~Y. Zhao, D.~Agrawal, and A.~El~Abbadi,
  ``Parallelizing skyline queries for scalable distribution,'' in {\em
  Proceedings of the 10th International Conference on Advances in Database
  Technology}, EDBT'06, (Berlin, Heidelberg), pp.~112--130, Springer-Verlag,
  2006.

\bibitem{Ratnasamy:2001:SCN:383059.383072}
S.~Ratnasamy, P.~Francis, M.~Handley, R.~Karp, and S.~Shenker, ``A scalable
  content-addressable network,'' in {\em Proceedings of the 2001 Conference on
  Applications, Technologies, Architectures, and Protocols for Computer
  Communications}, SIGCOMM '01, (New York, NY, USA), pp.~161--172, ACM, 2001.

\bibitem{conf/mdm/KimCT08}
K.~Kim, Y.~Cai, and W.~Tavanapong, ``Safe-time: Distributed real-time
  monitoring of cknn in mobile peer-to-peer networks.,'' in {\em Proceedings of
  the 9th IEEE International Conference on Mobile Data Management (MDM)},
  pp.~124--131, 2008.

\bibitem{Bentley:1978:ANM:322092.322095}
J.~L. Bentley, H.~T. Kung, M.~Schkolnick, and C.~D. Thompson, ``On the average
  number of maxima in a set of vectors and applications,'' {\em J. ACM},
  vol.~25, pp.~536--543, Oct. 1978.

\bibitem{4277081}
J.~C. Kuo and W.~Liao, ``Hop count distribution of multihop paths in wireless
  networks with arbitrary node density: Modeling and its applications,'' {\em
  IEEE Transactions on Vehicular Technology}, vol.~56, pp.~2321--2331, July
  2007.

\bibitem{749281}
C.~E. {Perkins} and E.~M. {Royer}, ``Ad-hoc on-demand distance vector
  routing,'' in {\em Proceedings WMCSA'99. Second IEEE Workshop on Mobile
  Computing Systems and Applications}, pp.~90--100, Feb 1999.

\bibitem{Mosquitto}
R.~A. Light, ``Mosquitto: server and client implementation of the mqtt
  protocol,'' {\em Journal of Open Source Software}, vol.~2, May 2017.

\bibitem{7496514}
I.~{Minakov}, R.~{Passerone}, A.~{Rizzardi}, and S.~{Sicari}, ``Routing
  behavior across wsn simulators: The aodv case study,'' in {\em 2016 IEEE
  World Conference on Factory Communication Systems (WFCS)}, pp.~1--8, May
  2016.

\bibitem{8642333}
C.~{Lai}, C.~{Chen}, and L.~{Wang}, ``On-demand density-aware uav base station
  3d placement for arbitrarily distributed users with guaranteed data rates,''
  {\em IEEE Wireless Communications Letters}, pp.~1--1, 2019.

\end{thebibliography}
%\begin{thebibliography}{1}


%\end{thebibliography}

%\vspace{-20pt}
\begin{IEEEbiography}[{\includegraphics[width=1in,height=1.25in,clip,keepaspectratio]{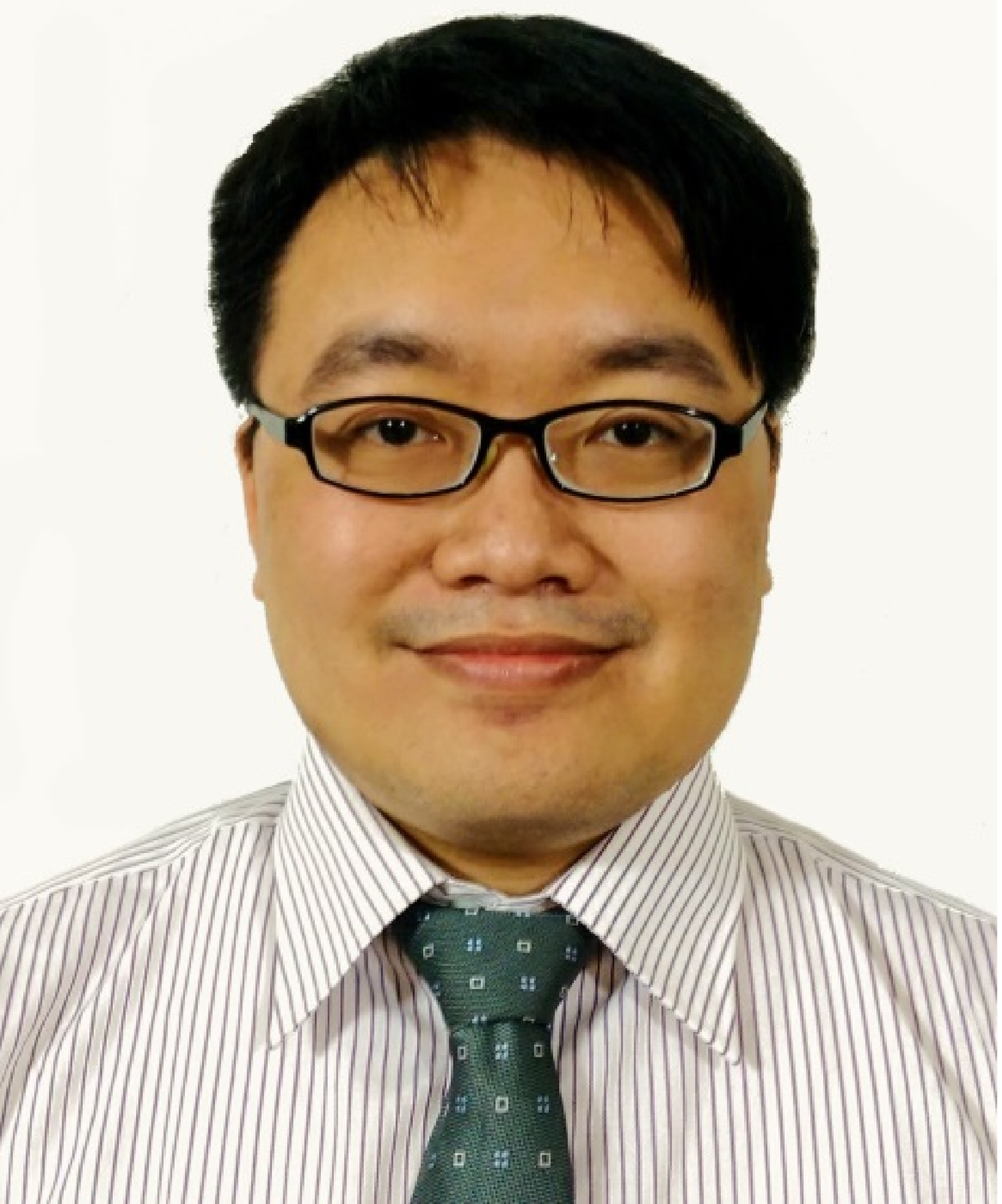}}]{Chuan-Chi Lai}
	is currently holding a post-doctoral position in the Department of Electrical and Computer Engineering at National Chiao Tung University, Taiwan, R.O.C. He received his Ph. D. in Computer Science and Information Engineering from National Taipei University of Technology (Taipei Tech), Taiwan in 2017. He won Excellent Paper Award and Best Paper Award in ICUFN 2015 and WOCC 2018 conferences, respectively. His current research interests are in the areas of data management and dissemination techniques in mobile wireless environments, mobile ad-hoc and sensor networks, distributed query processing over moving objects, and analysis and design of distributed algorithms.
\end{IEEEbiography}

%\vspace{-20pt}
\begin{IEEEbiography}[{\includegraphics[width=1in,height=1.25in,clip,keepaspectratio]{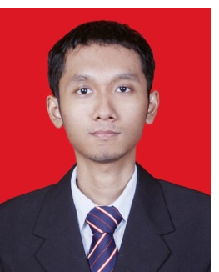}}]{Zulhaydar Fairozal Akbar}
	received the MSc. degree in Electrical Engineering and Computer Science from National Taipei University of Technology (NTUT), in 2016. Now, he is a Junior Lecturer in Informatics Engineering Department, Electronic Engineering Polytechnic Institute of Surabaya (PENS). His research interests include mobile computing, data mining and machine learning.
\end{IEEEbiography}

%\newpage

%\vspace{-20pt}
\begin{IEEEbiography}[{\includegraphics[width=1in,height=1.25in,clip,keepaspectratio]{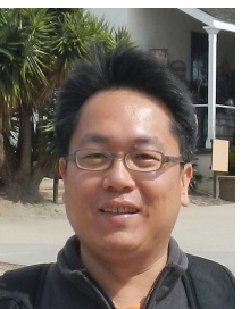}}]{Chuan-Ming Liu}
	is an associate professor in the Department of Computer Science and Information Engineering, National Taipei University of Technology (NTUT), TAIWAN. He received his Ph. D. in Computer Sciences from Purdue University in 2002 and B.S. and M.S. degrees both in Applied Mathematics from National Chung-Hsing University, Taiwan, in 1992 and 1994, respectively. In the summer of 2010 and 2011, he has held visiting appointments at Auburn University and Beijing Institute of Technology, respectively. Dr. Liu's research interests include data management and data dissemination in various emerging computing environments, query processing in moving objects, location-based services, ad-hoc and sensor networks, parallel and distributed computation, and analysis and design of algorithms.
\end{IEEEbiography}

%\vspace{-20pt}
\begin{IEEEbiography}[{\includegraphics[width=1in,height=1.25in,clip,keepaspectratio]{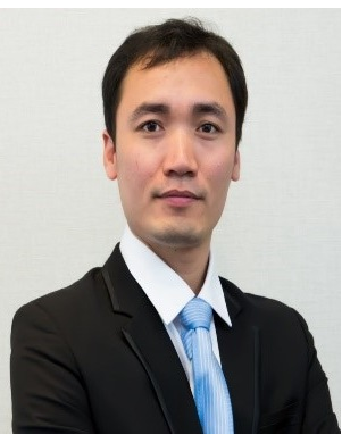}}]{Van-Dai Ta} received the MSc. degree in computer science from National Formosa University, Taiwan, in 2015. Now, he is currently a PhD student of National Taipei University of Technology, Taiwan. His current research interests are computer networks, wireless sensor networks, Data Mining.
\end{IEEEbiography}

%\vspace{-20pt}
\begin{IEEEbiography}[{\includegraphics[width=1in,height=1.25in,clip,keepaspectratio]{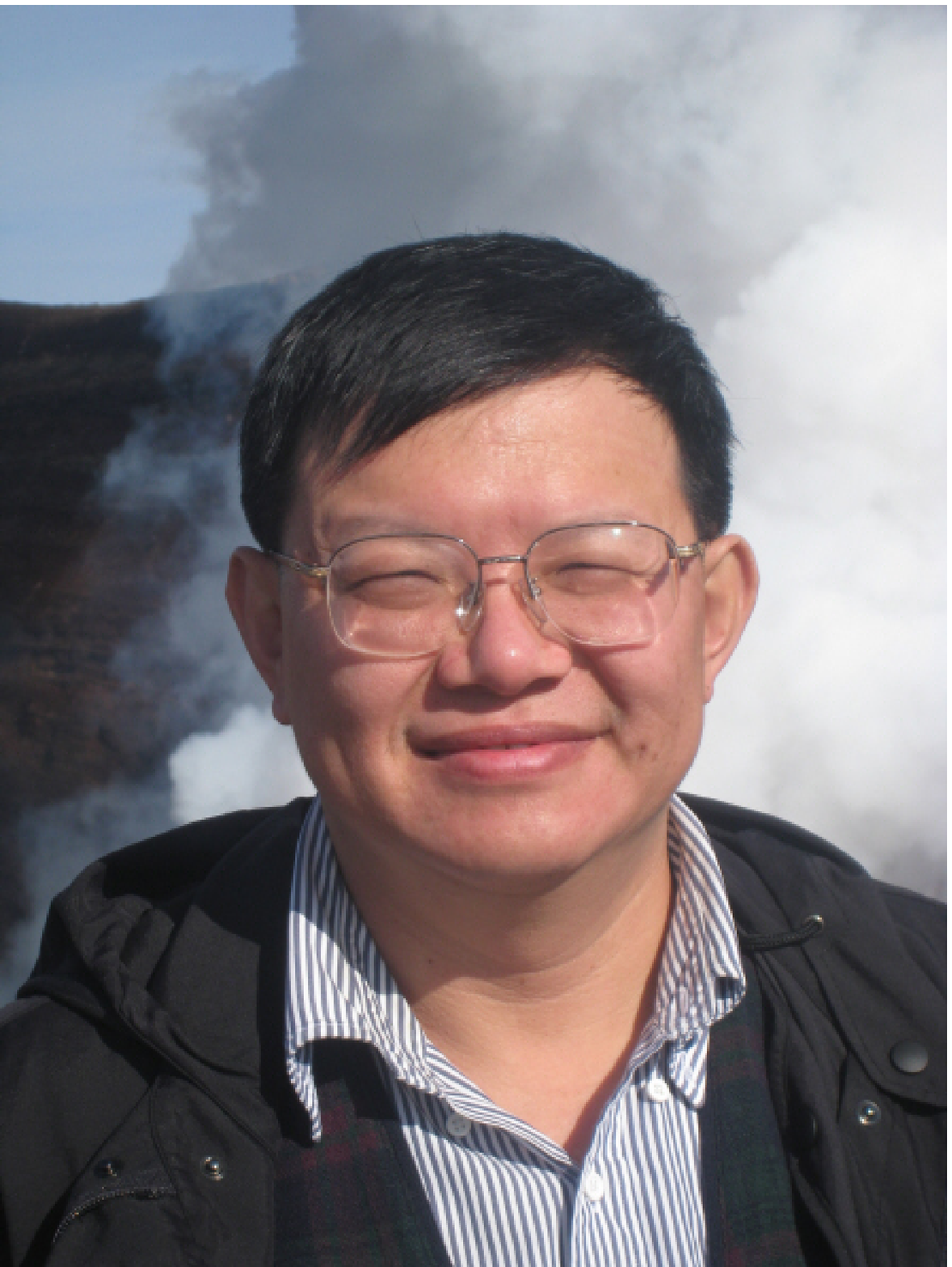}}]{Li-Chun Wang}
	(M'96 -- SM'06 -- F'11) received Ph. D. degree from the Georgia Institute of Technology, Atlanta, in 1996. From 1996 to 2000, he was with AT\&T Laboratories, where he was a Senior Technical Staff Member in the Wireless Communications Research Department. Since August 2000, he has joined the Department of Electrical and Computer Engineering of National Chiao Tung University in Taiwan and is jointly appointed by Department of Computer Science and Information Engineering of the same university.
	Dr. Wang was elected to the IEEE Fellow in 2011 for his contributions to cellular architectures and radio resource management in wireless networks. He won two Distinguished Research Awards of National Science Council, Taiwan in 2012 and 2017, respectively. He was the co-recipients of IEEE Communications Society Asia-Pacific Board Best Award (2015), Y. Z. Hsu Scientific Paper Award (2013), and IEEE Jack Neubauer Best Paper Award (1997).
	His current research interests are in the areas of software-defined mobile networks, heterogeneous networks, and data-driven intelligent wireless communications. He holds 19 US patents, and have published over 200 journal and conference papers, and co-edited a book, "Key Technologies for 5G Wireless Systems," (Cambridge University Press 2017).
\end{IEEEbiography}

%\vfill


\end{document}